\documentclass[aps,prc,preprint,amsfonts,amsmath,superscriptaddress,floatfix,nofootinbib]{revtex4-2}
\usepackage{graphicx,amsmath,epsfig}
\newcommand{\beqy}{\begin{eqnarray}}
\newcommand{\eeqy}{\end{eqnarray}}
\newcommand{\U}{\mathcal{U}}
\newcommand{\V}{\mathcal{V}}
\newcommand{\E}{\mathfrak{E}}
\newcommand{\D}{\Delta}

\newcommand{\rb}{\pmb{r}}
\newcommand{\rp}{\pmb{r^\prime}}

\newcommand{\sg}{\sigma}
\newcommand{\sgp}{\sigma^\prime}

\newcommand{\integ}{\int{\rm d}^3\pmb{r}}
\newcommand{\integp}{\int{\rm d}^3\pmb{r^\prime}}

\begin{document}

\title{Superfluid fraction and effective ion mass in the crystalline crust of a neutron star: role of interband response}

\author{N. Chamel}

\affiliation{Institut d'Astronomie et d'Astrophysique, CP-226, Universit\'e Libre de Bruxelles, 
1050 Brussels, Belgium}

\affiliation{Brussels Laboratory of the Universe, Belgium}

\begin{abstract}
Neutron superfluidity in the inner crust of a neutron star is further investigated, focusing on the role of the interband response in the superfluid fraction 
and the effective mass of crustal ions induced by their motion through the superfluid. Calculations are performed within the linear response theory of the 
self-consistent time-dependent Hartree-Fock-Bogoliubov equations with Skyrme nuclear energy density functionals in the Bardeen-Cooper-Schrieffer approximation. 
The absence of interband response in previous analyses is clarified. 
The neutron superfluid density is formally shown to be consistent with the entrainment matrix derived earlier in homogeneous neutron-proton superfluid
mixture, thus providing a unified description of entrainment effects in the inner crust and outer core of a neutron star within the same microscopic framework. 
The relative importance of the intraband and interband responses in different regions of the crust is numerically assessed from three-dimensional band-structure
calculations, taking into account quantum zero-point motion of ions about their equilibrium position. The neutron superfluid fraction is found to be enhanced by 
the interband response, resulting in effective ion masses that remain close to the mass of quantum mechanically bound nucleons for realistic neutron pairing gaps. 
Results are compared to predictions from classical hydrodynamics with different prescriptions for the permeability of ions to superfluidity. 
\end{abstract}


\maketitle

\section{Introduction}
\label{sec:introduction}

Detections of spin frequency glitches in pulsars such as Vela provide  evidence for the presence of a relative neutron 
superfluid flow in the inner crust of these neutron stars 
(see, e.g., Refs.~\cite{antonopoulou2022,zhou2022} for a review). This flow is sustained by the pinning of quantized vortices,
and is independently supported by observations of transiently accreting neutron stars~\cite{allard2024,allard2024b,allard2025}. 
The glitch activity depends on the ratio  between the coarse-grained neutron superfluid density $\rho_{n,s}$ (averaging spatial variations 
over scales large compared to the lattice spacing and the coherence length; see, e.g., Ref.~\cite{pethick2010}) 
and the average neutron mass density $\bar \rho_n$ in the different crustal layers~\cite{chamelcarter2006}. The neutron superfluid density relates
the average neutron mass current $\pmb{\bar \rho_n}$ to the average superfluid velocity $ \pmb{\bar V_{n,s}}$ through 
the equation $\pmb{\bar \rho_n}= \rho_{n,s}  \pmb{\bar V_{n,s}}$ in the rest frame of the crust~\cite{CCH06}. 
The superfluid density does not generally coincide with the mass density $\rho_{n,f}$ of ``free'' neutrons: some of them may be entrained by the crust
and therefore do not contribute to the superfluid flow. It is as if ``free'' neutrons were moving with an effective mass~\cite{CCH06} 
\begin{equation}\label{eq:effective-neutron-mass}
m_n^\star=m_n \frac{\rho_{n,f}}{\rho_{n,s}}\, .
\end{equation}

In general, the mass current may not be 
necessarily aligned with the superfluid velocity, in which case the superfluid density has to be replaced by a tensor. However, this tensor 
reduces to a scalar for the expected structure of the crust made of spherical clusters arranged in a body-centered cubic crystal lattice~\cite{chamelhaensel2008}. 
We will not consider here the possibility of a nuclear pasta mantle between the crust and the core, whose existence remains uncertain (see, e.g., Ref.~\cite{chamel2022} and references therein for a recent review). 
Although the superfluid response is no 
longer isotropic in such transition region, the influence of spatial inhomogeneities was found to be much smaller than in shallower
regions, considering the traditional phases of nuclear rods (``spaghetti''), slabs (``lasagna''), tubes (``bucatini''), and bubbles (``Swiss cheese'')\cite{CCH05a,chamel2005} (see also Refs.~\cite{kashiwaba2019,kenta2024,almirante2024} and Ref.~\cite{almirante2024b} for more recent calculations
in the lasagna and spaghetti phases respectively). Incidentally, the neutron-star region containing pasta has been recently found to be considerably 
reduced when quantum shell and pairing effects are added perturbatively on top of 4th-order extended Thomas-Fermi calculations~\cite{shchechilin2024}.

The same entrainment effects can be analyzed in the superfluid frame where the ions constituting the 
crust all move with a velocity $\pmb{v_I}=-\pmb{\bar V_{n,s}}$ and carry a mass current $n_I m_I^\star \pmb{v_I}$ as if they had an effective mass $m_I^\star$ 
($n_I$ is the ion number density). On the other hand, it follows from Galilean invariance that the mass density in this frame is also given by 
$(\rho_{n,s} -\bar \rho) \pmb{\bar V_{n,s}}$, where $\bar \rho$ denotes the total average mass density. Comparing the two expressions, we find 
\begin{equation}\label{eq:effective-ion-mass}
m_I^\star = \frac{\bar \rho- \rho_{n,s}}{n_I}\, , 
\end{equation}
which coincides with the expression given in Ref.~\cite{pethick2010}. The equivalence between the two points of view was formally shown in 
Ref.~\cite{CCH06}, where alternative definitions of effective masses and entrainment parameters were discussed (see Appendix~\ref{app:entrainment}). 
Unlike $m_n^\star$, both $\rho_{n,s}$ and $m_I^\star$ are well defined in that they do not depend on the 
fuzzy specification of ``free'' neutrons.

The determination of the superfluid density or the effective ion mass has 
implication for the pinning of quantized vortices to the crust~\cite{epsteinbaym1988},  the spin evolution of neutron stars~\cite{andersson2012,chamel2013,delsate2016},  their oscillations (see, e.g., Ref.~\cite{andersson2021} for a review), dynamical tides in binaries~\cite{passamonti2022}, the formation of the crust~\cite{dinhthi2023,dinhthi2023b}, pycnonuclear fusions~\cite{Meisel2018},  and neutron-star  cooling~\cite{chamelpagereddy2013,chamelpagereddy2016}. 
As a general feature of a spatially inhomogeneous superfluid~\cite{leggett1970} confirmed by terrestrial experiments with cold atoms~\cite{chauveau2023,tao2023,biagioni_measurement_2024}, $\rho_{n,s}<\bar \rho_n$ or equivalently $m_I^\star > \bar \rho_p/n_I$ using Eq.~\eqref{eq:effective-ion-mass}. 
The effective ion mass $m_I^\star$  
was first calculated analytically by Epstein and Baym from classical hydrodynamics considering a single spherical ion of radius $R$ in presence of an irrotational and incompressible 
uniform neutron flow and can be expressed as~\cite{epsteinbaym1988,epstein1988}
\begin{equation}\label{eq:effective-ion-mass-Epstein}
m^\star_I = m_I + \dfrac{(\lambda-1)^2}{\lambda+2}m_d\, .
\end{equation}
Here $m_d$ is the mass of the neutron superfluid displaced by the ion
\begin{equation}
m_d=\frac{4}{3}\pi R^3 \rho_{n,o} \, , 
\end{equation}
assuming a uniform neutron superfluid mass density $\rho_{n,o}$ outside the ion. 
The ion is taken to have uniform neutron and proton mass densities, $\rho_{n,i}$ and $\rho_{n,i}$ respectively. The ion mass is then defined as 
\begin{equation}\label{eq:ion-mass-transport}
m_I=m_I^0 - \frac{4}{3}\pi R^3  \rho_{s,i}\, ,
\end{equation}
allowing for some nucleons inside the ion with mass density $\rho_{s,i}\equiv \lambda \rho_{n,o}$ to participate in the superfluid motion. In other words, 
$m_I$ represents the mass of the nucleons transported by the ion. 
Here $m^0_I$ is the mass the nucleons located inside the ion 
\begin{equation}\label{eq:ion-mass-inside}
m^0_I=\frac{4}{3}\pi R^3 (\rho_{n,i}+\rho_{p,i})\, .
\end{equation}
In reality, ions do not have a sharp surface and this introduces some ambiguity in the extraction of $R$ from more  realistic nucleon distributions, as obtained from 
mean-field or semiclassical calculations. It is therefore more convenient to consider the 
dimensionless ratio 
\begin{equation}\label{eq:dimensionless-effective-ion-mass1}
\dfrac{m^\star_I}{m_I}=1 + \dfrac{(\lambda-1)^2}{\lambda+2}\dfrac{\rho_{n,o}}{\rho_{p,i}+\rho_{n,i}-\lambda\rho_{n,o}}\, .
\end{equation}
The mass density $\rho_{s,i}$ therefore the parameter $\lambda$ cannot be determined by hydrodynamics alone and require a more microscopic treatment. Epstein and Baym applied 
the Ginzburg-Landau theory~\cite{epsteinbaym1988} and found that the effective ion mass $m^\star_I$ is increased by a few percents at most compared to $m_I$. 
In subsequent studies, $\rho_{s,i}$ was treated as a free parameter. 
Sedrakian treated the ion as a rigid impenetrable obstacle~\cite{sedrakian1996}. Substituting $\rho_{s,i}=0$ in Eq.~\eqref{eq:effective-ion-mass-Epstein} remarking that $m_I = m^0_I$ yields
\begin{equation}\label{eq:effective-ion-mass-Sedrakian}
m^\star_I = m_I + \dfrac{1}{2}m_d= m_I\left(1+\dfrac{1}{2}\dfrac{\rho_{n,o}}{\rho_{n,i}+\rho_{p,i}}\right)\, . 
\end{equation}
Approximating $\rho_{n,i}+\rho_{p,i}$ by the nuclear saturation density leads to the same formula as given in Ref.~\cite{sedrakian1996}. 
On the other hand, Magierski \& Bulgac supposed that \emph{all} nucleons in the ion (including protons) contribute to the superflow~\cite{magierski2004}, i.e.
$\rho_{s,i}=\rho_{n,i}+\rho_{p,i}$. Introducing the parameter $\gamma=1/\lambda=\rho_{n,o}/(\rho_{n,i}+\rho_{p,i})$ in Eq.~\eqref{eq:effective-ion-mass-Epstein}, 
we recover the formula obtained in Ref.~\cite{magierski2004}, namely 
\begin{equation}\label{eq:effective-ion-mass-Magierski}
m^\star_I = m^0_I \dfrac{(1-\gamma)^2}{1+2\gamma} \, . 
\end{equation}
With this prescription, the mass of nucleons transported by the ion vanishes, $m_I=0$, while $m_I^0$ remains finite. 
However, neither Eq.~\eqref{eq:effective-ion-mass-Sedrakian} nor Eq.~\eqref{eq:effective-ion-mass-Magierski} are expected to be realistic: the neutron superfluid 
permeates the ion due to proximity effects (as shown from band-structure calculations in Ref.~\cite{chamel2010b}) whereas protons are deeply bound and moves with the ion. The mass density $\rho_{s,i}$ is therefore more plausibly equal or 
lower than the neutron mass density $\rho_{n,i}$. Martin \& Urban~\cite{martinurban2016} considered  $\rho_{s,i}=\rho_{n,i}$, in which case Eq.~\eqref{eq:effective-ion-mass-Epstein} reduces to 
\begin{equation}\label{eq:effective-ion-mass-MartinUrban}
m_I^\star = Z m_p + \frac{(1-\Upsilon)^2}{1+2\Upsilon} N m_n \, ,
\end{equation}
where $\Upsilon=\rho_{n,o}/\rho_{n,i}$. Here $Z$ and $N$ denote respectively the numbers of protons and neutrons located inside the ions and defined by
\begin{equation}\label{eq:nucleon-numbers}
Z=\frac{4}{3}\pi R^2 \frac{\rho_{p,i}}{m_p}\, , \hskip0.5cm N=\frac{4}{3}\pi R^2 \frac{\rho_{n,i}}{m_n} \, .
\end{equation}
Solving the hydrodynamics equations numerically for a neutron superfluid flowing through a perfect crystal lattice, they found that the effective ion mass remains given by Eq.~\eqref{eq:effective-ion-mass-MartinUrban} to a very good approximation, as anticipated in Ref.~\cite{epstein1988}.  The prescription $\rho_{s,i}=\rho_{n,i}$ implies that 
only protons are transported by ions, and the ion mass thus reduces to $m_I=Z m_p$.   However, it is not clear whether deeply bound neutrons could participate to the flow.  Martin \& Urban therefore extended Eq.~\eqref{eq:effective-ion-mass-MartinUrban} to allow for an arbitrary degree of permeability of the ions through the parameter $\delta=\rho_{s,i}/\rho_{n,i}$:  
\begin{equation}\label{eq:effective-ion-mass-MartinUrban2}
m_I^\star = Z m_p + N m_n \left(1-\delta+ \frac{(\delta-\Upsilon)^2}{\delta+2\Upsilon}\right) \, .
\end{equation}
This formula, which is equivalent to Eq.~\eqref{eq:effective-ion-mass-Epstein} with $\lambda=\delta/\Upsilon$, was later explicitly rederived in Ref.~\cite{dinhthi2023}. 
By applying numerically Eq.~\eqref{eq:effective-ion-mass-MartinUrban2}, Martin \& Urban showed that entrainment effects are quite sensitive to the assumed value for $\delta$. 
The different prescriptions proposed in the literature are summarized in Table~\ref{tab:hydrodynamics-ion-mass}. In any case, the hydrodynamic description predicts an enhancement of the effective ion mass $m_I^\star$ if normalized by $m_I$, as can be easily seen from Eq.~\eqref{eq:dimensionless-effective-ion-mass1}.

The validity of the hydrodynamics description implicitly requires sufficiently strong pairing for the coherence length to be much smaller than the ion size, implying that neutron pairs outside ions are not correlated to neutron pairs inside so that superfluidity does not permeate the ions as originally assumed in Ref.~\cite{sedrakian1996}. Setting $\rho_{s,i}=0$ would thus be the most consistent prescription with the hydrodynamics description. The corresponding neutron superfluid fraction $\rho_{n,s}/\bar \rho_n$ as obtained from Ref.~\cite{martinurban2016} ranges from about 0.6 to 0.8 depending on the crustal layer. In the opposite limit of vanishingly small pairing, three-dimensional band-structure
 calculations predicted a very small superfluid fraction of about 0.1 in the intermediate region of the inner crust~\cite{chamel2005,chamel2012}. This result was expected to remain unchanged when Bardeen-Cooper-Schrieffer (BCS) pairing is included~\cite{CCH05b}, as confirmed by recent numerical calculations~\cite{chamel2025}. The validity of the BCS approximation was questioned in Refs.~\cite{watanabe2017,minami2022} based on one-dimensional toy models (but see the discussion in Ref.~\cite{chamel2025}).
The suppression of the superfluid fraction found in band-structure calculations translates into a large effective ion mass~\cite{chamel2025b} at variance with the effective mass extracted dynamically from three-dimensional time-dependent Hartree-Fock-Bogoliubov (HFB) simulations of a single ion moving through a neutron superfluid~\cite{pecak2024}. Almirante \& Urban~\cite{almirante2025,almirante2026} have recently shown that this discrepancy does not originate from the BCS approximation but from the previously neglected interband response, as confirmed by HFB calculations of a stationary neutron superflow in a three-dimensional rigid crystal lattice~\cite{almirantekaskitsiurban2026}. 
These calculations have also revealed that the ions are not impenetrable as would be expected from a strict application of the hydrodynamics description: the superfluid is found to flow through the ions (with a reduced velocity) rather than around them. The weaker entrainment effects resulting in a reduced ion effective mass suggest that lattice vibrations could be more important than estimated in Ref.~\cite{chamel2025b}.

This paper is the third in a series about the neutron superfluid density in the inner crust of a neutron star. The first paper~\cite{chamel2025} focused 
on the role of BCS pairing. The second paper~\cite{chamel2025b} addressed the importance of quantum zero-point fluctuations of ions about their equilibrium 
position and their effective mass. In the present study, the interband response is examined. The associated contribution to the neutron superfluid density and the effective ion mass is numerically evaluated in different crustal layers taking into account lattice vibrations. 

The model of neutron-star crust and the calculations of the superfluid fraction are described in Sec.~\ref{sec:micro}. The numerical implementation and the results 
are presented and discussed in Sec.~\ref{sec:results}. Concluding remarks are given in Sec.~\ref{sec:conclusion}.

\begin{table}
\centering
\caption{Prescriptions proposed in the literature for the mass density $\rho_{s,i}$ of nucleons inside ions participating to the superflow in the hydrodynamic description of superfluidity in neutron-star crusts. $m_I$ is the corresponding ``bare'' ion mass as calculated from Eq.~\eqref{eq:ion-mass-transport}. $Z$ and $N$ denote respectively the numbers of protons and neutrons inside ions, as defined by Eq.~\eqref{eq:nucleon-numbers}. }
\label{tab:hydrodynamics-ion-mass}
\begin{tabular} {|ccc|}
\hline
                             & $\rho_{s,i}$            & $m_I$    \\
\hline
Sedrakian (1996)             & 0                       & $Z m_p+N m_n$   \\
Magierski \& Bulgac (2004)   & $\rho_{n,i}+\rho_{p,i}$ & 0         \\ 
Martin \& Urban (2016)       & $\rho_{n,i}$            & $Z m_p$   \\
\hline
\end{tabular}
\end{table}

\section{Microscopic framework}
\label{sec:micro}

\subsection{Time-dependent Hartree-Fock-Bogoliubov theory}
\label{sec:TDHFB}

In the traditional approach presented in classical textbooks~\cite{ringschuck1980,blaizot1986}, the energy $E$ of a nucleon-matter element of volume $\Omega$ is expressed as a function of the one-body density matrix $n_q^{ij}$ and pairing tensor $\kappa_q^{ij}$ for nucleons of charge type $q=n,p$ for neutron, proton respectively (using the symbol $*$ for complex conjugation) 
\begin{equation}\label{eq:density-matrix-def}
	n_q^{ij}=\langle \Psi \vert c_q^{j\dagger} c_q^i \vert \Psi \rangle = n_q^{ji*} \, ,
\end{equation}
\begin{equation}
	\kappa_q^{ij}=\langle \Psi \vert  c_q^j c_q^i \vert \Psi \rangle=-\kappa_q^{ji} \, ,
\end{equation}
defined in terms of the creation and destruction operators $c_q^{i\dagger}$ and $c_q^i$ (using the symbol $\dagger$ for Hermitian conjugation) of a single particle in a state characterized by quantum numbers $i$.  Here $\Psi$ is the many-nucleon state.  

Introducing the Bogoliubov transformation
\beqy
\begin{pmatrix} b_q^i \\ b_q^{i\dagger}\end{pmatrix} = \sum_{j}\begin{pmatrix}\U^{(q)*}_{ij} & \V^{(q)*}_{ij} \\ \V^{(q)}_{ij} & \U^{(q)}_{ij}\end{pmatrix}\begin{pmatrix} c_q^j \\ c_q^{j\dagger}\end{pmatrix} \, , 
\eeqy
where $b_q^{i\dagger}$ and $b_q^i$ are creation and destruction operators of a quasiparticle in a quantum state $i$, the TDHFB equations can be written as ($\delta^{ij}$ is the Kronecker's symbol)
\beqy
i\hbar\frac{\partial \U^{(q)}_{ki}}{\partial t}=\sum_{j}\bigl[(h_q^{ij}-\mu_q\delta^{ij})\U^{(q)}_{kj}+\D^{ij}\V^{(q)}_{kj}\bigr]\, ,
\label{eq:TDHFB-Uij}
\eeqy
\beqy
i\hbar\frac{\partial \V^{(q)}_{ki}}{\partial t}=\sum_{j}\bigl[-\D^{ij*}\U^{(q)}_{kj}-(h_q^{ij*}-\mu_q \delta^{ij})\V^{(q)}_{kj}\bigr] \, ,
\label{eq:TDHFB-Vij}
\eeqy
where $\mu_q$ denotes the chemical potential of nucleon species $q$, which is fixed by the requirement of fixed particle number $N_q$: 
\begin{equation}\label{eq:particle-number}
N_q=\sum_i n_q^{ii} \, .
\end{equation}

The fermionic algebra of the (quasi)particle operators leads to the following identities
\beqy \label{eq:constraint1}
 \sum_k  \left(\U^{(q)}_{ik}\U^{(q)*}_{jk} + \V^{(q)}_{ik}\V^{(q)*}_{jk}\right)=\delta_{ij}\, , \qquad \sum_k \left(\U^{(q)}_{ik}\V^{(q)}_{jk} + \V^{(q)}_{ik}\U^{(q)}_{jk}\right)=0\, , 
\eeqy 
\beqy \label{eq:constraint2}
 \sum_{k}\left(\U^{(q)*}_{ki}\U^{(q)}_{kj}+\V^{(q)}_{ki}\V^{(q)*}_{kj}\right)=\delta_{ij} , \qquad \sum_{k}\left(\U^{(q)*}_{ki}\V^{(q)}_{kj}+\V^{(q)}_{ki}\U^{(q)*}_{kj}\right)=0 .
\eeqy 
The one-body density matrix and the pairing tensor can be expressed in terms of the quasiparticle components as
\beqy\label{eq:DensityMatrix}
n_q^{ij} = \sum_k  \V^{(q)*}_{ki}\V^{(q)}_{kj}\, ,
\eeqy
\beqy
\kappa_q^{ij} =\sum_k \V^{(q)*}_{ki}\U^{(q)}_{kj} \, .
\eeqy

The matrices  $h_q^{ij}$ and  $\D_q^{ij}$ of the single-particle Hamiltonian and the pair potential, respectively, are defined as 
\beqy\label{eq:Hamiltonian-matrix}
h_q^{ij}=\frac{\partial E}{\partial n_q^{ji}}=h_q^{ji*}
\, , 
\eeqy
\beqy\label{eq:pairing-matrix}
\D_q^{ij}=\frac{\partial  E}{\partial \kappa_q^{ij*}}=-\D_q^{ji}
\, .
\eeqy

The TDHFB Eqs.~\eqref{eq:TDHFB-Uij} and \eqref{eq:TDHFB-Vij} can be equivalently expressed as
\beqy
\label{eq:TDHFB1}
i\hbar \frac{\partial n_q^{ij}}{\partial t}=\sum_k \left(h_q^{ik}n_q^{kj}-n_q^{ik}h_q^{kj}+\kappa_q^{ik}\Delta_q^{kj*} -\Delta_q^{ik}\kappa_q^{kj*}\right) \, ,
\eeqy
\beqy
\label{eq:TDHFB2}
i \hbar \frac{\partial \kappa_q^{ij}}{\partial t}=\sum_k  \left[(h_q^{ik}-\mu_q\delta^{ik})\kappa_q^{kj}+\kappa_q^{ik}(h_q^{kj*}-\mu_q \delta^{kj})-\Delta_q^{ik}n_q^{kj*} - n_q^{ik}\Delta_q^{kj}\right]+ \Delta_q^{ij}\, .
\eeqy

The TDHFB theory can be alternatively formulated in coordinate space using the particle density and pair density  matrices introduced in Refs.~\cite{dobaczewski1984,dobaczewski1996} and defined by
\begin{equation}\label{eq:ParticleDensityMatrixDefinition}
n_q(\rb, \sigma; \pmb{r^\prime}, \sigma^\prime; t) = <c_q(\pmb{r^\prime},\sigma^\prime;t)^\dagger c_q(\rb,\sigma;t)>\, ,
\end{equation}
\begin{equation}\label{eq:PairDensityMatrixDefinition}
\widetilde{n}_q(\rb, \sigma; \pmb{r^\prime}, \sigma^\prime;t) = -\sigma^\prime <c_q(\pmb{r^\prime},-\sigma^\prime;t) c_q(\rb,\sigma;t)>\, , 
\end{equation}
respectively where $c_q(\rb,\sigma; t)^\dagger$ and $c_q(\rb,\sigma; t)$ are the creation and destruction operators for nucleons of charge type $q$  at position $\rb$ with spin projection $\sigma=\pm1$ (in units of $\hbar/2$) at time $t$.  

Introducing single-particle basis wavefunctions $\varphi^{(q)}_i(\rb,\sg)$ with a set of quantum numbers $i$, the creation and annihilation operators in coordinate space can be expressed as 
\begin{align}
&c_q(\rb,\sigma)=\sum_i\varphi^{(q)}_{i} (\rb,\sg)c_q \, , 
&c_q(\rb,\sigma)^\dagger=\sum_i\varphi^{(q)}_{i} (\rb,\sg)^* c_q^\dagger \, .
\end{align}
The two different representations of the particle and pair density matrices are thus related by 
\begin{equation}\label{eq:DensityMatrixCoordinateSpaceDef}
n_q(\rb, \sg; \rp, \sgp; t)=\sum_{i,j}n^{ij}_{q}(t)\varphi^{(q)}_i (\rb,\sg)\varphi^{(q)}_j (\rp,\sgp)^{*}\, ,
\end{equation}
\begin{align}\label{eq:AbnormalDensityMatrixCoordinateSpaceDef}
\widetilde{n}_q(\rb, \sg; \rp, \sgp; t)=\sum_{i,j}\kappa^{ij}_{q}(t)\varphi^{(q)}_i(\rb,\sg)\varphi^{(q)}_{\bar j}(\rp,\sgp)^* \, ,
\end{align}
\begin{equation}\label{eq:DensityMatrixCoordinateSpaceDef2}
n^{ij}_{q}(t)=\sum_{\sg,\sgp}\integ\integp\ n_q(\rb, \sg; \rp, \sgp; t) \varphi^{(q)}_i (\rb,\sg)^*\varphi^{(q)}_j (\rp,\sgp)\, ,
\end{equation}
\begin{align}
\kappa^{ij}_{q}(t)=\sum_{\sg,\sgp}  \integ\integp\ \widetilde{n}_q(\rb, \sg; \rp, \sgp; t)\varphi^{(q)}_i (\rb,\sg)^*\varphi^{(q)}_{\bar j} (\rp,\sgp) \, ,
\end{align}
where we have introduced the time-reversed state of $j$ denoted by $\bar j$ and whose single-particle wavefunction is given by 
\begin{equation}\label{eq:OneParticleBasisTimeReversed}
\varphi^{(q)}_{\bar j} (\rb,\sg)=- \sg \varphi^{(q)}_{j} (\rb,-\sg)^*\, .
\end{equation}
Note that $\varphi^{(q)}_{\bar{\bar{j}}}(\rb,\sg)=-\varphi^{(q)}_{j}(\rb,\sg)$. 
Using Eq.~\eqref{eq:DensityMatrixCoordinateSpaceDef}, the hermiticity of the density matrix $n_q^{ij}=n_q^{ji*}$ is easily transposed in coordinate space
\begin{equation}
n_q(\rb, \sg; \rp, \sgp; t)=n_q(\rp, \sgp;\rb, \sg;  t)^*\, .
\end{equation}
The antisymmetry of the pairing tensor $\kappa_q^{ij}=-\kappa_q^{ji}$  leads to the following relation
\begin{equation}
\widetilde{n}_q(\rb, \sg; \rp, \sgp; t)=\sg\sgp \widetilde{n}_q(\rp, -\sgp;\rb, -\sg;  t)\, ,
\end{equation}
as can be seen from Eq.~\eqref{eq:AbnormalDensityMatrixCoordinateSpaceDef}.

For the Skyrme type functionals considered here (see, e.g., Ref.~\cite{bender03}), the energy $E$ depends on the following local densities and currents:

(i) the nucleon number density at position $\rb$ and time $t$, 
\beqy\label{eq:local-density}
n_q(\rb,t)=\sum_{\sg=\pm 1} n_q(\rb,\sg;\rb,\sg;t) \, ,
\eeqy

(ii) the kinetic-energy density (in units of $\hbar^2/2m_q)$ at position $\rb$ and time $t$, 
\beqy\label{eq:kinetic-density}
\tau_q(\rb,t)=\sum_{\sg=\pm 1}\int\text{d}^3\rp\; \delta (\rb-\rp) \pmb{\nabla}\cdot\pmb{\nabla^\prime} n_q(\rb,\sg;\rp,\sg;t)\, ,
\eeqy

(iii) the momentum density (in units of $\hbar$) at position $\rb$  and time $t$,
\beqy
\label{eq:momentum-density}
\pmb{j_q}(\rb,t)=-\frac{ i}{2}\sum_{\sigma=\pm 1}\int\,{\rm d}^3\rp\,\delta(\rb-\rp) (\pmb{\nabla} -\pmb{\nabla^\prime})n_q(\rb, \sigma; \rp, \sigma;t)\, ,
\eeqy 

(iv) the pair density at position $\rb$ and time $t$,
\beqy
\widetilde{n}_q(\rb,t)=\sum_{\sg = \pm 1}\widetilde{n}_q (\rb,\sg ; \rb,\sg;t )\, .
\eeqy 

Using Eq.~\eqref{eq:DensityMatrixCoordinateSpaceDef}, as well as 
the orthonormality property of the single-particle wavefunctions 
\beqy\label{eq:orthonormality}
\sum_\sg \integ\, \varphi^{(q)}_i(\rb,\sg)\varphi^{(q)}_j(\rb,\sg)^* = \delta_{ij}\, ,
\eeqy 
the condition~\eqref{eq:particle-number}  ensures that 
\begin{equation}
\integ\,  n_q(\rb;t) = N_q\, .
\end{equation}

The TDHFB equations can be written in coordinate space as 
\beqy
{\rm i}\hbar\frac{\partial}{\partial t} \begin{pmatrix} \psi^{(q)}_{1k}(\rb,\sigma; t) \\ \psi^{(q)}_{2k}(\rb,\sigma; t)\end{pmatrix} =\begin{pmatrix}h_q(\rb,t)-\mu_q & \Delta_q(\rb,t) \\ \Delta_q(\rb,t)^* & - h_q(\rb,t)^* + \mu_q \end{pmatrix}\begin{pmatrix} \psi^{(q)}_{1k}(\rb,\sg; t) \\ \psi^{(q)}_{2k}(\rb,\sg; t)\end{pmatrix}\, ,
\label{eq:TDHFB-Russian}
\eeqy
with the single-particle Hamiltonian  
\begin{align}
\label{eq:Hamiltonian}
&h_q(\rb,t)= -\pmb{\nabla}\cdot\frac{\hbar^2}{2m_q^{\oplus}(\rb,t)}\pmb{\nabla}+U_q(\rb,t)-\frac{i}{2}\left[\pmb{I_q}(\rb,t)\cdot\pmb{\nabla}+\pmb{\nabla}\cdot \pmb{I_q}(\rb,t)\right]
\end{align}
\begin{align}\label{eq:Fields1}
 &\frac{\hbar^2}{2m_q^{\oplus}(\rb,t)}=\frac{\delta E}{\delta \tau_q(\rb,t)}\,,  \qquad U_q(\rb,t)=\frac{\delta E}{\delta n_q(\rb,t)}\,, \qquad \pmb{I_q}(\rb,t)=\frac{\delta E}{\delta \pmb{j_q}(\rb,t)}\,, 
\end{align}
and the pair potential
\beqy
\label{eq:pair-pot}
\D_q(\rb,t) =2\frac{\delta E}{\delta \widetilde{n}_q(\rb,t)^*}\, .
\eeqy
The spin-orbit coupling is much weaker in the inner crust of a neutron star than in ordinary nuclei (see, e.g., Ref.~\cite{pearson2018}) and has therefore been neglected for simplicity, as in previous HFB calculations~\cite{kenta2024,almirante2024,almirante2024b,almirante2025}. 

The single-particle Hamiltonian matrix~(\ref{eq:Hamiltonian-matrix}) and pairing matrix~(\ref{eq:pairing-matrix}) are given by the matrix elements of the corresponding fields \eqref{eq:Hamiltonian} and \eqref{eq:pair-pot} 
\begin{align}\label{eq:Hamiltonian-matrix2}
h_q^{ij}(t)=\sum_{\sg}\int {\rm d}^3\rb \,  \varphi^{(q)}_i(\rb,\sigma^\prime)^*h_q(\rb,t)\,\varphi^{(q)}_j(\rb,\sigma)\, , 
\end{align}
\begin{align}\label{eq:pairing-matrix2}
\D^{ij}(t)=\sum_{\sg} \int {\rm d}^3\rb \,  \varphi^{(q)}_i(\rb,\sigma)^*\D_q(\rb,t)\varphi^{(q)}_{\bar j}(\rb,\sigma)\, .
\end{align}

The quasiparticle wavefunction is related to the matrices introduced previously through 
\begin{equation}\label{eq:psi1}
\psi^{(q)}_{1k}(\rb,\sg; t)=\sum_j \varphi^{(q)}_j(\rb,\sg)\, \U^{(q)}_{kj} \, ,
\end{equation}
\begin{equation}\label{eq:psi2}
\psi^{(q)}_{2k}(\rb,\sg; t)=\sum_j (-\sg)\varphi^{(q)}_j(\rb,-\sg)^*\, \V^{(q)}_{kj} =\sum_j \varphi^{(q)}_{\bar j}(\rb,\sg)\, \V^{(q)}_{kj} \, . 
\end{equation}
Using the orthonormality property~\eqref{eq:orthonormality}, the fermionic algebra conditions~\eqref{eq:constraint1} and \eqref{eq:constraint2} read in coordinate space: 
\begin{equation}
\sum_\sg \integ\,  \left[\psi_{1i}^{(q)}(\rb,\sg; t) \psi_{1j}^{(q)}(\rb,\sg; t)^* 
+\psi_{2i}^{(q)}(\rb,\sg; t) \psi_{2j}^{(q)}(\rb,\sg; t)^* \right] =\delta_{ij} \, ,
\end{equation}
\begin{equation}
\sum_\sg \integ\, \left[ \psi_{1\bar{i}}^{(q)}(\rb,\sg; t)^* \psi_{2j}^{(q)}(\rb,\sg; t) 
+\psi_{1\bar{j}}^{(q)}(\rb,\sg; t)^* \psi_{2i}^{(q)}(\rb,\sg; t) \right] =0 \, , 
\end{equation}
\begin{equation}
\sum_k \left[\psi_{1k}^{(q)}(\rb,\sg; t)^* \psi_{1k}^{(q)}(\rp,\sgp; t) 
+\psi_{2\bar{k}}^{(q)}(\rb,\sg; t)^* \psi_{2\bar{k}}^{(q)}(\rp,\sgp; t) \right] =\delta(\rb-\rp)\delta_{\sg\sgp} \, ,
\end{equation}
\begin{equation}
\sum_k \left[\psi_{1k}^{(q)}(\rb,\sg; t)^* \psi_{2k}^{(q)}(\rp,\sgp; t) 
-\psi_{2\bar{k}}^{(q)}(\rb,\sg; t)^* \psi_{1\bar{k}}^{(q)}(\rp,\sgp; t) \right] =0 \, .
\end{equation}

The particle and pair density matrices can be equivalently written as
\beqy\label{eq:density-matrix2}
n_q(\rb, \sg; \rp, \sgp; t)=\sum_k \sg\sgp\psi_{2k}^{(q)}(\rb,-\sg; t)^* \psi_{2k}^{(q)}(\rp,-\sgp; t)\, ,
\eeqy
\beqy\label{eq:pair-density-matrix2}
\widetilde{n}_q(\rb, \sg; \rp, \sgp; t)=\sum_k \sg\sgp\psi_{2k}^{(q)}(\rb,-\sg; t)^* \psi_{1k}^{(q)}(\rp,-\sgp; t) \, .
\eeqy

The pair density $\widetilde{n}_q(\rb,t)=\vert\widetilde{n}_q(\rb,t)\vert \exp[i\phi_q(\rb,t)]$ physically represents the local (complex) order parameter of the superfluid phase. It defines the superfluid velocity as follows (see, e.g., Ref.~\cite{allard2021}) 
\begin{align}
	\pmb{V_{q,s}}(\rb,t)=\frac{\hbar}{2m_q} \pmb{\nabla}\phi_q(\rb,t)\, .
\end{align}
The mass current $\pmb{\rho_q}(\rb,t)$ is given by~\cite{ChamelAllard2019,allard2021} 
\beqy\label{eq:local-mass-current}
\pmb{\rho_q}(\rb,t) =\frac{m_q}{m_q^\oplus(\rb,t)} \hbar \pmb{j_q}(\rb,t)+\rho_q(\rb,t)\frac{\pmb{I_q}(\rb,t)}{\hbar} \, ,
\eeqy
where $\rho_q(\rb,t)=m_q n_q(\rb,t)$ is the local mass density, 
\beqy\label{eq:local-mass-density}
\rho_q(\rb,t) =m_q \sum_k \sum_\sg \psi_{2k}^{(q)}(\rb,\sg; t)^* \psi_{2k}^{(q)}(\rb,\sg; t) \, .
\eeqy
The average mass density is thus given by 
\beqy 
\bar \rho_q &=&\frac{1}{\Omega}\int {\rm d}^3\rb\,\rho_q(\rb,t) \notag \\ 
&=& \frac{m_q}{\Omega} \sum_k \sum_\sg \int {\rm d}^3\rb\,  \psi^{(q)}_{2k}(\rb,\sg; t)  \psi^{(q)}_{2k}(\rb,\sg; t)^* \notag \\ 
&=& \frac{m_q}{\Omega}\sum_{i,j}\sum_k  \V^{(q)*}_{k i} \V^{(q)}_{k j} \sum_\sg \int {\rm d}^3\rb\, \varphi^{(q)}_{\bar j}(\rb,\sg)\, \varphi^{(q)}_{\bar i}(\rb,\sg)^* \notag \\
&=& \frac{m_q}{\Omega}\sum_{i,j}n_q^{ij} \sum_\sg \int {\rm d}^3\rb\, \varphi^{(q)}_{ j}(\rb,\sg)^*\, \varphi^{(q)}_{i}(\rb,\sg) \notag \\
&=&  \frac{m_q}{\Omega} \sum_{i} n_q^{ii}\, ,
\eeqy 
using Eqs.~\eqref{eq:psi2} in the third line,  Eqs.~\eqref{eq:DensityMatrix} and \eqref{eq:OneParticleBasisTimeReversed} in the fourth line,  changing $\sg\rightarrow -\sg$ in the summation over spin recalling $\sg^2=1$ and using Eq.~\eqref{eq:orthonormality}  in the last line.  
From the definition~\eqref{eq:momentum-density} and Eq.~\eqref{eq:density-matrix2}, the momentum density can be equivalently expressed as  
\beqy
\pmb{j_q}(\rb,t)&=&-\frac{ i}{2}\sum_k \sum_{\sg} \left[  \psi_{2k}^{(q)}(\rb,\sg; t) \pmb{\nabla}\psi_{2k}^{(q)}(\rb,\sg; t)^*- \psi_{2k}^{(q)}(\rb,\sg; t)^*  \pmb{\nabla}\psi_{2k}^{(q)}(\rb,\sg; t) \right] \notag \\
&=&-\frac{ i}{2}\sum_{i,j,k}\V^{(q)}_{k j}\V^{(q)*}_{k i} \sum_{\sg} \left[ \varphi^{(q)}_{\bar j}(\rb,\sg) \pmb{\nabla}\varphi^{(q)}_{\bar i}(\rb,\sg)^*  - \varphi^{(q)}_{\bar i}(\rb,\sg)^*\pmb{\nabla}\varphi^{(q)}_{\bar j}(\rb,\sg) \right]  \notag \\ 
&=&-\frac{ i}{2}\sum_{i,j}n_q^{ij} \sum_{\sg}  \left[(-\sg) \varphi^{(q)}_{j}(\rb,-\sg)^* \pmb{\nabla}(-\sg)\varphi^{(q)}_{i}(\rb,-\sg)  - (-\sg)\varphi^{(q)}_{i}(\rb,-\sg)\pmb{\nabla}(-\sg)\varphi^{(q)}_{j}(\rb,-\sg)^* \right]\notag \\ 
&=&\frac{1}{2}\sum_{i,j}n_q^{ij} \sum_{\sg} \left[  \varphi^{(q)}_{j}(\rb,\sg)^* \frac{\pmb{p}}{\hbar}\varphi^{(q)}_{i}(\rb,\sg)  - \varphi^{(q)}_{i}(\rb,\sg)\frac{\pmb{p}}{\hbar}\varphi^{(q)}_{j}(\rb,\sg)^* \right] \, ,
\eeqy 
where we have used Eq.~\eqref{eq:psi2} in the second line,  Eqs.~\eqref{eq:DensityMatrix} and \eqref{eq:OneParticleBasisTimeReversed} in the third line, and changed in the last line $\sg\rightarrow-\sg$ in the summation over spin and introduced the momentum operator $\pmb{p}=-i\hbar \pmb{\nabla}$. 
The coarse-grained average mass current is given by 
\beqy\label{eq:average-mass-current1}
\pmb{\bar \rho_q} &=&\frac{1}{\Omega}\int {\rm d}^3\rb\,\pmb{\rho_q}(\rb,t) \notag \\ 
&=&\frac{m_q}{\Omega} \sum_k \sum_\sigma \int {\rm d}^3\rb\,  \psi^{(q)}_{2k}(\rb,\sg; t) \pmb{v_q}(\rb,t) \psi^{(q)}_{2k}(\rb,\sg; t)^* 
\eeqy 
in the coordinate-space representation, or 
\beqy\label{eq:average-mass-current2} 
\pmb{\bar \rho_q} &=&\frac{m_q}{\Omega} \sum_{i,j,k} \V^{(q)}_{k j}\V^{(q)*}_{k i} \sum_\sg \int {\rm d}^3\rb\,  \varphi^{(q)}_{\bar j}(\rb,\sg)\pmb{v_q}(\rb,t)\varphi^{(q)}_{\bar i}(\rb,\sg)^* \, , \\
&=& \frac{m_q}{\Omega} \sum_{i,j} n_q^{ij} \pmb{v_q^{ji}}
\eeqy 
in the traditional matrix formulation,
where 
\beqy
\pmb{v_q^{ji}} = \sum_\sg \int {\rm d}^3\rb\,  \varphi^{(q)}_{j}(\rb,\sg)^*\pmb{v_q}(\rb,t)\varphi^{(q)}_{i}(\rb,\sg) 
\eeqy 
denotes the matrix elements of the velocity operator
\beqy
\pmb{v_q}(\rb,t)= \frac{1}{i \hbar}\left[\rb h_q(\rb,t)-h_q(\rb,t)\rb\right] = \frac{1}{2m_q^\oplus(\rb,t)}\pmb{p} + \pmb{p}\frac{1}{2m_q^\oplus(\rb,t)}+\frac{1}{\hbar}\pmb{I_q}(\rb,t) \, .
\eeqy

\subsection{Application to neutron superfluidity in neutron-star crusts}
\label{sec:TDHFB-crust}

Let us consider a stationary neutron superfluid flow in the rest frame of the crust,  assumed to be a perfect crystal characterized by primitive translation vectors $\pmb{a_1}$, $\pmb{a_2}$, and $\pmb{a_3}$. All the densities and currents hence also the mean fields are time-independent in the crust frame.  Therefore,  it is enough to solve the stationary HFB equations (substituting  $i\hbar \partial/\partial t$ by the quasiparticle energies $\E$) 
\beqy\label{eq:HFB-crust-frame}
\begin{pmatrix}h_n(\rb)-\mu_n & \Delta_n(\rb) \\ \Delta_n(\rb)^* & - h_n(\rb)^* + \mu_n \end{pmatrix}\begin{pmatrix} \psi_{1}(\rb,\sg) \\ \psi_{2}(\rb,\sg)\end{pmatrix}=\E \begin{pmatrix} \psi_{1}(\rb,\sigma) \\ \psi_{2}(\rb,\sigma)\end{pmatrix}\, ,
\eeqy
in a sufficiently large volume $\Omega$ (containing a macroscopic number of lattice sites) with Born-von K\'arm\'an periodic boundary conditions.  

The single-particle Hamiltonian $h_n(\rb)$ is invariant under lattice translations, i.e.,  $h_n(\rb+\pmb{\ell})=h_n(\rb)$ for any lattice translation vector $\pmb{\ell}=n_1\pmb{a_1}+n_2\pmb{a_2}+n_3\pmb{a_3}$, where $(n_1,n_2,n_3)$ are arbitrary integers.  The pair potential is not periodic,  but can be expressed as $\D_n(\rb)=\D_{\pmb{Q}}(\rb)e^{2 i\pmb{Q}\cdot\rb}$ where $\D_{\pmb{Q}}(\rb)$ is periodic (but not necessarily real). 
The average superfluid velocity is thus given by 
\beqy\label{eq:average-superfluid-velocity}
\pmb{\bar V_{n,s}}\equiv\frac{1}{\Omega}\int {\rm d}^3\rb\, \pmb{V_{n,s}}(\rb)=\frac{\hbar \pmb{Q}}{m_n} \, .
\eeqy

Introducing the shifted quasiparticle wavefunctions
\begin{equation}\label{eq:psi1tilde}
\widehat{\psi}_{1}(\rb,\sg)=\psi_{1}(\rb,\sg)e^{-i \pmb{Q}\cdot \rb}\, ,
\end{equation}
\begin{equation}\label{eq:psi2tilde}
\widehat{\psi}_{2}(\rb,\sg)=\psi_{2}(\rb,\sg) e^{ i \pmb{Q}\cdot \rb}\, ,
\end{equation}
the HFB equations~\eqref{eq:HFB-crust-frame} reduce to 
\beqy\label{eq:HFB-superframe}
\begin{pmatrix}h_{\pmb{Q}}(\rb) -\mu_n & \D_{\pmb{Q}}(\rb) \\ \D_{\pmb{Q}}(\rb)^* & - h_{\pmb{Q}}(\rb)^* + \mu_n \end{pmatrix}\begin{pmatrix} \widehat{\psi}_{1}(\rb,\sg) \\ \widehat{\psi}_{2}(\rb,\sg)\end{pmatrix}=\E\begin{pmatrix} \widehat{\psi}_{1}(\rb,\sigma) \\ \widehat{\psi}_{2}(\rb,\sigma)\end{pmatrix} \, ,
\eeqy
where $h_{\pmb{Q}}(\rb)\equiv e^{-i \pmb{Q}\cdot \rb} h_n(\rb) e^{i \pmb{Q}\cdot \rb}$.  Since now both $h_{\pmb{Q}}(\rb)$ and $\D_{\pmb{Q}}(\rb)$ are invariant under lattice translations,
the quasiparticle wavefunction must obey the Floquet-Bloch theorem~\cite{kittel}: 
\begin{equation}\label{eq:psi1-Floquet-Bloch}
\widehat{\psi}_{1\pmb{k}}(\rb+\pmb{\ell},\sg)=e^{ i \pmb{k}\cdot \pmb{\ell}}\widehat{\psi}_{1\pmb{k}}(\rb,\sg)\, ,
\end{equation}
\begin{equation}\label{eq:psi2-Floquet-Bloch}
\widehat{\psi}_{2\pmb{k}}(\rb+\pmb{\ell},\sg)=e^{ i \pmb{k}\cdot \pmb{\ell}} \widehat{\psi}_{2\pmb{k}}(\rb,\sg)\, ,
\end{equation}
for any lattice translation vector $\pmb{\ell}$.  For each Bloch wave vector,  $\pmb{k}$, which can 
be restricted to the first Brillouin zone of the reciprocal lattice, the HFB equations will admit discrete 
states labelled by the band index $\alpha$. 

Following Eqs.~\eqref{eq:psi1} and \eqref{eq:psi2}, the quasiparticle wavefunctions can be expanded as
\begin{equation}\label{eq:psi1-inter}
\widehat{\psi}_{1\alpha \pmb{k}}(\rb,\sg)=\sum_{\beta}\mathcal{U}_{\alpha \pmb{k}, \beta \pmb{k}}\, \varphi_{\beta \pmb{k}}(\rb,\sg)\, ,
\end{equation}
\begin{equation}\label{eq:psi2-inter}
\widehat{\psi}_{2\alpha \pmb{k}}(\rb,\sg)=-\sum_\beta \mathcal{V}_{\alpha \pmb{k}, \beta \pmb{\bar k}}\, \varphi_{\beta \pmb{k}}(\rb,\sg)\, .
\end{equation}
Here the single-particle wavefunction $\varphi_{\alpha \pmb{k}}(\rb,\sg)$ represents a Bloch state with wave vector $\pmb{k}$ in a band $\alpha$,   
normalized as
\beqy \label{eq:normalization-Bloch-states}
\sum_\sg \int {\rm d}^3\rb \vert \varphi_{\alpha \pmb{k}}(\rb,\sg)\vert^2 = 1\, .
\eeqy 
We have used the fact that $\varphi_{\beta \pmb{\bar k}}(\rb,\sg)\equiv(-\sg)\varphi_{\beta \pmb{ k}}(\rb,-\sg)^*$ is also a Bloch state with wave vector $-\pmb{k}$ and opposite spin~\cite{kittel}.  
Note that the quasiparticle wave function includes only single-particle states with Bloch wave vector $\pmb{k}$ as required by the Floquet-Bloch theorem. However, the quasiparticle wave function may involve single-particle states from different bands.

In this basis, the HFB equations~\eqref{eq:HFB-superframe} read 
\beqy\label{eq:HFB-matrix-superfluid-frame}
\sum_\beta 
\begin{pmatrix}
h_{\alpha\pmb{k},\beta\pmb{k}}-\mu_n \delta_{\alpha\beta} & \Delta_{\alpha\pmb{k},\beta\pmb{\bar k}}  \\ 
\Delta_{\beta\pmb{k},\alpha\pmb{\bar k}}^* & -h^*_{\alpha\pmb{\bar k},\beta\pmb{\bar k}}+\mu_n \delta_{\alpha\beta}
\end{pmatrix}\begin{pmatrix} \U_{\gamma \pmb{k}, \beta \pmb{k}} \\ \V_{\gamma \pmb{k}, \beta \pmb{\bar k}}\end{pmatrix}=\E_{\alpha \pmb{k}}\begin{pmatrix} \U_{\gamma \pmb{k}, \alpha \pmb{k}} \\ \V_{\gamma \pmb{k}, \alpha \pmb{\bar k}}\end{pmatrix}\, .
\eeqy
The matrix elements of the single-particle Hamiltonian and pair potential are given respectively by
\beqy 
h_{\alpha\pmb{k},\beta\pmb{k}} =\sum_\sg \int {\rm d}^3\rb\, \varphi_{\alpha \pmb{k}}(\rb,\sg)^* h_{\pmb{Q}}(\rb)\varphi_{\beta \pmb{k}}(\rb,\sg)\, , 
\eeqy 
\beqy\label{eq:pairing-matrix-elements} 
\Delta_{\alpha\pmb{k},\beta\pmb{\bar k}} =-\sum_\sg \int {\rm d}^3\rb\, \varphi_{\alpha \pmb{k}}(\rb,\sg)^* \Delta_{\pmb{Q}}(\rb)\varphi_{\beta \pmb{k}}(\rb,\sg)\, .
\eeqy

Going back to the original quasiparticle wavefunctions using Eqs.~\eqref{eq:psi1tilde} and \eqref{eq:psi2tilde}, 
and substituting into Eq.~\eqref{eq:average-mass-current1}, the average neutron mass current in the crust frame is thus expressible as 
(the factor of 2 accounting for the spin degeneracy)
\beqy \label{eq:average-neutron-mass-current}
\pmb{\bar \rho_n} &=&\frac{2 m_n}{\Omega} \sum_{\pmb{k}} \sum_{\alpha,\beta} n_{\alpha\pmb{k},\beta\pmb{k}} \pmb{v}_{\beta\pmb{k},\alpha\pmb{k}}
\eeqy 
with 
\beqy 
n_{\alpha\pmb{k},\beta\pmb{k}}= \sum_\gamma \V^*_{\gamma \pmb{\bar k},\alpha \pmb{k}} \V_{\gamma \pmb{\bar k},\beta \pmb{k}}\, ,
\eeqy 
\beqy\label{eq:velocity-matrix-element}
\pmb{v}_{\beta\pmb{k},\alpha\pmb{k}} = \sum_\sigma \int {\rm d}^3\rb\,  \varphi_{\beta\pmb{k}}(\rb,\sg)^*\pmb{v}_{\pmb{Q}}(\rb)\varphi_{\alpha\pmb{k}}(\rb,\sg) \, ,
\eeqy 
and 
\beqy\label{eq:velocityQ}
\pmb{v}_{\pmb{Q}}(\rb)=e^{-i\pmb{Q}\cdot\rb}\pmb{v_n}(\rb)e^{i\pmb{Q}\cdot\rb} \, .
\eeqy

Note that 
\beqy 
\pmb{v}_{\pmb{Q}}(\rb)=\frac{1}{i\hbar}\left[\rb h_{\pmb{Q}}(\rb) - h_{\pmb{Q}}(\rb)\rb\right] = \pmb{v_n}(\rb) + \frac{\hbar \pmb{Q}}{m_n^\oplus(\rb)}\, .
\eeqy 

So far no approximation has been made and the expressions derived in this section are therefore very general.

\subsection{Linear response and superfluid fraction}
\label{sec:linear-response}

In practice,  the determination of the superfluid fraction requires the calculation of the average neutron mass current induced by given superfluid velocities,  varying not only the norm of $\pmb{\bar V_{n,s}}$ but also its direction.  This means solving the highly-nonlinear HFB equations~\eqref{eq:HFB-matrix-superfluid-frame} for millions of wavefunctions on a sufficiently fine spatial grid with a spacing of about 1 fm to properly resolve the effects of the nuclear interactions (see Ref.~\cite{chamel2025} for numerical comparisons of different grid spacings).  Such calculations are very challenging, especially in the shallow layers of the crust where neighboring ions are typically separated by about 100 fm.  A computationally more tractable treatment is therefore highly desirable. 

In most isolated pulsars,  the superfluid velocity is estimated to be much smaller than Landau's velocity $V_{n,L}$ above which quasiparticles are excited~\cite{allard2023}, as inferred from the analysis of the glitch data (note, however, that this may not be the case in rapidly rotating neutron stars spun up by accretion from a stellar companion, as evidenced by astrophysical observations of soft X-ray transient sources~\cite{allard2024,allard2024b,allard2025}).  This suggests a perturbative calculation of the 
superfluid fraction within the linear response theory.  The computational cost can be further reduced by adopting the BCS approximation,  which consists in neglecting the off-diagonal matrix elements of the pair potential namely  $\Delta_{\alpha\pmb{k},\beta\pmb{\bar k}}  \approx \Delta_{\alpha\pmb{k}}  \delta_{\alpha\beta}$.  It follows from the antisymmetry property~\eqref{eq:pairing-matrix} that $\Delta_{\alpha\pmb{\bar k}}=-\Delta_{\alpha\pmb{k}}$. 

In principle, the HFB equations can be formulated in any basis of single-particle states. Let us choose the eigenstates  $\varphi^0_{\alpha \pmb{k}}(\rb,\sg)$ 
of the single-particle Hamiltonian $h^0_n(\rb)$ with single-particle energy $\epsilon^0_{\alpha \pmb{k}}$  in the absence of current. 
Without any loss of generality, the pairing gaps $\Delta^0_{\alpha\pmb{k}}$ can be chosen real since the superfluid velocity vanishes locally.  
In this static case, the solutions of Eq.~\eqref{eq:HFB-matrix-superfluid-frame} satisfying Eqs.~\eqref{eq:constraint1} and \eqref{eq:constraint2} are readily obtained: 
\beqy\label{eq:QuasiparticleEnergy}
\E^0_{\alpha \pmb{k}} = \frac{\xi^0_{\alpha \pmb{k}}-\xi^0_{\alpha \pmb{\bar k}}}{2} +E^0_{\alpha\pmb{k}}\,,
\eeqy
\beqy\label{eq:UTerms}
\U^0_{\alpha\pmb{k},\alpha \pmb{k}}=\U^0_{\alpha\pmb{\bar k},\alpha \pmb{\bar k}}=\frac{1}{\sqrt{2}}\left(1+ \frac{\varepsilon^0_{\alpha \pmb{k}}}{E^0_{\alpha \pmb{k}}}  \right)^{1/2}\,,
\eeqy
\beqy\label{eq:VTerms}
\V^0_{\alpha\pmb{k},\alpha \pmb{\bar k}}=-\V^0_{\alpha\pmb{\bar k},\alpha \pmb{k}}=\frac{1}{\sqrt{2}}{\rm sgn}(\Delta^0_{\alpha\pmb{\bar k},\alpha\pmb{k}})\left(1- \frac{\varepsilon^0_{\alpha \pmb{k}}}{E_{\alpha\pmb{k}}}  \right)^{1/2}\,,
\eeqy
where 
\beqy\label{eq:E0} 
E^0_{\alpha\pmb{k}}=\sqrt{\varepsilon^0_{\alpha\pmb{k}} + \vert\Delta^0_{\alpha \pmb{k}}\vert^2}\, ,
\eeqy 
\begin{equation}\label{eq:eps0}
\varepsilon^0_{\alpha\pmb{k}}\equiv \frac{\xi^0_{\alpha\pmb{k}}+\xi^0_{\alpha\pmb{\bar k}}}{2} = \varepsilon^0_{\alpha \pmb{\bar k}}\, ,
\end{equation}
\beqy 
\xi^0_{\alpha \pmb{k}}=\epsilon^0_{\alpha\pmb{k}}-\mu_n\, ,
\eeqy 
\begin{equation}
\kappa^0_{\alpha\pmb{k},\alpha\pmb{\bar k}}=-\kappa^0_{\alpha\pmb{\bar k},\alpha\pmb{k}}
=\V^{0 *}_{\alpha\pmb{\bar k},\alpha \pmb{k}}\U^0_{\alpha\pmb{\bar k},\alpha \pmb{\bar k}}
= -{\rm sgn}(\Delta^0_{\alpha\pmb{\bar k},\alpha\pmb{k}})\frac{\vert\Delta^0_{\alpha\pmb{k}}\vert}{2E^0_{\alpha\pmb{k}}} \, ,
\end{equation}
\begin{equation}\label{eq:density-matrix-static}
n^0_{\alpha\pmb{k},\alpha\pmb{k}}=n^0_{\alpha\pmb{\bar k},\alpha\pmb{\bar k}}=\vert\V^0_{\alpha\pmb{k},\alpha \pmb{\bar k}} \vert^2=\frac{1}{2}\left(1- \frac{\varepsilon^0_{\alpha \pmb{k}}}{E_{\alpha\pmb{k}}}\right) \, .
\end{equation}
Since the static single-particle Hamiltonian  $h^0_n(\rb)$ is invariant under time reversal,  one has from Kramers theorem (see,  e.g., Ref.~\cite{kittel})  $\xi^0_{\alpha \pmb{\bar k}}=\xi^0_{\alpha \pmb{k}}$ therefore $\E^0_{\alpha \pmb{k}}$ reduces to $E^0_{\alpha\pmb{k}}$. 

Let us now consider small current perturbations, in the sense that $\bar V_{n,s}\ll V_{n,L}$.  Expanding the matrix elements of the single-particle Hamiltonian to first order yields $h_{\alpha\pmb{k},\beta\pmb{k}} = \epsilon^0_{\alpha \pmb{k}}\delta_{\alpha\beta} + \delta h_{\alpha\pmb{k},\beta\pmb{k}}$, in which the correction  
\begin{equation}\label{eq:linearized-Hamiltonian}
\delta h_{\alpha\pmb{k},\beta\pmb{k}} = \delta h^{\pmb{Q}}_{\alpha\pmb{k},\beta\pmb{k}}+\delta h^{\pmb{I}}_{\alpha\pmb{k},\beta\pmb{k}}
\end{equation}
consists of two terms. The first one is given by 
\begin{equation}
\delta h^{\pmb{Q}}_{\alpha\pmb{k},\beta\pmb{k}}=\hbar \pmb{Q}\cdot \pmb{v}^0_{\alpha\pmb{k},\beta\pmb{k}}
\end{equation} 
where 
\begin{equation}
\pmb{v}^0_{\alpha\pmb{k},\beta\pmb{k}}=\sum_\sigma\int d^3\rb\, \varphi^0_{\alpha\pmb{k}}(\rb,\sigma)^*\pmb{v}_{\pmb n}^{0}(\rb)\varphi^0_{\beta\pmb{k}}(\rb,\sigma)
\end{equation} 
denotes the matrix element of the \emph{unperturbed} velocity operator
\begin{equation}\label{eq:velocity-operator-unperturbed}
\pmb{v}_{\pmb n}^{0}(\rb)\equiv \frac{1}{i\hbar}\left[\rb,h_n^0(\rb)\right]=\frac{1}{2m_n^\oplus(\rb)}\pmb{p}+\pmb{p}\frac{1}{2m_n^\oplus(\rb)}
\end{equation}
between \emph{unperturbed} Bloch states. The second correction  
\begin{equation}
\delta h^{\pmb{I}}_{\alpha\pmb{k},\beta\pmb{k}}=\frac{-i}{2} \sum_\sigma\int d^3\rb\, \varphi^0_{\alpha\pmb{k}}(\rb,\sigma)^* \left(\delta\pmb{I_n}\cdot\pmb{\nabla}+\pmb{\nabla}\cdot\delta\pmb{I_n}\right) \varphi^0_{\beta\pmb{k}}(\rb,\sigma)
\end{equation} 
arises because of the velocity dependence of the perturbed single-particle Hamiltonian. Here $\delta\pmb{I_n}=\pmb{I_n}$ since the unperturbed configuration is assumed to be static. 
As shown in Appendix~\ref{app:perturbed-Hamiltonian}, the perturbed single-particle Hamiltonian satisfies the following identities: $-\delta h^*_{\alpha\pmb{\bar k},\beta\pmb{\bar k}}=-\delta h_{\beta\pmb{\bar k},\alpha\pmb{\bar k}}=\delta h_{\alpha\pmb{k},\beta\pmb{k}}$. Note that the chemical potential $\mu_n$ depends on $\pmb{Q}$ but only from second order. Therefore, $\mu_n\approx \mu^0_n$ to first order. Neglecting the first-order corrections to the pairing gaps, the perturbed HFB equations finally read 
\beqy\label{eq:linearized-HFB-superfluid-frame}
\sum_\beta 
\begin{pmatrix}
\xi^0_{\alpha \pmb{k}}\delta_{\alpha\beta} + \delta h_{\alpha\pmb{k},\beta\pmb{k}}  & \Delta^0_{\alpha\pmb{k}}\delta_{\alpha\beta}  \\ 
\Delta^0_{\alpha\pmb{k}}\delta_{\alpha\beta} & -\xi^0_{\alpha \pmb{k}}\delta_{\alpha\beta} +\delta h_{\alpha\pmb{k},\beta\pmb{k}}
\end{pmatrix}\begin{pmatrix} \U_{\gamma \pmb{k}, \beta \pmb{k}} \\ \V_{\gamma \pmb{k}, \beta \pmb{\bar k}}\end{pmatrix}=\E_{\alpha \pmb{k}}\begin{pmatrix} \U_{\gamma \pmb{k}, \alpha \pmb{k}} \\ \V_{\gamma \pmb{k}, \alpha \pmb{\bar k}}\end{pmatrix}\, .
\eeqy
This equation reduces to Eq.~(4) of Ref.~\cite{almirante2025} in the limiting case $m_n^\oplus(\rb)=m_n$ and $I_n(\rb)=0$. 

Instead of solving this equation for $\U_{\alpha \pmb{k}, \beta \pmb{k}}$ and $\V_{\alpha \pmb{k}, \beta \pmb{\bar k}}$ and substitute into Eq.~\eqref{eq:DensityMatrix}, the perturbed density matrix can be more directly obtained from Eqs.~\eqref{eq:TDHFB1} and \eqref{eq:TDHFB2}. Linearizing 
these equations yields 
\beqy
\label{eq:perturbed_TDHFB1}
\sum_{\beta} \delta h_{\alpha \pmb{k},\beta \pmb{k}} (n^0_{\beta \pmb{k},\beta \pmb{k}}-n^0_{\alpha \pmb{k},\alpha \pmb{k}}) + (\xi^0_{\alpha \pmb{k}} - \xi^0_{\beta \pmb{k}}) \delta n_{\alpha \pmb{k},\beta \pmb{k}} - \delta  \kappa_{\alpha \pmb{k},\beta \pmb{\bar{k}}} \Delta^0_{\beta\pmb{k}} -  \Delta^0_{\alpha\pmb{k}}\delta  \kappa_{\alpha \pmb{\bar{k}},\beta \pmb{k}}^* = 0 \, ,
\eeqy
\beqy
\label{eq:perturbed_TDHFB2}
\sum_{\beta} \delta h_{\alpha \pmb{k},\beta \pmb{k}} (\kappa^0_{\beta \pmb{k},\beta \pmb{\bar{k}}} - \kappa^0_{\alpha \pmb{k},\alpha \pmb{\bar{k}}} ) + (\xi^0_{\alpha \pmb{k}} + \xi^0_{\beta \pmb{k}}) \delta \kappa_{\alpha \pmb{k},\beta \pmb{\bar k}}
- \Delta^0_{\alpha\pmb{k}} \delta n_{\alpha \pmb{\bar{k}},\beta \pmb{\bar k}}^* - \Delta^0_{\beta\pmb{k}} \delta n_{\alpha \pmb{k},\beta \pmb{k}}=0\, .
\eeqy
Solving Eq.~\eqref{eq:perturbed_TDHFB2} for $\delta \kappa_{\alpha\pmb{k},\beta\pmb{k}}$ substituting into \eqref{eq:perturbed_TDHFB1} making use of the specific form of the unperturbed static solution leads after some algebraic simplifications to 
\beqy\label{eq:perturbed-density-matrix}
\delta n_{\alpha\pmb{k},\beta\pmb{k}} = \frac{\xi^0_{\alpha\pmb{k}}\xi^0_{\beta\pmb{k}}-E^0_{\alpha\pmb{k}}E^0_{\beta\pmb{k}}+\Delta^0_{\alpha\pmb{k}}\Delta^0_{\beta\pmb{k}}}{2E^0_{\alpha\pmb{k}}E^0_{\beta\pmb{k}}(E^0_{\alpha\pmb{k}}+E^0_{\beta\pmb{k}})}\delta h_{\alpha\pmb{k},\beta\pmb{k}} \, .
\eeqy
It follows from Eqs.~\eqref{eq:E0} and \eqref{eq:eps0} that $\delta n_{\alpha\pmb{k},\alpha\pmb{k}}=0$. The perturbed density matrix satisfies similar identities as 
$\delta h_{\alpha\pmb{k},\beta\pmb{k}}$ (see Appendix~\ref{app:perturbed-Hamiltonian}), namely $-\delta n^*_{\alpha\pmb{\bar k},\beta\pmb{\bar k}}=-\delta n_{\beta\pmb{\bar k},\alpha\pmb{\bar k}}=\delta n_{\alpha\pmb{k},\beta\pmb{k}}$. 

As expected,  the local neutron mass density is not modified by the flow (at first order) and is given by 
\beqy \label{eq:local-neutron-mass-density}
\rho_n(\rb) =\rho^0_n(\rb)=2 m_n\sum_{\alpha,\pmb{k}} n^0_{\alpha\pmb{k},\alpha\pmb{k}} \sum_\sigma \vert\varphi^0_{\alpha\pmb{k}}(\rb,\sigma)\vert^2 \, .
\eeqy 
The explicit proof is given in Appendix~\ref{app:perturbed-density}. This means that the effective mass $m_n^\oplus(\rb)$ and the mean field potential $U_n(\rb)$, which 
depend on the local densities only for Skyrme functionals, remain also unchanged, as we implicitly assumed in the perturbation expansion.  Integrating Eq.~\eqref{eq:local-neutron-mass-density} using \eqref{eq:normalization-Bloch-states},  the average neutron mass density is given by 
\beqy \label{eq:average-neutron-mass-density-static}
\bar \rho_n = \frac{2 m_n}{\Omega}\sum_{\alpha,\pmb{k}} n^0_{\alpha\pmb{k},\alpha\pmb{k}} \, .
\eeqy 
The average neutron mass current can be expressed as $\pmb{\bar \rho_n} = \delta\pmb{\bar \rho_n}=\rho_{n,s}\pmb{\bar V_{n,s}}$ with 
\beqy \label{eq:superfluid-density-full}
\rho_{n,s}&=&\frac{1}{\Omega} \int{\rm d}^3\rb\,\rho_n(\rb)\left[ \frac{m_n}{m_n^\oplus(\rb)}+\frac{m_n}{3\hbar^2} \pmb{\nabla_Q}\cdot \pmb{I_n}(\rb)\right] \notag \\
&& +\frac{1}{3}\frac{m_n^2}{\Omega}\sum_{\pmb{k}} \sum_{\alpha, \beta}\frac{\xi^0_{\alpha\pmb{k}}\xi^0_{\beta\pmb{k}}-E^0_{\alpha\pmb{k}}E^0_{\beta\pmb{k}}+\Delta^0_{\alpha\pmb{k}}\Delta^0_{\beta\pmb{k}}}{E^0_{\alpha\pmb{k}}E^0_{\beta\pmb{k}}(E^0_{\alpha\pmb{k}}+E^0_{\beta\pmb{k}})} \pmb{v}^0_{\beta\pmb{k},\alpha\pmb{k}} \cdot \Biggl\{\pmb{v}^0_{\alpha\pmb{k},\beta\pmb{k}} \biggr.\notag \\ 
&&\left. +\frac{1}{6\hbar^2} \sum_\sigma \int{\rm d}^3\rb\,\varphi^0_{\alpha\pmb{k}}(\rb,\sigma)^*[ \pmb{\nabla_Q}\cdot \pmb{I_n}(\rb) \pmb{p} +  \pmb{p}  \pmb{\nabla_Q}\cdot\pmb{I_n}(\rb)] \varphi^0_{\beta\pmb{k}}(\rb,\sigma)\right\}
\eeqy 
(see Appendix \ref{app:superfluid-density}). 
This generalizes the expression obtained in Ref.~\cite{almirante2025} by taking into account the local effective mass $m_n^\oplus(\rb)$ and the potential $\pmb{I_n}(\rb)$. 
 
For Skyrme like nuclear energy density functionals, both $m_n^\oplus(\rb)$ and $\pmb{I_n}(\rb)$ arise from the same momentum- and possibly density-dependent terms in the underlying effective interactions and can be expressed as (see Ref.~\cite{chamel2009} for the relation between the coupling coefficients $C_0^\tau$ and $C_1^\tau$ and the usual Skyrme parameters)
\beqy\label{eq:def-Bn}
\frac{m_n}{m_n^\oplus(\rb)} = 1+ \frac{2}{\hbar^2}(C_0^\tau - C_1^\tau)\rho(\rb) + \frac{4}{\hbar^2}C_1^\tau \rho_n(\rb) \, ,
\eeqy
\beqy\label{eq:def-In}
\pmb{I_n}(\rb)=-2 \pmb{j}(\rb) (C_0^\tau - C_1^\tau)  - 4 \pmb{j_n} C_1^\tau \, ,
\eeqy
where $\rho(\rb)=\rho_n(\rb)+\rho_p(\rb)$ and $\pmb{j}(\rb)=\pmb{j_n}(\rb)+\pmb{j_p}(\rb)$. Note that these expressions are very general and are not limited to the 
linear response. 
Since $m_n^\oplus(\rb)$ depends solely on the densities, it is unaffected by the superfluid flow at first order. Although each neutron Cooper pair carries effectively a momentum $2\hbar\pmb{Q}$, the neutron momentum density $\pmb{j_n}(\rb)$ is not simply given by $n_n(\rb)\pmb{Q}$ 
due to the partial depletion of the superfluid reservoir (see Appendix \ref{app:momentum-density}): 
\beqy\label{eq:linearized-momentum}
\pmb{j_n}(\rb) =\delta\pmb{j_n}(\rb) = n_n(\rb)\pmb{Q}+2\sum_{\alpha,\beta}\sum_{\pmb{k}} \delta n_{\alpha\pmb{k},\beta\pmb{k}} \sum_\sigma \varphi^0_{\beta\pmb{k}}(\rb,\sg)^* \frac{\pmb{p}}{\hbar}\varphi^0_{\alpha\pmb{k}}(\rb,\sg)\, .
\eeqy 
Because of mutual entrainment effects, the proton momentum density does not vanish even though the proton mass current does. Setting $\pmb{\rho_p}(\rb)=\pmb{0}$ in Eq.~(16) of Ref.~\cite{ChamelAllard2019}, we find 
\beqy\label{eq:linearized-proton-momentum}
\pmb{j_p}(\rb)=\pmb{j_n}(\rb) \rho_p(\rb) \frac{2}{\hbar^2}(C_0^\tau - C_1^\tau)\left[1+ \frac{2}{\hbar^2}(C_0^\tau - C_1^\tau)\rho_n(\rb) \right]^{-1} \, .
\eeqy 
Note that $\pmb{j_p}(\rb)$ is proportional to $ \rho_p(\rb)$ and is therefore confined inside clusters. 
The proton momentum density vanishes everywhere only for functionals with $C_0^\tau=C_1^\tau=0$, in which case $m_n^\oplus(\rb)=m_n$ and $\pmb{I_n}(\rb)=\pmb{0}$. The motion of protons is then locally decoupled from that of neutrons in the sense that $\pmb{\rho_p}(\rb)=\hbar \pmb{j_p}(\rb)=\pmb{0}$ and $\pmb{\rho_n}(\rb)=\hbar \pmb{j_n}(\rb)$ from Eq.~\eqref{eq:local-mass-current}. However, the neutron superfluid flow is still influenced by the crust due to spatial inhomogeneities leading to $\rho_{n,s}<\bar \rho_n$.

Substituting Eq.~\eqref{eq:linearized-proton-momentum} into \eqref{eq:def-In}, $\pmb{I_n}(\rb)$ can be expressed in terms of $\pmb{j_n}(\rb)$ only: 
\beqy\label{eq:In-crust}
\pmb{I_n}(\rb)=-2 \pmb{j_n}(\rb) \biggl\{ C_0^\tau + C_1^\tau +  \rho_p(\rb) \frac{2}{\hbar^2}(C_0^\tau - C_1^\tau)^2\left[1+ \frac{2}{\hbar^2}(C_0^\tau - C_1^\tau)\rho_n(\rb) \right]^{-1} \biggr\} \, .
\eeqy
Since $\pmb{j_n}$ can only be proportional to $\pmb{Q}$, we have $\nabla^i_Q j_n^j = (1/3) \delta^{ij} \pmb{\nabla_Q}\cdot\pmb{j_n}$. In turn, taking the divergence of Eq.~\eqref{eq:linearized-momentum} using \eqref{eq:perturbed-density-matrix} yields
\beqy \label{eq:divergenceQ-j}
&&\pmb{\nabla_Q}\cdot\pmb{j_n}(\rb)
=3n_n(\rb) + \sum_{\alpha,\beta}\sum_{\pmb k} \frac{\xi^0_{\alpha\pmb{k}}\xi^0_{\beta\pmb{k}}-E^0_{\alpha\pmb{k}}E^0_{\beta\pmb{k}}+\Delta^0_{\alpha\pmb{k}}\Delta^0_{\beta\pmb{k}}}{E^0_{\alpha\pmb{k}}E^0_{\beta\pmb{k}}(E^0_{\alpha\pmb{k}}+E^0_{\beta\pmb{k}})} \notag \\ 
&&\times \sum_\sigma \varphi^0_{\alpha\pmb{k}}(\rb,\sigma)^* \frac{\pmb{p}}{\hbar}\varphi^0_{\beta\pmb{k}}(\rb,\sigma) \cdot \biggl\{ \hbar \pmb{v}^0_{\beta\pmb{k},\alpha\pmb{k}} \biggr. \notag \\ 
&&\biggl. +\frac{1}{6} \sum_{\sigma^\prime} \int{\rm d}^3\rb^\prime\, \pmb{\nabla_Q}\cdot\pmb{I_n}(\rb^\prime) [\varphi^0_{\beta\pmb{k}}(\pmb{r^\prime},\sigma^\prime)^*\frac{\pmb{p^\prime}}{\hbar}  \varphi^0_{\alpha\pmb{k}}(\pmb{r^\prime},\sigma^\prime)
-[\varphi^0_{\alpha\pmb{k}}(\pmb{r^\prime},\sigma^\prime)\frac{\pmb{p^\prime}}{\hbar}  \varphi^0_{\beta\pmb{k}}(\pmb{r^\prime},\sigma^\prime)^* ]
\biggr\}\, .
\eeqy 
Together with Eq.~\eqref{eq:In-crust}, this leads to a self-consistency integral equation for determining the field $\pmb{\nabla_Q}\cdot\pmb{I_n}(\rb)$ appearing in Eq.~\eqref{eq:superfluid-density-full}.

\subsection{Entrainment effects from the inner crust to the outer core}

In principle, our derivation of entrainment effects in the inner crust of a neutron star remains valid in the region beneath, where clusters dissolve into an homogeneous mixture of neutrons, protons and electrons. The outer core thus appears as a limiting case of an empty lattice. 

The Bloch wave functions $\varphi^0_{\alpha\pmb{k}}(\rb,\sg)$ then reduce to pure plane waves~\cite{shockley1937}, 
\beqy\label{eq:Bloch-state-empty-lattice}
\varphi^0_{\alpha\pmb{k}}(\rb,\sg)=\frac{1}{\sqrt{\Omega}}e^{i(\pmb{k}+\pmb{G}_\alpha)\cdot\rb}\chi(\sg) \, ,
\eeqy 
where $\pmb{G}_\alpha$ is a reciprocal lattice vector and $\chi(\sg)$ denotes the Pauli spinor. 
In homogeneous matter, the unperturbed velocity operator~\eqref{eq:velocity-operator-unperturbed} is simply proportional to the momentum operator, namely $\pmb{v}^0_{\pmb{n}}=\pmb{p}/m_n^\oplus$. Since the Bloch states~\eqref{eq:Bloch-state-empty-lattice} are eigenstates of $\pmb{p}$, the matrices $\pmb{v}^0_{\alpha\pmb{k},\beta\pmb{k}}$ and $\delta h_{\alpha\pmb{k},\beta\pmb{k}}$ are purely diagonal. As a consequence, $\delta n_{\alpha\pmb{k},\beta\pmb{k}}=0$. It can also be easily seen from  Eq.~\eqref{eq:pairing-matrix-elements} that $\Delta_{\alpha\pmb{k},\beta\pmb{\bar k}}$ is also diagonal, as follows from the fact that $\Delta_{\pmb Q}(\rb)=\Delta_{\pmb Q}$ and the orthonormality condition~\eqref{eq:orthonormality}. In other words, the BCS ansatz is exact in this case. 

It follows from Eq.~\eqref{eq:linearized-momentum} that 
the neutron momentum density is now entirely carried by Cooper pairs and is given by $\pmb{j_n}=n_n\pmb{Q}$ in perfect agreement with Eq.~(62) of Ref.~\cite{allard2021}.  
However, the neutron mass current is not simply given by $\rho_n\pmb{V_{n,s}}$ because neutrons can still be entrained by protons due to the momentum-dependence of the Skyrme effective interaction. 
The neutron superfluid density~\eqref{eq:superfluid-density-full} reduces to 
\beqy\label{eq:rhons-hom}
\rho_{n,s}&=&\rho_n \left( \frac{m_n}{m_n^\oplus}+\frac{m_n}{3\hbar^2} \pmb{\nabla_Q}\cdot \pmb{I_n}\right) \, ,
\eeqy 
and from Eqs.~\eqref{eq:In-crust} and \eqref{eq:divergenceQ-j}, we have 
\beqy\label{eq:divQ-I-hom}
\frac{m_n}{3\hbar^2}\pmb{\nabla_Q}\cdot \pmb{I_n}=- \frac{2 \rho_n}\hbar^2 \biggl\{ C_0^\tau + C_1^\tau +  \rho_p \frac{2}{\hbar^2}(C_0^\tau - C_1^\tau)^2\left[1+ \frac{2}{\hbar^2}(C_0^\tau - C_1^\tau)\rho_n \right]^{-1} \biggr\} \, .
\eeqy
Substituting Eqs.~\eqref{eq:def-Bn} and \eqref{eq:divQ-I-hom} into \eqref{eq:rhons-hom}, we find 
\beqy\label{eq:rhons-hom-final}
\rho_{n,s}= \rho_n \left\{ 1 + \frac{2}{\hbar^2} (C_0^\tau - C_1^\tau) \rho_p\left[1+\frac{2}{\hbar^2} (C_0^\tau - C_1^\tau) \rho_n\right]^{-1} \right\} \, .
\eeqy

This can be compared with the general expression of the entrainment matrix of a neutron-proton superfluid mixture derived in Refs.~\cite{ChamelAllard2019,allard2021}, and defined as 
\beqy\label{eq:entrainment-matrix} 
\pmb{\rho_q}=\sum_{q^\prime} \rho_{qq^\prime} \pmb{V_{q^\prime,s}}\, ,
\eeqy  
where 
\beqy\label{eq:rhonn}
\rho_{nn}=\rho_n\left[1+ \frac{2}{\hbar^2}\left(C_0^\tau-C_1^\tau\right)\rho_p\right]\, , 
\eeqy 
\beqy\label{eq:rhopp}
\rho_{pp}=\rho_p\left[1+ \frac{2}{\hbar^2}\left(C_0^\tau-C_1^\tau\right)\rho_n\right]\, , 
\eeqy 
\beqy\label{eq:rhonp}
\rho_{np}=\rho_{pn}=\rho_n\rho_p\frac{2}{\hbar^2}\left(C_1^\tau-C_0^\tau\right)\, .
\eeqy 
Recalling that the neutron superfluid density~\eqref{eq:superfluid-density-full} was derived in the crust frame, i.e. in the proton rest frame where $\pmb{\rho_p}=\pmb{0}$ and $\pmb{\rho_n}\equiv\rho_{n,s} \pmb{V_{n,s}}$, 
we find from Eq.~\eqref{eq:entrainment-matrix} 
\beqy 
\rho_{n,s} = \left(\rho_{nn}-\frac{\rho_{np}\rho_{pn}}{\rho_{pp}}\right) \, .
\eeqy 
Substituting Eqs.~\eqref{eq:rhonn}, \eqref{eq:rhopp} and \eqref{eq:rhonp} into the above expression leads to the same formula \eqref{eq:rhons-hom-final}. 

Therefore Eq.~\eqref{eq:superfluid-density-full} ensures a unified description of entrainment effects in both the inner crust and the outer core of a neutron star. Note that entrainment effects in the core are entirely due to the momentum-dependent terms of the effective interaction. If such terms are absent, i.e. $C_0^\tau=C_1^\tau=0$, we will have $\pmb{\rho_n}=\rho_n \pmb{V_n}=\hbar \pmb{j_n}$.

\subsection{BCS approximation and interband response}
\label{sec:BCS}

In Ref.~\cite{chamel2025}, the quasiparticle wavefunction was expanded in the crust frame following a similar treatment in homogeneous neutron-proton superfluid 
mixtures~\cite{allard2021}, as
\begin{equation}\label{eq:psi1-intraband}
\psi_{1\alpha k}(\rb,\sg)=\sum_{\beta}\mathcal{U}_{\alpha k, \beta k}\, \varphi_{\beta k}(\rb,\sg)\, ,
\end{equation}
\begin{equation}\label{eq:psi2-intraband}
\psi_{2\alpha k}(\rb,\sg)=-\sg \sum_\beta \mathcal{V}_{\alpha k, \beta \bar k}\, \varphi_{\beta \bar k}(\rb,-\sg)^*\, ,
\end{equation}
where $\varphi_{\beta k}(\rb,\sg)$ and $\varphi_{\beta \bar k}(\rb,\sg)$ were taken as the Bloch eigenstates of $h_n(\rb)$ with Bloch wave 
vectors $\pmb{k}+\pmb{Q}$ and $-\pmb{k}+\pmb{Q}$,  and opposite spins respectively. In this section, we use the same shorthand notation
as in Ref.~\cite{chamel2025}
\beqy
k\equiv (\pmb{k}+\pmb{Q},\sigma)\,,\qquad\qquad \bar{k}\equiv (-\pmb{k}+\pmb{Q},-\sigma)\,  .
\eeqy

The HFB equations~\eqref{eq:HFB-crust-frame} can be expressed as 
\beqy\label{eq:HFB-matrix-crust-frame}
\sum_\beta 
\begin{pmatrix}
h_{\alpha k,\beta k} -\mu_n\delta_{\alpha\beta} & \Delta_{\alpha k,\beta\bar k}  \\ 
\Delta_{\beta k,\alpha \bar k}^* & -h^*_{\alpha \bar k,\beta \bar k} + \mu_n\delta_{\alpha\beta}
\end{pmatrix}\begin{pmatrix} \U_{\gamma k, \beta k} \\ \V_{\gamma k, \beta \bar{k}}\end{pmatrix}=\E_{\alpha k}\begin{pmatrix} \U_{\gamma k, \alpha k} \\ \V_{\gamma k, \alpha \bar{k}}\end{pmatrix}\, ,
\eeqy
with 
\beqy 
h_{\alpha k,\beta k} =\sum_\sg \int {\rm d}^3\rb\, \varphi_{\alpha k}(\rb,\sg)^* h_n(\rb)\varphi_{\beta k}(\rb,\sg)=\epsilon_{\alpha k}\delta_{\alpha\beta}\, ,
\eeqy
\beqy \label{eq:pair-matrix-old}
\Delta_{\alpha k,\beta\bar{k}} =-\sum_\sg \sg \int {\rm d}^3\rb\, \varphi_{\alpha k}(\rb,\sg)^* \Delta_n(\rb)\varphi_{\beta \bar{k}}(\rb,-\sg)^*\, .
\eeqy
The Hamiltonian matrix is now diagonal independently of $\pmb{Q}$. Adopting the BCS approximation $\Delta_{\alpha k,\beta\bar k} \approx \Delta_{\alpha k} \delta_{\alpha\beta}$, the HFB equations 
with and without current are then formally similar, and can thus be readily solved for arbitrary average neutron superfluid velocity:  
\beqy
\E_{\alpha k} = \frac{\xi_{\alpha k}-\xi_{\alpha \bar k}}{2} +\sqrt{\varepsilon_{\alpha k} + \vert\Delta_{\alpha k}\vert^2}\,,
\eeqy
\beqy
\vert\U_{\alpha k,\alpha k}\vert^2=\frac{1}{2}\left(1+ \frac{\varepsilon_{\alpha k}}{\sqrt{\varepsilon_{\alpha k}^2 + \vert\Delta_{\alpha k}\vert^2}}  \right)\,,
\eeqy
\beqy
\vert\V_{\alpha k,\alpha \bar{k}}\vert^2=\frac{1}{2}\left(1- \frac{\varepsilon_{\alpha k}}{\sqrt{\varepsilon_{\alpha k}^2 + \vert\Delta_{\alpha k}\vert^2}}  \right)\,,
\eeqy
where $\xi_{\alpha k}\equiv \epsilon_{\alpha k}-\mu_n$ and  
\begin{equation}
\varepsilon_{\alpha k}\equiv \frac{\xi_{\alpha k}+\xi_{\alpha \bar k}}{2} = \varepsilon_{\alpha \bar k}\, .
\end{equation}

The density matrix is diagonal and is simply given by $n_{\alpha k,\alpha k}=\vert\V_{\alpha k,\alpha \bar{k}}\vert^2$. 
The average neutron mass density and current thus reduce to 
\beqy\label{eq:average-neutron-mass-density}
\bar \rho_n = \frac{m_n}{\Omega}\sum_{\alpha,k} \vert\V_{\alpha k,\alpha \bar{k}}\vert^2\, ,
\eeqy 
\beqy\label{eq:average-neutron-mass-current-intraband}
\pmb{\bar \rho_n} = \frac{m_n}{\Omega}\sum_{\alpha,k} \vert\V_{\alpha k,\alpha \bar{k}}\vert^2\pmb{v}_{\alpha k}\, ,
\eeqy 
where
\beqy
\label{eq:group-velocity}
\pmb{v}_{\alpha\pmb{k}}=\sum_\sigma\int d^3\rb\,  \varphi_{\alpha k}(\rb,\sigma)^* \pmb{v_n}(\rb)\varphi_{\alpha k}(\rb,\sigma)=\frac{1}{\hbar}\pmb{\nabla_k} \epsilon_{\alpha k} \, ,
\eeqy
the latter identity following from the application of the Hellmann-Feynman theorem~\cite{feynman1939} (see Appendix C of Ref.~\cite{ChamelAllard2019} for the explicit proof in the present context). 

Expanding Eqs.~\eqref{eq:average-neutron-mass-current-intraband} and \eqref{eq:group-velocity} to first order in the superfluid velocity writing $\delta\pmb{v}_{\alpha k}= (\pmb{Q}\cdot \pmb{\nabla_Q})\pmb{v}_{\alpha k}$, ignoring the dependence of $\Delta_{\alpha k}$ on $\pmb{Q}$ and remarking that $\varepsilon_{\alpha k}$ hence also $\V_{\alpha k,\alpha \bar{k}}$ cannot contain terms of the form $\pmb{k}\cdot\pmb{Q}$, the neutron superfluid density can be readily extracted 
\beqy\label{eq:superfluid-density-app}
\rho_{n,s}=\frac{2 m_n^2}{3 \hbar \Omega}\sum_{\alpha,\pmb{k}}\vert\V^0_{\alpha \pmb{k},\alpha \pmb{k}}\vert^2 \pmb{\nabla_Q}\cdot \pmb{v}_{\alpha k}\, ,
\eeqy 
in agreement with Eq.~(47) of Ref.~\cite{chamel2025}. 
The absence of interband current response thus originates from the BCS approximation with the decomposition~\eqref{eq:psi1-intraband} and \eqref{eq:psi2-intraband}. 

This analysis shows that this approximation depends on the choice of the 
single-particle wave functions used to express the quasiparticle wave functions. With the alternative decomposition \eqref{eq:psi1-inter} and \eqref{eq:psi2-inter} adopted earlier, 
the BCS approximation is equivalent to neglecting the spatial variations of the periodic part of the pair potential, $\Delta_{\pmb Q}(\rb)\approx \Delta$, as can be seen from Eq.~\eqref{eq:pairing-matrix-elements}  recalling the orthonormality of the Bloch states.  
Adopting this same approximation here, the matrix of the pair potential is no longer diagonal in the presence of current.  
The correction ignored in Ref.~\cite{chamel2025} appears to be of first order and is given by (see Appendix~\ref{app:pair-matrix})
\beqy 
\Delta_{\alpha k,\beta\bar k} \approx  2\Delta \frac{\hbar \pmb{Q}\cdot\pmb{v}^0_{\alpha\pmb{k},\beta\pmb{k}}}{\epsilon^0_{\alpha \pmb{k}}-\epsilon^0_{\beta\pmb{k}}}\ \mathrm{if~} \alpha\neq\beta.
\eeqy 
These off-diagonal terms are negligible if the pairing gap $\Delta$ is much smaller than the band gaps $\vert \epsilon^0_{\alpha \pmb{k}}-\epsilon^0_{\beta\pmb{k}}\vert$. However, this condition is generally not fulfilled for realistic pairing interactions.

In homogeneous matter, the Bloch states reduce to pure plane waves, i.e. eigenstates of $\pmb{p}$, so that $\pmb{v}^0_{\alpha\pmb{k},\beta\pmb{k}}$ becomes purely diagonal and $\Delta_{\alpha k,\beta\bar k}$ vanishes identically for $\alpha\neq \beta$ independently of $\Delta$. The BCS treatment is therefore exact in this limiting case. 
The single-particle energies are then given by~\cite{allard2021}
\beqy
\epsilon_{k}=\frac{\hbar^2}{2m_n^{\oplus}}\left(\pmb{k}+\pmb{Q}\right)^2 + U_n+\pmb{I_n}\cdot \left(\pmb{k}+\pmb{Q}\right) \, .
\eeqy
Taking the gradient in $\pmb{k}$-space yields 
\beqy 
\pmb{v}_k = \frac{\hbar}{m_n^{\oplus}}\left(\pmb{k}+\pmb{Q}\right) +\frac{\pmb{I_n}}{\hbar} \, .
\eeqy 
Taking the divergence with respect to $\pmb{Q}$ and substituting into Eq.~\eqref{eq:superfluid-density-app} leads to
\beqy
\rho_{n,s}=\rho_{n}\left(\frac{m_n}{m_n^\oplus}+\frac{m_n}{3\hbar^2}\pmb{\nabla}_{\pmb{Q}} \cdot \pmb{I_n} \right)  \, .
\eeqy 
This coincides with the exact result~\eqref{eq:rhons-hom}.

\section{Results and discussion}
\label{sec:results}

\subsection{Numerical implementation}
\label{sec:numerical-implementation}

For comparison with our previous studies~\cite{chamel2012,chamel2025,chamel2025b}, we will take the same internal constitution of neutron-star crust from the calculations of Ref.~\cite{onsi2008}. The equilibrium composition was determined under the assumption of cold catalyzed matter by solving approximately the HFB equations in spherical Wigner-Seitz cells using the 4th-order extended Thomas Fermi method including shell effects perturbatively through the Strutinsky integral theorem. The key results are summarized in Table~\ref{tab:crust-composition}. The nucleon distributions obtained from this approach have been shown to be very close to those predicted by full three-dimensional self-consistent HFB calculations~\cite{pecak2024}.

\begin{table}
\centering
\caption{Composition of the inner crust of a neutron star at different average baryon number densities $\bar n$: proton number $Z$ and nucleon number $A_\mathrm{cell}$ in the Wigner-Seitz cell, number of nucleons in ions $A=m_I/m_n$ with $m_I$ calculated from Eq.~\eqref{eq:bare-ion-mass}, number $A_\mathrm{bound}$ of quantum mechanically bound nucleons, effective number $A^\star=m_I^\star/m_n$ of nucleons in ions, $\rho_{n,i}$ neutron mass density inside ions, $\rho_{p,i}$ proton mass density inside ions, $\rho_{n,o}$ neutron mass density outside ions.}
\label{tab:crust-composition}
\begin{tabular} {|ccccccccc|}
\hline
$\bar n$ [fm$^{-3}$]  & $Z$ & $A_\mathrm{cell}$ & $A$ & $A_\mathrm{bound}$ & $A^\star$  & $\rho_{n,i}/m_n$ & $\rho_{p,i}/m_p$ & $\rho_{n,o}/m_n$ \\
\hline 
0.03                  & 40  &  1590             & 181  &      110          &  98.4      & 0.1021           &  0.02778         &  0.02659       \\
0.04                  & 40  &  1610             & 179  &      110          &  96.7      & 0.1006           &  0.02349         &  0.03555         \\
0.05                  & 20  &   800             &  89  &      42           &  52.9      & 0.1022           &  0.02144         &  0.04446        \\ 
0.06                  & 20  &   780             &  85  &      40           &  53.7      & 0.1005           &  0.01720         &  0.05346       \\ 
0.07                  & 20  &   714             &  75  &      40           &  49.4      & 0.09948          &  0.01361         &  0.06242        \\
\hline
\end{tabular}
\end{table}

The crust is assumed to be a perfect body-centered cubic lattice crystal. The primitive basis vectors are given by 
\beqy \label{eq:primitive-vectors}
 \pmb{a_1}&=&\frac{a}{2}\left( -\pmb{\hat x} + \pmb{\hat y} + \pmb{\hat z}\right) \, ,\notag \\ 
 \pmb{a_2}&=&\frac{a}{2}\left( \pmb{\hat x} - \pmb{\hat y} + \pmb{\hat z}\right) \, ,\notag \\ 
 \pmb{a_3}&=&\frac{a}{2}\left( \pmb{\hat x} + \pmb{\hat y} - \pmb{\hat z}\right) \, ,
\eeqy  
where $\pmb{\hat x}$, $\pmb{\hat y}$, $\pmb{\hat z}$ are the Cartesian unit vectors and $a$ is the 
size of the conventional cubic cell. The volume of a primitive cell  
is therefore $\Omega_{\rm cell}=a^3/2$. Requiring $\Omega_{\rm cell}$ to be the same as the volume of the approximate spherical cell of radius $R_\mathrm{cell}$
fixes the lattice spacing $a=R_{\rm cell}(8\pi/3)^{1/3}$. The periodic mean fields are constructed by adding those calculated in a spherical 
Wigner-Seitz cell around each lattice site (see Ref.~\cite{chamel2025b} for further details).  We will take into account the Skyrme effective mass $m_n^\oplus(\rb)$ and the Skyrme potential vector $\pmb{I_n}(\rb)$ through a suitable readjustment of the scalar potential $U_n(\rb)$, as discussed in Ref.~\cite{chamel2025}. 

The Floquet-Bloch boundary conditions $\varphi_{\alpha\pmb{k}}(\rb+\pmb{\ell},\sigma)=e^{i \pmb{k}\cdot \pmb{\ell}}\varphi_{\alpha\pmb{k}}(\rb,\sigma)$
for any lattice translation vector $\pmb{\ell}$, is automatically enforced by expanding the wavefunction into Fourier 
series,
\beqy
\label{eq:plane-wave-expansion}
\varphi_{\pmb{k}}(\rb,\sigma)=\frac{1}{\sqrt{\Omega}}\exp({\rm i}\, \pmb{k}\cdot\rb)\sum_{\pmb{G}} \widetilde{\varphi}_{\pmb{k}}(\pmb{G}) \exp({\rm i} \pmb{G}\cdot\rb)\chi(\sigma)\, ,
\eeqy
with the normalization 
\beqy
\label{eq:norm-wavefunction}
\sum_{\pmb{G}} |\widetilde{\varphi}_{\pmb{k}}(\pmb{G})|^2=1\, .
\eeqy
Here $\chi$ denotes Pauli spinor and the summation is over reciprocal lattice vectors $\pmb{G}=m_1 \pmb{b_1}+m_2 \pmb{b_2}+m_3 \pmb{b_3}$ with ($m_1$, $m_2$, $m_3$) arbitrary integers and $\pmb{b_1}$, $\pmb{b_2}$, and $\pmb{b_3}$ are the corresponding primitive basis vectors satisfying $\pmb{a_i}\cdot \pmb{b_j}=2\pi \delta_{ij}$ ($\delta_{ij}$ is Kronecker's symbol), namely 
\beqy \label{eq:reciprical-primitive-vectors}
 \pmb{b_1}&=&\frac{2\pi}{a}\left(\pmb{\hat y} + \pmb{\hat z}\right) \, ,\notag \\ 
 \pmb{b_2}&=&\frac{2\pi}{a}\left( \pmb{\hat x}  + \pmb{\hat z}\right) \, ,\notag \\ 
 \pmb{b_3}&=&\frac{2\pi}{a}\left( \pmb{\hat x} + \pmb{\hat y} \right) \, .
\eeqy  
In this way, the HF equation reduces to a matrix eigenvalue problem

As in our previous studies, we assume a constant pairing gap for all bands. This is a reasonable approximation,  at least for the density-dependent contact pairing interactions that are generally associated with Skyrme functionals, (see, e.g., Fig.  3 in Ref.~\cite{chamel2010b}).  Although the gaps exhibit some variations for bound states, the contribution of these states to the superfluid fraction is negligible, as we have explicitly checked and will be discussed in the next section. As shown in Ref.~\cite{chamel2010b}, the gaps for unbound states can be fairly well estimated from the pairing gap in infinite homogeneous neutron matter at the same average neutron density. We have considered here the pairing gaps obtained from the diagrammatic many-body calculations of Ref.\cite{cao2006} taking into account both polarization and self-energy effects, which can be conveniently fitted as \cite{chamel2024}
\beqy \label{eq:pairing-gap-fit}
\Delta(\bar n_n) = H(k_m - k_{Fn})\Delta_0\frac{k_{Fn}^2}{k_{Fn}^2 + k_1^2}\,\frac{(k_{Fn} - k_2)^2}{(k_{Fn} - k_2)^2 + k_3^2} \, ,
\eeqy
where $k_{Fn} = (3\pi^2 \bar n_n)^{1/3}$, $H$ is the unit-step Heaviside distribution and the parameters $\Delta_0, k_1, k_2, k_3$ and $k_m$ are given in
Table~\ref{tab1}. To assess the role of pairing, we will also consider lower values of $\Delta$.

\begin{table}
\centering
\caption{Parameters for Eq.\eqref{eq:pairing-gap-fit}.}
\label{tab1}
\begin{tabular} {|ccccc|}
\hline
$\Delta_0$ [MeV] & $k_1$ [fm$^{-1}$] & $k_2$ [fm$^{-1}$] & $k_3$ [fm$^{-1}$] & $k_m$ [fm$^{-1}$] \\
\hline
3.37968 & 0.556092 & 1.38236 & 0.327517 & 1.38  \\
\hline
\end{tabular}
\end{table}

Quantum zero-point motion of ions about their equilibrium position is taken into account through the introduction of the Debye-Waller factor in the Fourier transform of the mean field
\beqy\label{eq:U+DW}
\widetilde{U}_n(\pmb{G})\equiv\frac{1}{\Omega}\int  d^3\rb\, U_n(\rb) \exp(- i\, \pmb{G}\cdot\rb)
 \rightarrow \widetilde{U}_n(\pmb{G})\exp\left(-\frac{1}{6} G^2\langle \delta \rb(t)^2\rangle\right)\, ,
\eeqy
where 
\begin{equation}
\label{eq:zero-point-motion}
\langle \delta \rb(t)^2\rangle \approx  \frac{3 \hbar \langle \omega^\star_I/\omega_{ph}\rangle }{2 m^\star_I \omega^\star_I} \, , 
\end{equation}
is the mean square displacement of ions,  
\beqy \label{eq:ion-plasma-frequency}
\omega^\star_I = \sqrt{\frac{4\pi Z^2 e^2 n_I}{m^\star_I}}
\eeqy 
is the ion plasma frequency, 
and $\langle \omega^\star_I/\omega_{ph}\rangle \simeq  2.798 55$ is the thermal average of the inverse phonon frequencies~\cite{Baiko2001} (see Ref.~\cite{chamel2025b} for further details). 
In principle, the Debye-Waller factor should be calculated self-consistently as the effective ion mass itself defined by Eq.~\eqref{eq:effective-ion-mass} depends on the neutron superfluid fraction, which in turn is determined by the neutron band structure therefore by the mean field \eqref{eq:U+DW}. Because $\rho_{n,s}$ can also depend on pairing, calculations should be repeated for each value of $\Delta$. To reduce the computational cost, we replace $m_I^\star$ by the ``bare'' ion mass $m_I$ to evaluate the mean square displacement~\eqref{eq:zero-point-motion} entering the Debye-Waller factor in Eq.~\eqref{eq:U+DW}.

As shown from band-structure calculations,  unbound neutrons are also present inside ions (see,  e.g.,  Fig.~2 from Ref.~\cite{chamel2007}).  This leads to some ambiguity in the distinction between ``free'' and ``confined'' neutrons,  therefore also in the specification of the ion mass $m_I$, as discussed in Ref.~\cite{CCH06}.  To compare with the hydrodynamic description, we have thus adopted two different points of view.  First,  $m_I$ can be defined as the mass the ions would have if they were unaffected by the motion of ``free'' neutrons,  whose mass would therefore be the same as in vacuum.   
It follows from Eq.~\eqref{eq:effective-neutron-mass} that we should define the bare ion mass by setting $\rho_{n,s}=\rho_{n,f}$ in Eq.~\eqref{eq:effective-neutron-mass} leading to 
\beqy \label{eq:bare-ion-mass}
m_I = \frac{\bar \rho - \rho_{n,f}}{n_I} 
\eeqy  
with the ion number density simply given by $n_I=1/\Omega_\mathrm{cell}$ (the Wigner-Seitz cell containing one ion). 
This is the definition adopted in our previous study~\cite{chamel2025b}.  
Alternatively,  $m_I$ can be defined as the mass of bound nucleons in the quantum mechanical sense, namely nucleons whose single-particle energies lie below the maximum value of the mean-field potential $U_n(\rb)$. These states correspond to essentially flat bands in $\pmb{k}$ space and therefore their contribution to the 
average neutron mass current is expected to be negligible.  With this definition, $m_I$ thus represents the mass of nucleons that remain at rest in the crust frame and do not flow with the superfluid. This definition is more consistent with the ion mass~\eqref{eq:ion-mass-transport} introduced in the hydrodynamic description, as confirmed by numerical calculations. To assess the sensitivity of entrainment effects to lattice vibrations, we have repeated the band-structure calculations using the Debye-Waller factor considering these two different definitions of the ion mass. 


In practice, the HF matrix is diagonalized for each given Bloch wave vector $\pmb{k}$ in the first Brillouin zone keeping a finite number $\mathcal{N}^3$ of reciprocal lattice vectors
($\mathcal{N}$ along each reciprocal basis vector). The HF equation is thus discretized 
 in a spatial grid inside the primitive cell defined by the primitive basis vectors of the lattice, with $\mathcal{N}^3$ points defined by $\rb=(i_1/\mathcal{N})\pmb{a_1}+(i_2/\mathcal{N})\pmb{a_2}+(i_3/\mathcal{N})\pmb{a_3}$, where $i_1$, $i_2$, and $i_3$ are positive integers less or equal than $\mathcal{N}-1$. 
The HF matrix can be efficiently computed using fast Fourier transforms. For each Bloch wave vector, the HF equation is solved on grid of up to $25\times 25\times 25$ points in the primitive cell to ensure a spatial resolution $\delta x=\delta y=\delta z$ of about 1 fm or below (the spacing along each primitive axis is $\sqrt{3}$ times larger). See Ref.~\cite{chamel2025} for further discussion and convergence tests.

Summations over $\pmb{k}$ are performed using the special-point method~\cite{hama1992} (see, Ref. \cite{chamel2025} for further details and numerical tests). 


\subsection{Superfluid fraction}
\label{sec:results-superfluid-fraction}

As shown in Ref.~\cite{almirante2024}, the superfluid density~\eqref{eq:superfluid-density-full} can be decomposed into two terms:  
\beqy \label{eq:superfluid-density-decomposition}
\rho_{n,s}= \frac{m_n^2}{3 \Omega} \sum_{\pmb{k}} \Biggl[\sum_{\alpha} \vert\pmb{v}^0_{\alpha\pmb{k}}\vert^2 \frac{\Delta^2}{(E^0_{\alpha \pmb{k}})^3}  
+ 2\sum_{\alpha\neq\beta} \frac{\Delta^2}{E^0_{\alpha \pmb{k}}E^0_{\beta \pmb{k}}(E^0_{\alpha \pmb{k}}+E^0_{\beta \pmb{k}})} \pmb{v}^0_{\alpha\pmb{k},\beta\pmb{k}}\cdot\pmb{v}^0_{\beta\pmb{k},\alpha\pmb{k}}\Biggr ]\, .
\eeqy 
The first term accounting for the intraband contribution was originally derived in Ref.~\cite{CCH05b}. It was shown to be very weakly dependent on the gap and remains finite 
in the limit $\Delta\rightarrow 0$~\cite{chamel2025}. On the contrary, the second term vanishes in this limit and becomes increasingly important with increasing $\Delta$. This interband contribution only exists due to the presence of ions: in homogeneous matter, the single-particle states are pure plane waves and the velocity matrix $\pmb{v}^0_{\beta\pmb{k},\alpha\pmb{k}}=\delta_{\alpha\beta} \pmb{v}^0_{\alpha\pmb{k}}$ is therefore diagonal. 

The superfluid fraction $\rho_{n,s}/\bar \rho_{n}$ is computed from Eq.~\eqref{eq:superfluid-density-full} with up to 440 special points in the irreducible domain (15400 wave vectors $\pmb{k}$ in the first Brillouin zone) and up to about 2400 bands for different pairing gaps $\Delta$. To assess the relative importance of the intraband and  interband contributions, we have also computed each of them separately. As discussed in Ref.~\cite{chamel2025}, 
the convergence of the special-point method for the calculation of the intraband contribution deteriorates for very low values of $\Delta$. This is because the integrand is singular on the Fermi surface (summations could still be performed by smearing the Fermi-Dirac distribution, as discussed in Ref.~\cite{chamel2025}). For this reason, we have found it more reliable to infer the intraband contribution by subtracting the interband contribution to the total superfluid fraction. 
We have evaluated the superfluid fraction in the limit $\Delta=0$ ($\rho_{n,s}/\bar \rho_{n}$  is then entirely due to the interband response) using the semianalytical method of Gilat and Raubenheimer~\cite{gilat1966} including up to to 65280 Bloch wave vectors in the first Brillouin zone (see Ref.~\cite{chamel2025} for further details and numerical tests). As discussed in the previous section, we have performed three series of calculations: (i) assuming a perfectly rigid lattice, (ii) including the Debye-Waller factor with the ion mass~\eqref{eq:bare-ion-mass}, (iii) including the Debye-Waller with the ion mass defined from quantum-mechanically bound nucleons.

\begin{figure}
	\includegraphics[scale=0.6]{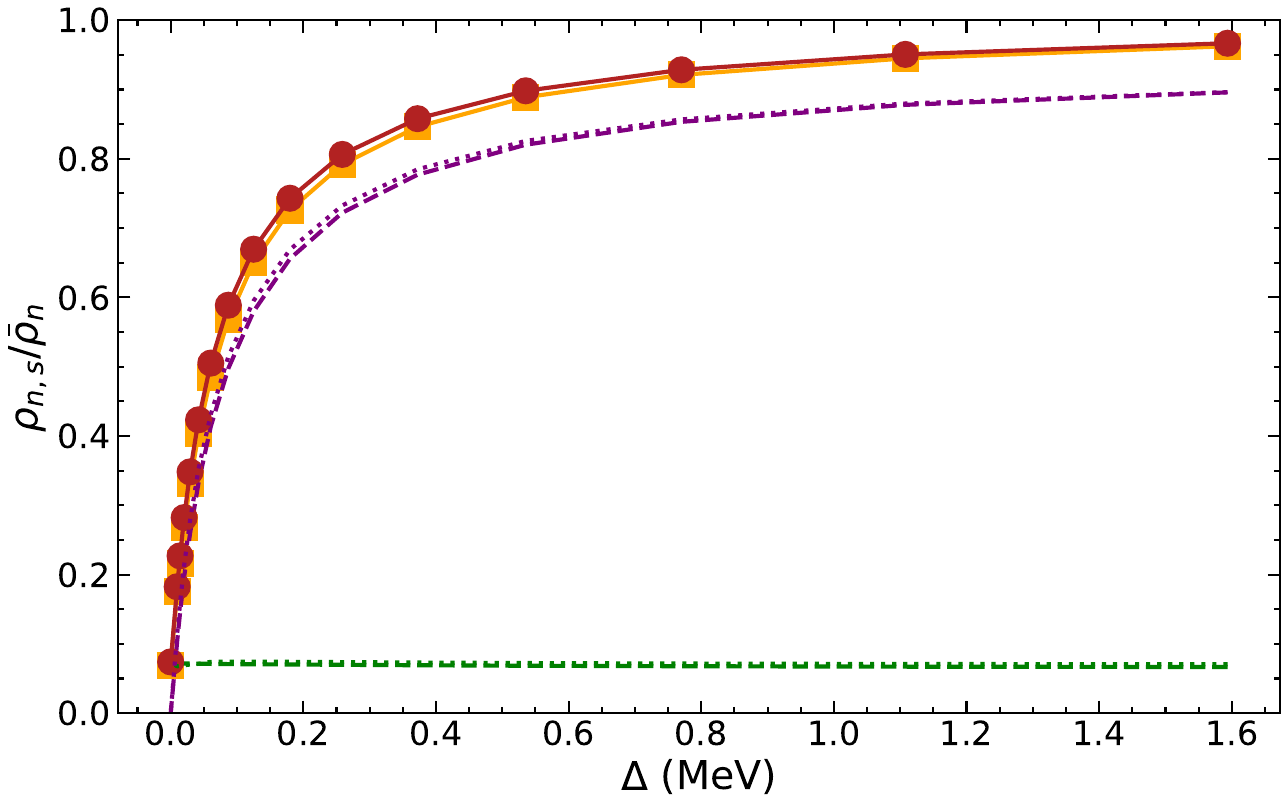}
	\vskip -0.5cm
	\caption{Neutron superfluid fraction $\rho_{n,s}/\bar \rho_n$ as a function of the pairing gap $\Delta$ (in MeV) in the inner crust of a neutron star at the average baryon number density $\bar n=0.03$~fm$^{-3}$: full response (orange solid line with squares), interband contribution (purple dashed line) and intraband contribution (green dashed line). Results obtained including lattice vibrations are represented by the red solid line with circles and dotted lines.}
	\label{fig:superfluid-fraction-Z40A1590}
\end{figure}
\begin{figure}
	\includegraphics[scale=0.6]{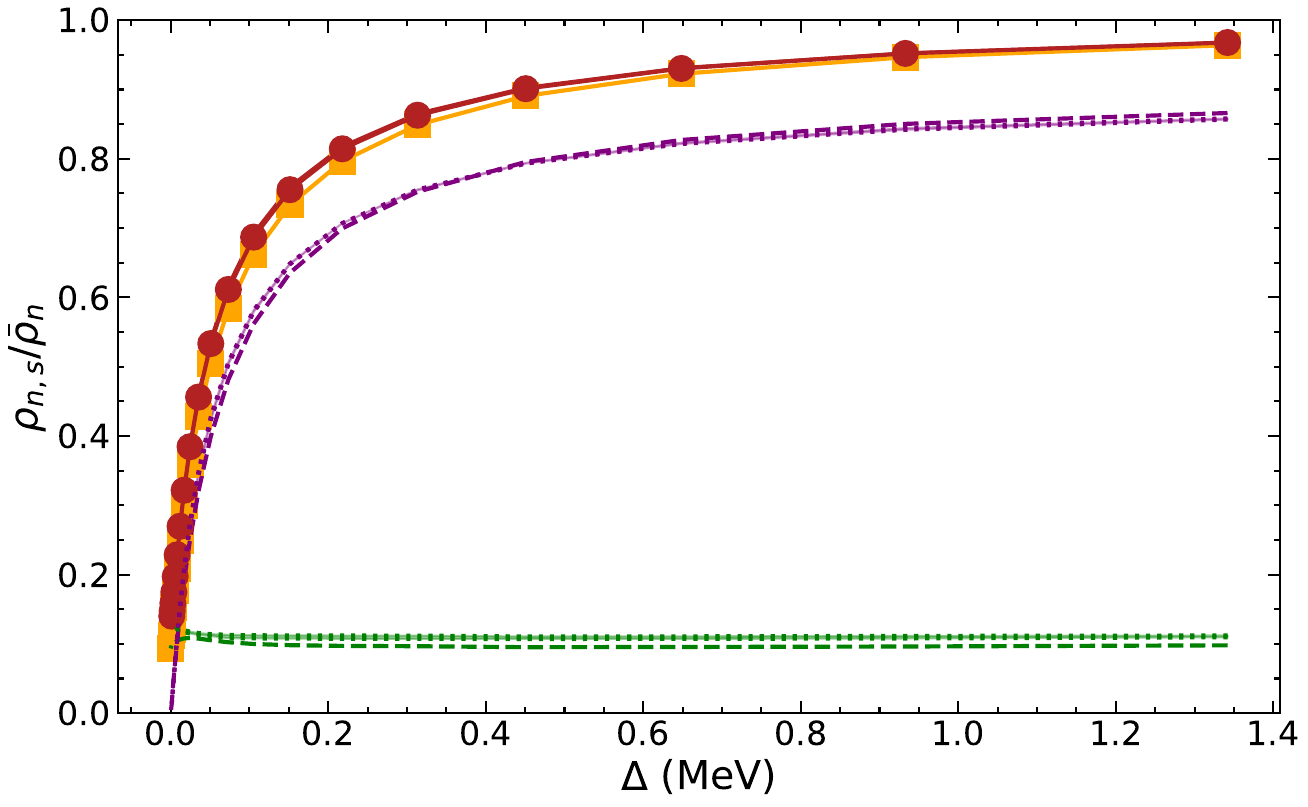}
	\vskip -0.5cm
	\caption{Same as Fig.~\ref{fig:superfluid-fraction-Z40A1590} at the average baryon number density $\bar n=0.04$~fm$^{-3}$. 
	}
	\label{fig:superfluid-fraction-Z40A1610}
\end{figure}
\begin{figure}
	\includegraphics[scale=0.6]{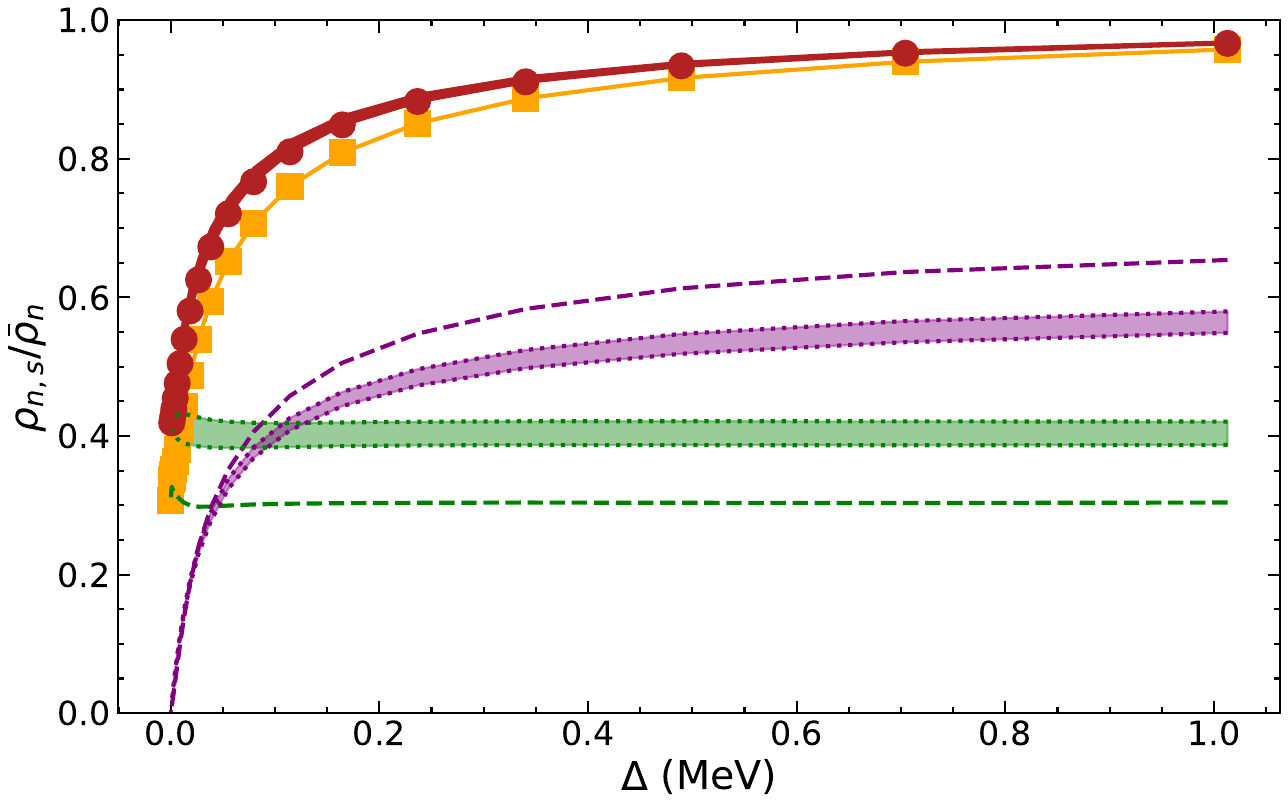}
	\vskip -0.5cm
	\caption{Same as Fig.~\ref{fig:superfluid-fraction-Z40A1590} at the average baryon number density $\bar n=0.05$~fm$^{-3}$. 
}
	\label{fig:superfluid-fraction-Z20A800}
\end{figure}
\begin{figure}
	\includegraphics[scale=0.6]{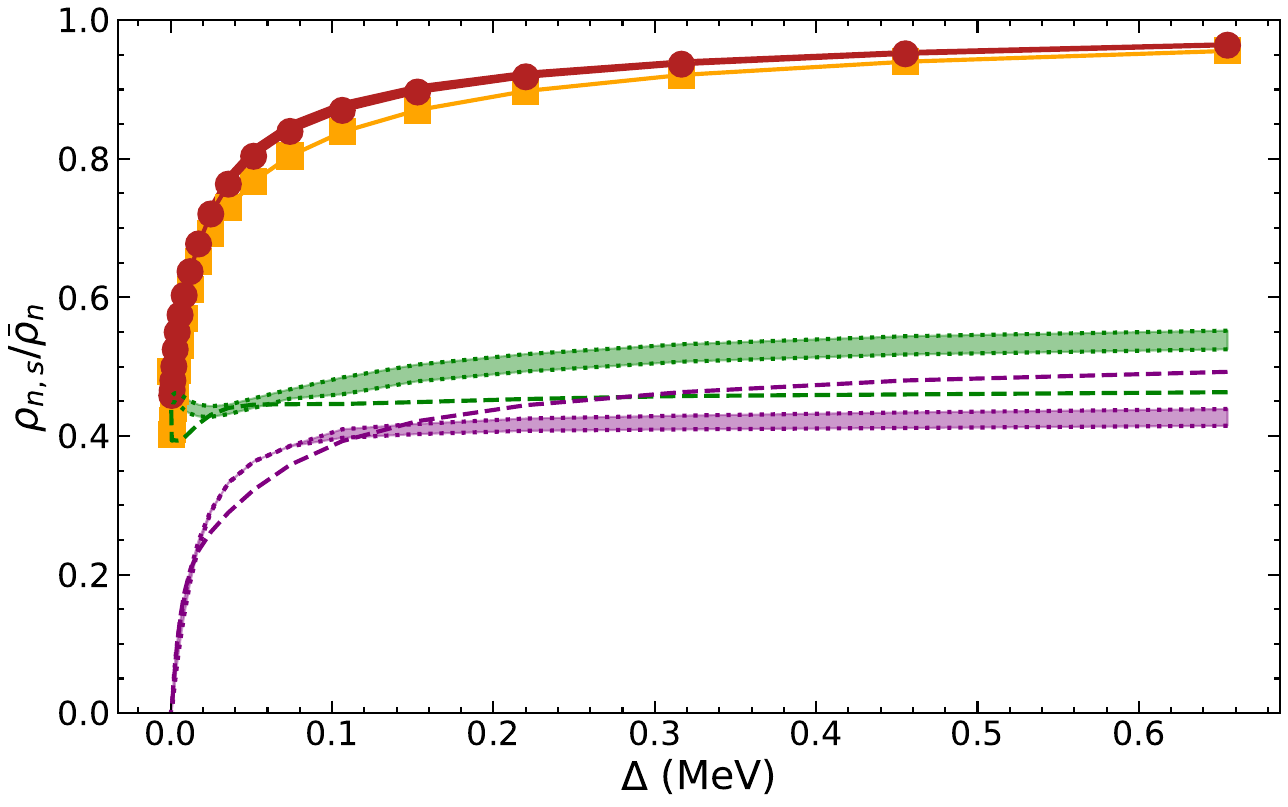}
	\vskip -0.5cm
	\caption{Same as Fig.~\ref{fig:superfluid-fraction-Z40A1590} at the average baryon number density $\bar n=0.06$~fm$^{-3}$. 
}
	\label{fig:superfluid-fraction-Z20A780}
\end{figure}
\begin{figure}
	\includegraphics[scale=0.6]{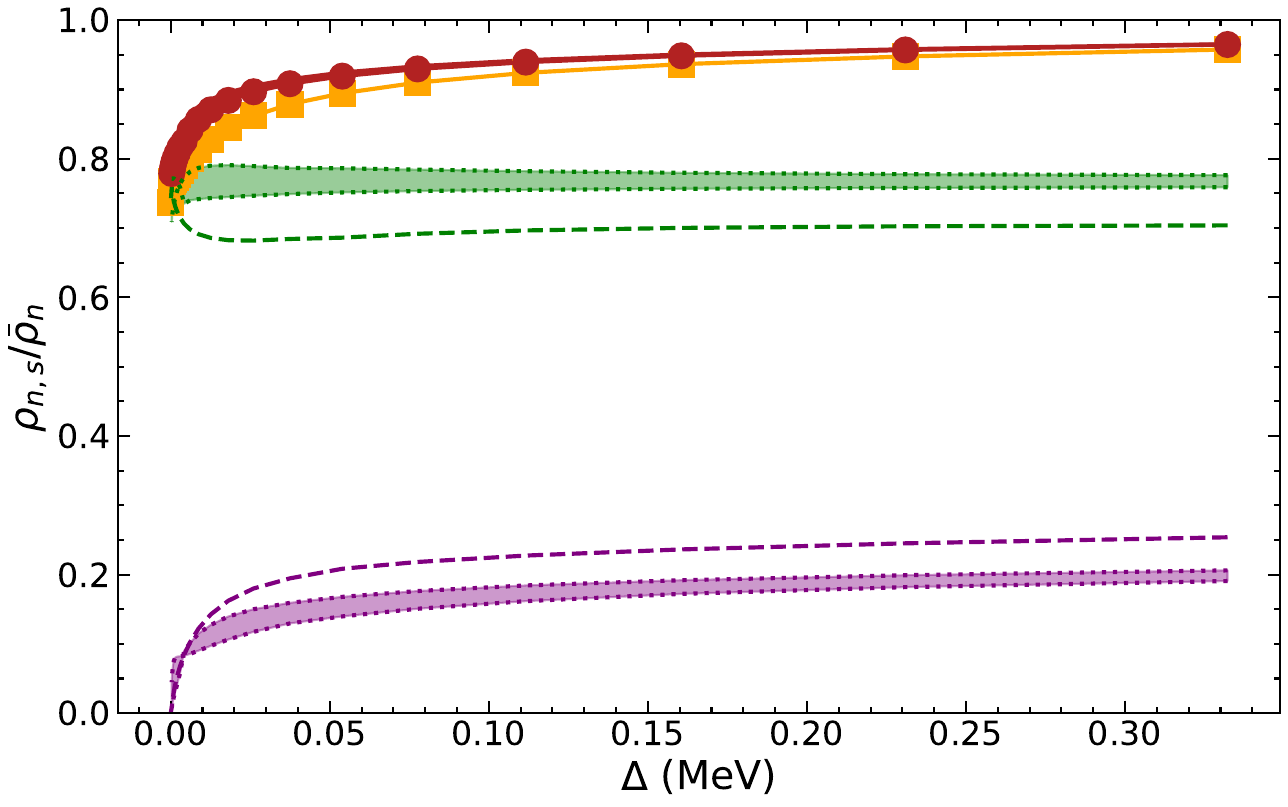}
	\vskip -0.5cm
	\caption{Same as Fig.~\ref{fig:superfluid-fraction-Z40A1590} at the average baryon number density $\bar n=0.07$~fm$^{-3}$. 
}
	\label{fig:superfluid-fraction-Z20A714}
\end{figure}

Results for the superfluid fraction at the average baryon number density $\bar n=0.03$~fm$^{-3}$ are plotted in Fig.~\ref{fig:superfluid-fraction-Z40A1590}. The superfluid fraction rapidly increases with $\Delta$ and reaches about 96\% for the ``realistic'' pairing gap $\Delta=1.59$~MeV. 
This very large superfluid fraction is almost entirely due to the interband response except in the limit $\Delta \rightarrow 0$,  thus confirming the results of Ref.~\cite{almirante2025}. 
Going deeper in the inner crust, the intraband response increases while the interband response decreases, as can be seen comparing Figs.~\ref{fig:superfluid-fraction-Z40A1610}, \ref{fig:superfluid-fraction-Z20A800}, \ref{fig:superfluid-fraction-Z20A780} and \ref{fig:superfluid-fraction-Z20A714}. The two contributions become comparable at the average baryon number density $\bar n=0.06$~fm$^{-3}$. At higher densities, the superfluid fraction is mainly determined by the intraband response for all values of $\Delta$. The intraband contributions to the superfluid fraction calculated here are larger than those reported in Ref.~\cite{chamel2012}, especially in the densest layers. The deviations come from a better interpolation of the fields using cubic splines and an improved implementation of the Fourier transforms. 

The values of the ion masses in the different crustal layers are collected in Table~\ref{tab:crust-composition}. The definition~\eqref{eq:bare-ion-mass} yields masses that are systematically larger than the mass of quantum-mechanically bound nucleons, thus leading to smaller mean-square displacements of ions~\eqref{eq:zero-point-motion}. 
The influence of lattice vibrations on  $\rho_{n,s}/\bar \rho_{n}$ becomes significant above $0.04$~fm$^{-3}$, reducing the interband response while enhancing the intraband one. This can be understood from the fact that lattice vibrations effectively render the superfluid more homogeneous: the Fourier components of the mean-field potential with finite $G$ are suppressed by the Debye-Waller factor~\eqref{eq:U+DW} thus reducing the effects of Bragg-scattering (if all components $\widetilde{U}_n(\pmb{G})$ except $\widetilde{U}_n(\pmb{0})$ were to vanish, $U_n(\rb)$ would be uniform). More surprisingly, the overall superfluid fraction remains essentially unchanged for realistic pairing gaps.

The role of the interband response was originally discussed in condensed-matter physics in the context of flat-band superconductivity~\cite{peotta2015}. We have thus examined more closely the contribution from bound states for which bands are essentially independent of $\pmb{k}$. This is because the wave functions are localized inside the clusters, and are therefore insensitive to the boundary conditions. Their contributions to both the intraband and interband parts of the superfluid fraction turn out to be completely negligible. This stems from the fact that both the intraband and interband responses scale roughly like $\Delta^2/E^3$ and are therefore mainly determined by single-particle states around the Fermi level: $\Delta^2/E^3$ is sharply peaked around $\xi=0$ and is reduced by a factor hundred for $\xi_{\rm max}\approx  4.53 \Delta$. Keeping only bands lying within this range leads to a very good estimate of the intraband and interband contributions to the superfluid fraction within a few \% for realistic pairing gaps.  However, the deviations are expected to be larger in shallower regions of the crust where the chemical potential lies slightly above the maximum value of the potential.

\subsection{Effective ion mass}
\label{sec:results-effective-ion-mass}

\begin{figure}
	\includegraphics[scale=0.6]{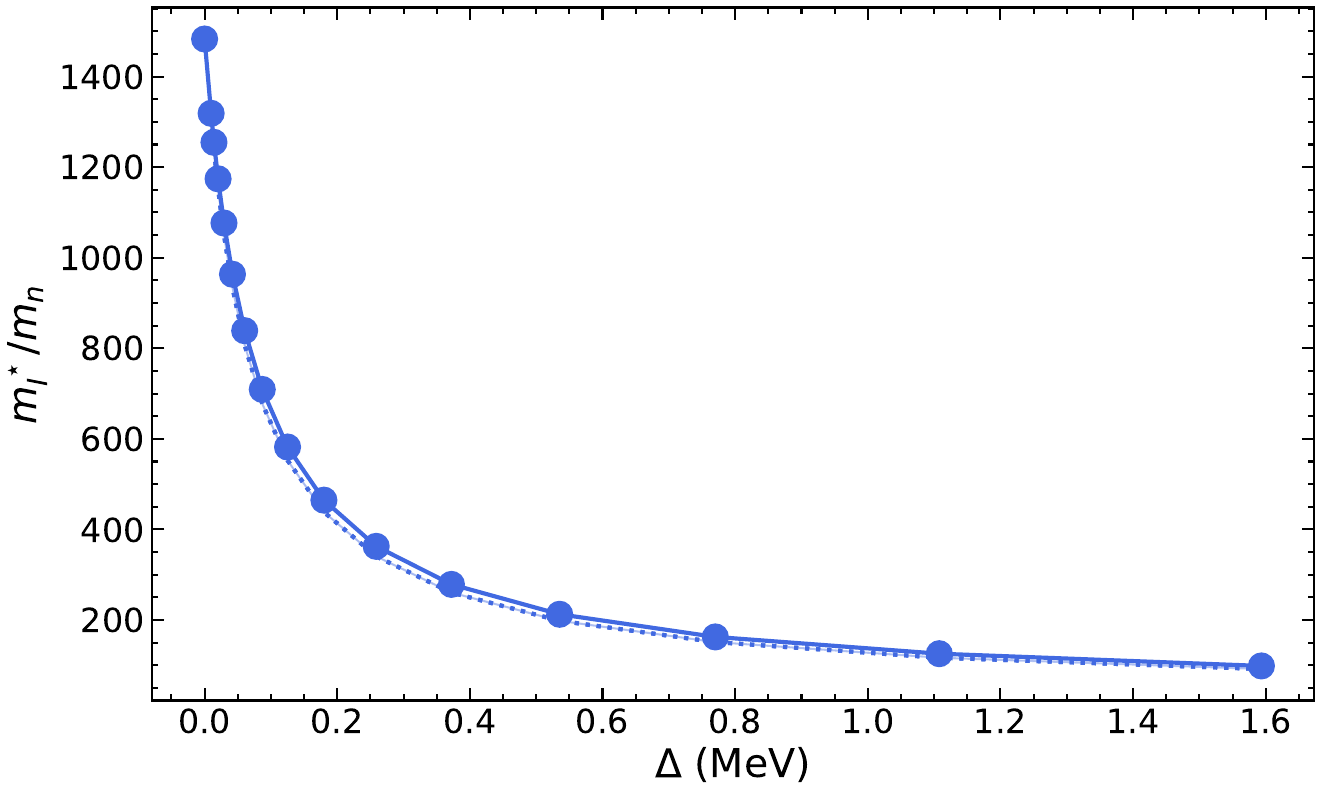}
	\vskip -0.5cm
	\caption{Effective ion mass $m_I^\star$ in units of the neutron mass $m_n$ as a function of the neutron pairing gap $\Delta$ (in MeV) in the inner crust of a neutron star at the average baryon number density $\bar n=0.03$~fm$^{-3}$. The solid line with circles was calculated for a perfectly rigid lattice. Results obtained including lattice vibrations are represented by dotted lines. 
	}
	\label{fig:effective-ion-mass-Z40A1590}
\end{figure}

\begin{figure}
	\includegraphics[scale=0.6]{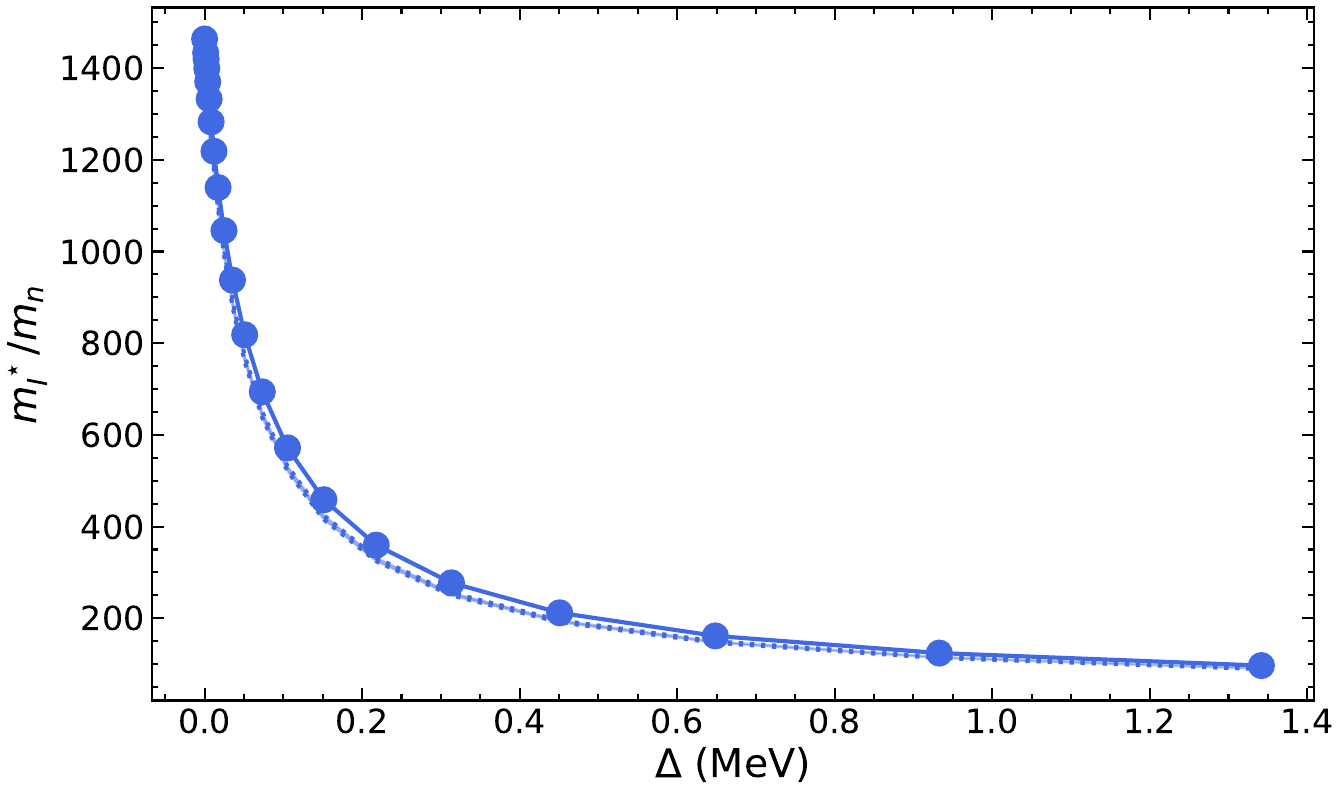}
	\vskip -0.5cm
	\caption{Same as Fig.~\ref{fig:effective-ion-mass-Z40A1590} at the average baryon number density $\bar n=0.04$~fm$^{-3}$.   }
	\label{fig:effective-ion-mass-Z40A1610}
\end{figure}

\begin{figure}
	\includegraphics[scale=0.6]{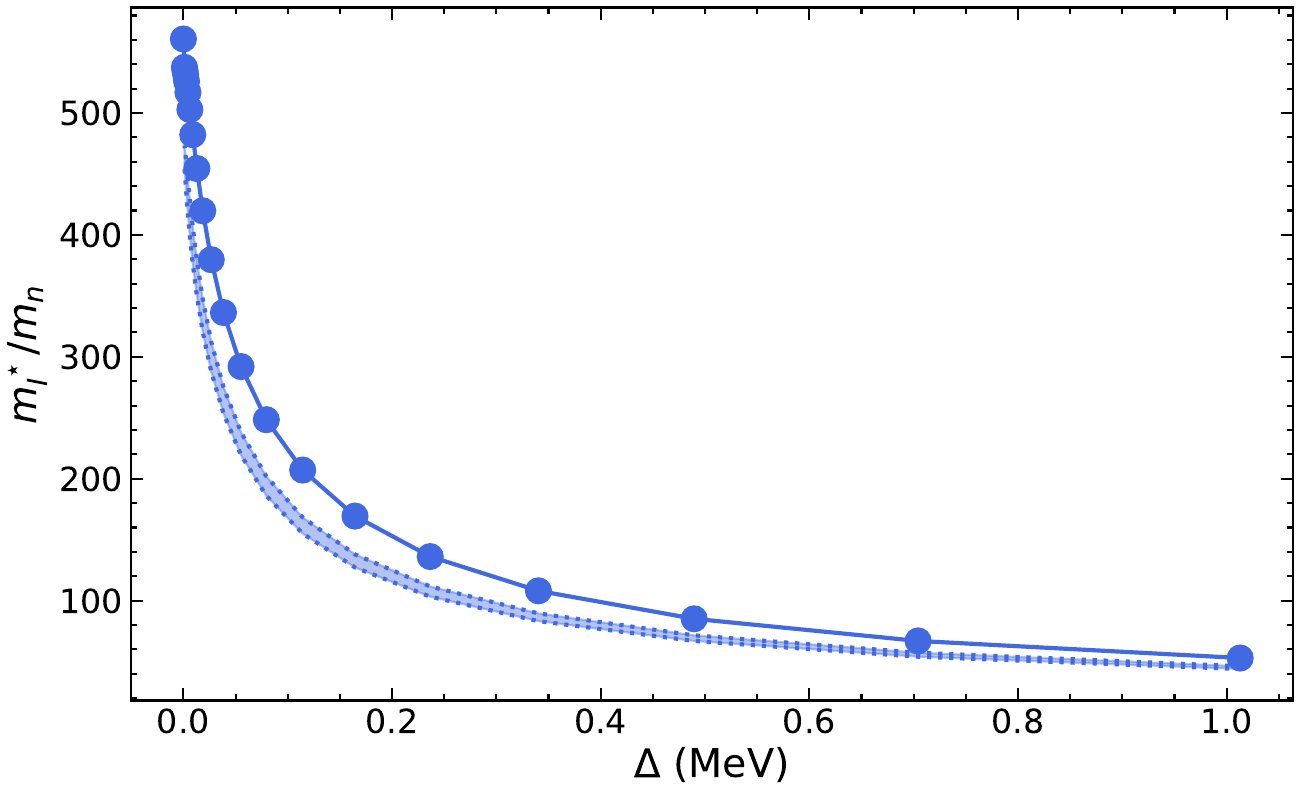}
	\vskip -0.5cm
	\caption{Same as Fig.~\ref{fig:effective-ion-mass-Z40A1590}  at the average baryon number density $\bar n=0.05$~fm$^{-3}$.  }
	\label{fig:effective-ion-mass-Z20A800}
\end{figure}

\begin{figure}
	\includegraphics[scale=0.6]{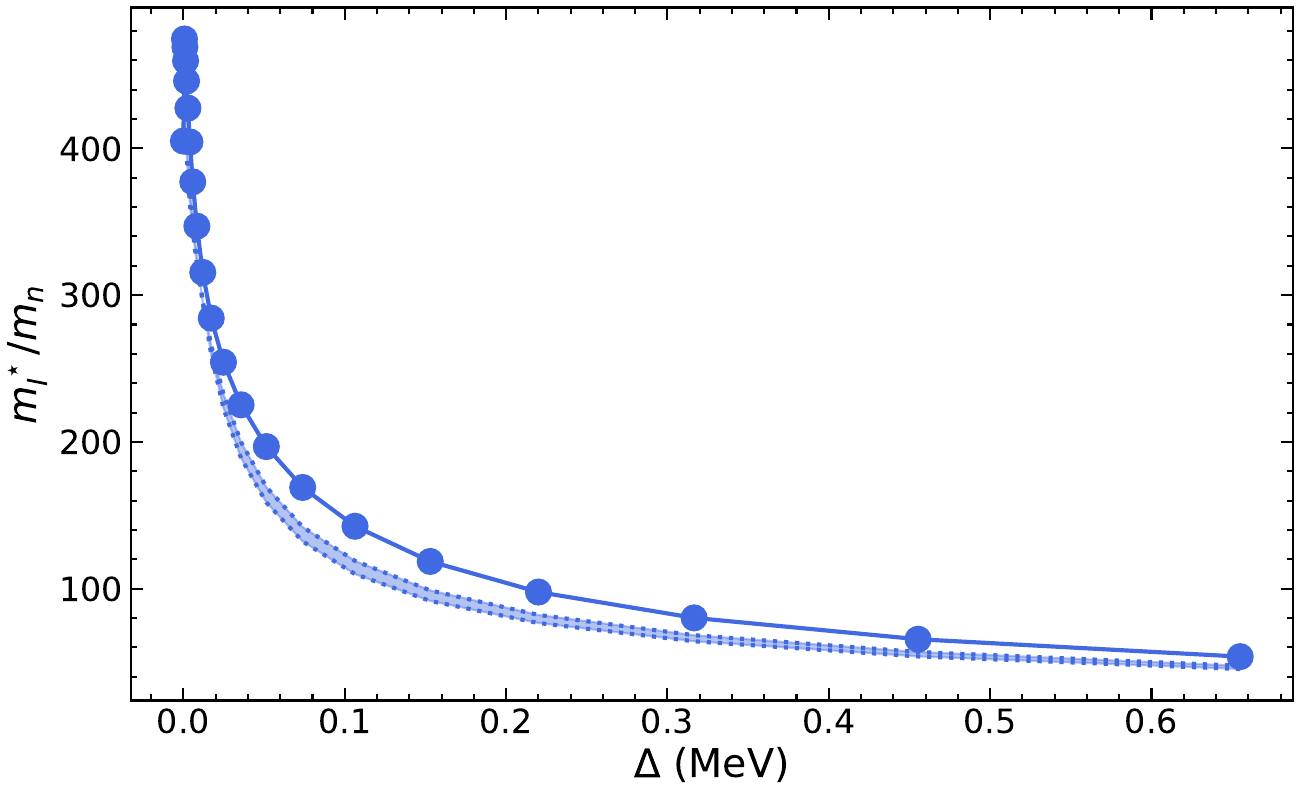}
	\vskip -0.5cm
	\caption{Same as Fig.~\ref{fig:effective-ion-mass-Z40A1590}  at the average baryon number density $\bar n=0.06$~fm$^{-3}$.  }
	\label{fig:effective-ion-mass-Z20A780}
\end{figure}

\begin{figure}
	\includegraphics[scale=0.6]{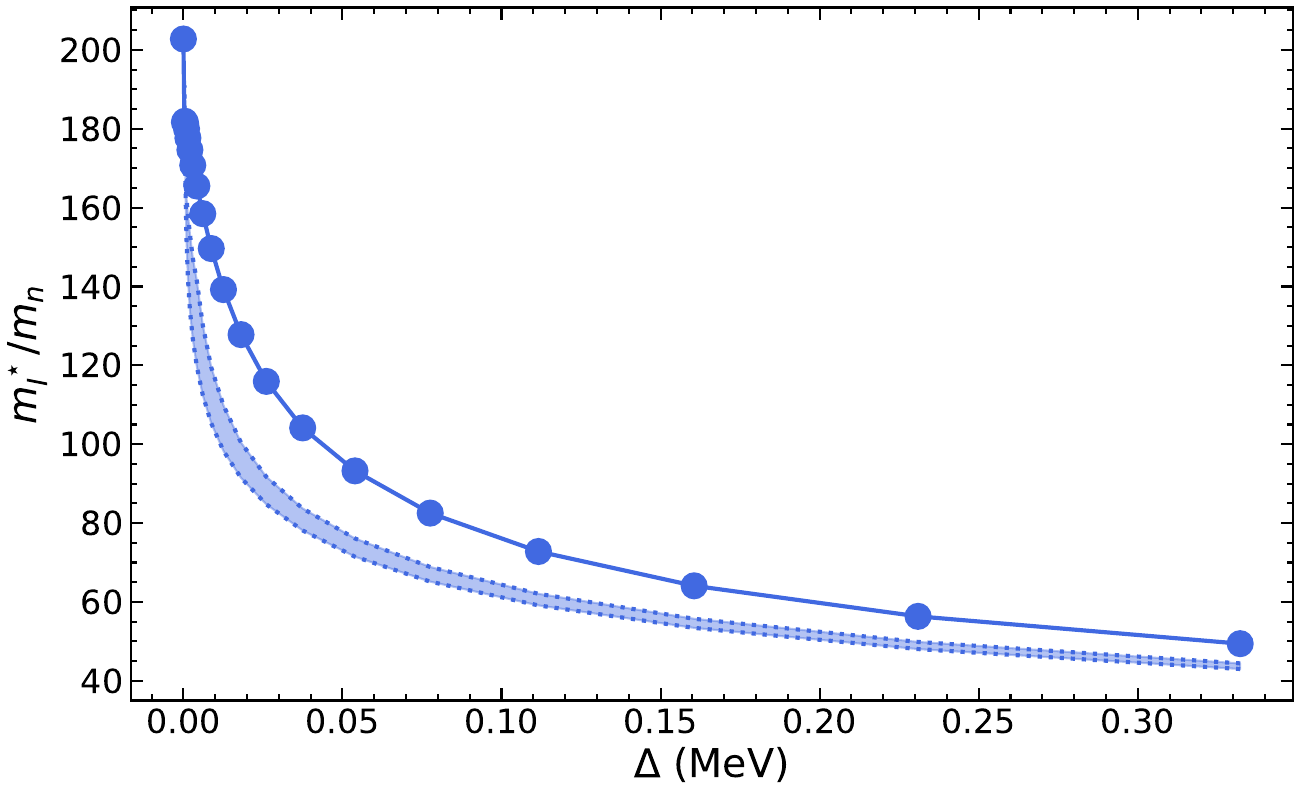}
	\vskip -0.5cm
	\caption{Same as Fig.~\ref{fig:effective-ion-mass-Z40A1590} at the average baryon number density $\bar n=0.07$~fm$^{-3}$.  }
	\label{fig:effective-ion-mass-Z20A714}
\end{figure}

After having determined the superfluid fraction,  we have evaluated the effective ion mass  from  Eq.~\eqref{eq:effective-neutron-mass}, which can be equivalently written as
\beqy 
m^\star_I=Z m_p + (A_\mathrm{cell}-Z)\left(1-\frac{\rho_{n,s}}{\bar\rho_n}\right)m_n \, .
\eeqy 
Results obtained at different baryon densities, displayed in Figs.~\ref{fig:effective-ion-mass-Z40A1590}, \ref{fig:effective-ion-mass-Z40A1610}, \ref{fig:effective-ion-mass-Z20A800}, \ref{fig:effective-ion-mass-Z20A780} and \ref{fig:effective-ion-mass-Z20A714}, are qualitatively similar.  The effective ion mass is found to be very sensitive to the pairing gap,  being enhanced by an order of magnitude for vanishing gap,  but decreasing sharply with increasing $\Delta$. The effects of lattice vibrations are significant above 0.04~fm$^{-3}$ and lead to lower effective ion masses. But the reduction remains small for realistic gaps. The values of the effective ion mass for realistic gaps are collected in Table~\ref{tab:crust-composition}. 


\begin{figure}
	\includegraphics[scale=0.6]{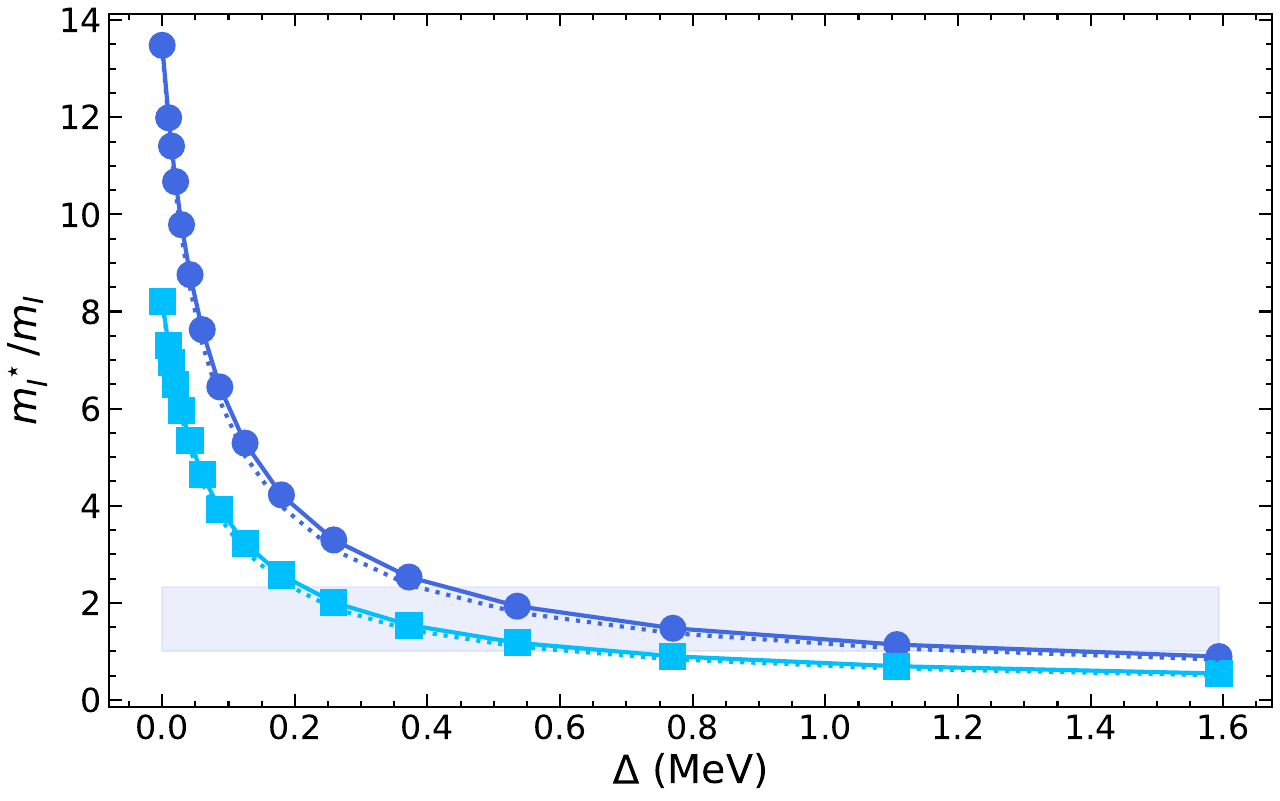}
	\vskip -0.5cm
	\caption{Effective ion mass ratio $m_I^\star/m_I$ as a function of the neutron pairing gap $\Delta$ (in MeV) in the inner crust of a neutron star at the average baryon number density $\bar n=0.03$~fm$^{-3}$.  The solid lines with squares and circles were calculated with $m_I$ defined by Eq.~\eqref{eq:bare-ion-mass} and by counting bound states,  respectively.  Results obtained including lattice vibrations are represented by dotted lines.  The shaded area represent the range of values predicted by classical hydrodynamics from Eqs.~\eqref{eq:dimensionless-effective-ion-mass1},  the lower and upper limits corresponding to $\rho_{s,i}=\rho_{n,o}$ and $\rho_{s,i}=\rho_{n,i}$ respectively.   
	}
	\label{fig:effective-ion-mass-ratio-Z40A1590}
\end{figure}

\begin{figure}
	\includegraphics[scale=0.6]{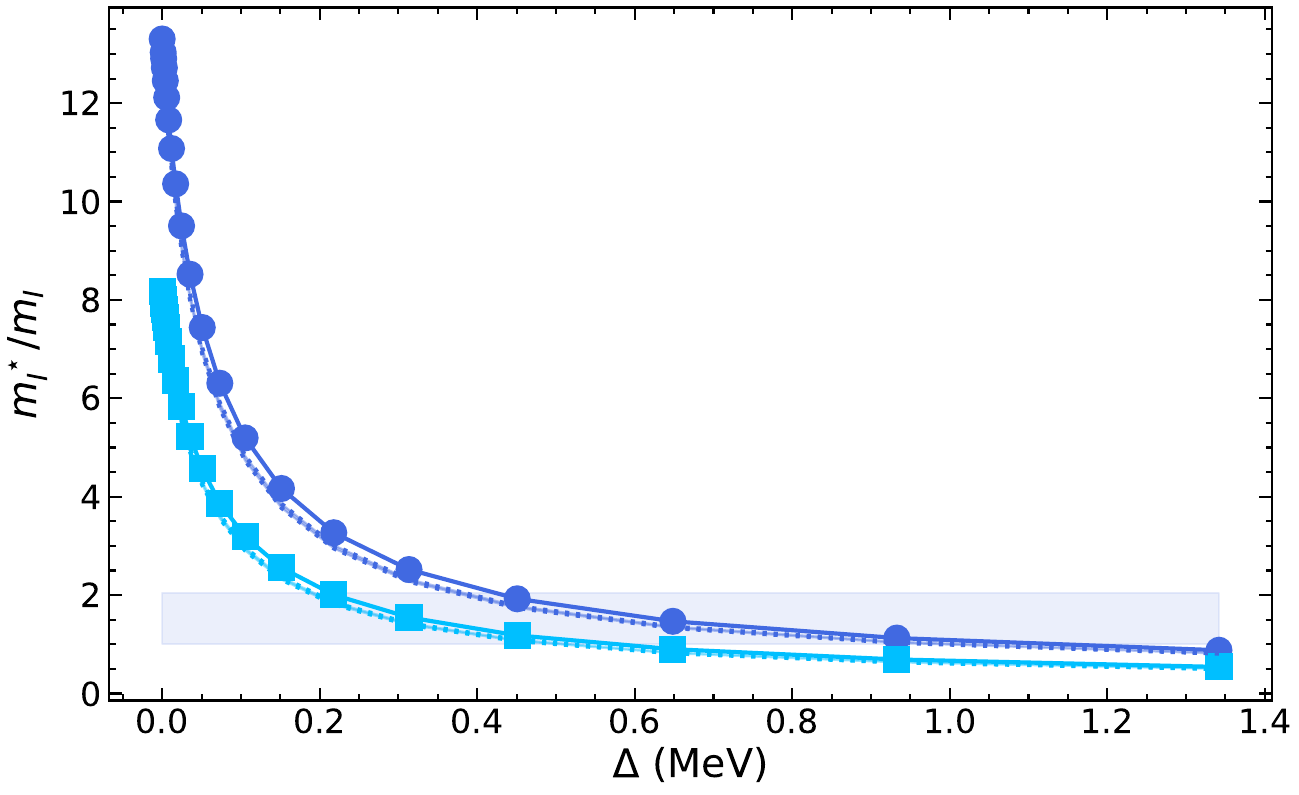}
	\vskip -0.5cm
	\caption{Same as Fig.~\ref{fig:effective-ion-mass-ratio-Z40A1590} at the average baryon number density $\bar n=0.04$~fm$^{-3}$.   }
	\label{fig:effective-ion-mass-ratio-Z40A1610}
\end{figure}

\begin{figure}
	\includegraphics[scale=0.6]{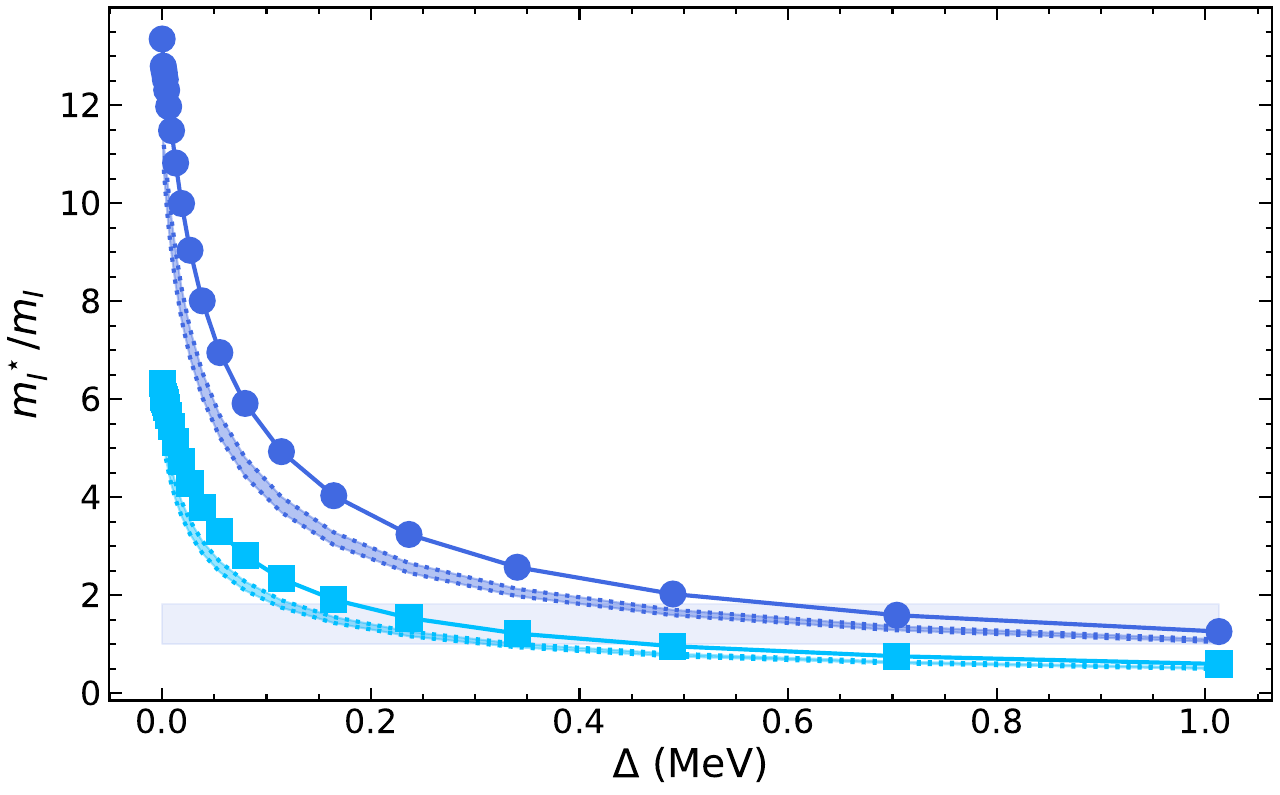}
	\vskip -0.5cm
	\caption{Same as Fig.~\ref{fig:effective-ion-mass-ratio-Z40A1590}  at the average baryon number density $\bar n=0.05$~fm$^{-3}$.  }
	\label{fig:effective-ion-mass-ratio-Z20A800}
\end{figure}

\begin{figure}
	\includegraphics[scale=0.6]{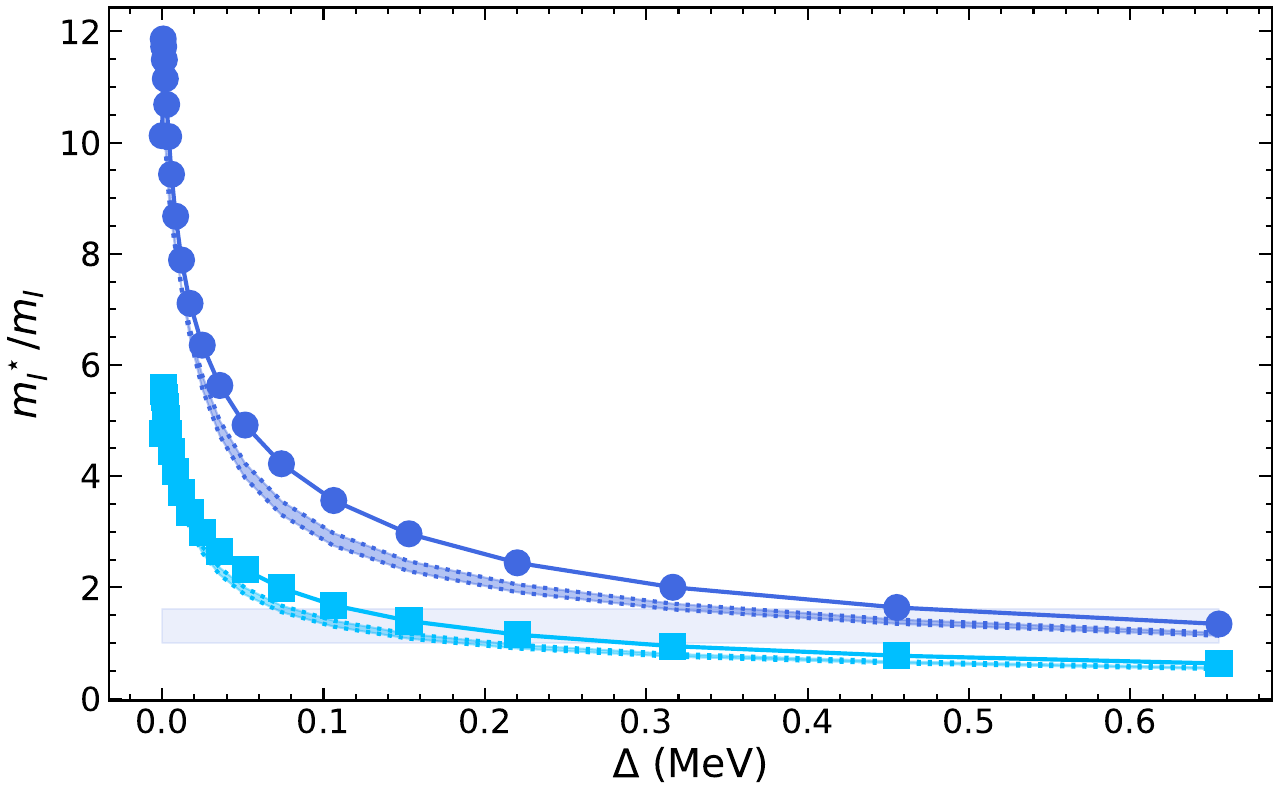}
	\vskip -0.5cm
	\caption{Same as Fig.~\ref{fig:effective-ion-mass-ratio-Z40A1590}  at the average baryon number density $\bar n=0.06$~fm$^{-3}$.  }
	\label{fig:effective-ion-mass-ratio-Z20A780}
\end{figure}

\begin{figure}
	\includegraphics[scale=0.6]{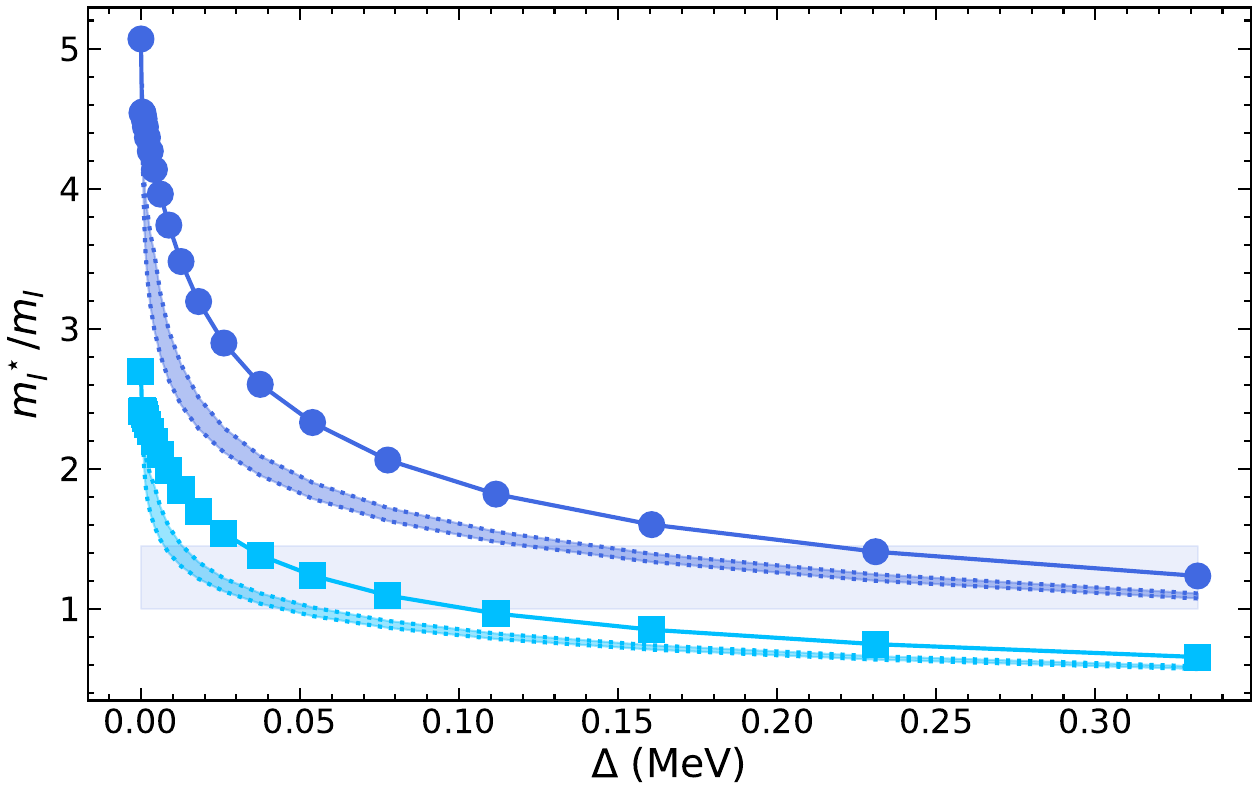}
	\vskip -0.5cm
	\caption{Same as Fig.~\ref{fig:effective-ion-mass-ratio-Z40A1590} at the average baryon number density $\bar n=0.07$~fm$^{-3}$.  }
	\label{fig:effective-ion-mass-ratio-Z20A714}
\end{figure}

To compare with the predictions from classical hydrodynamics, we have also computed the ratio $m^\star_I/m_I$ considering different definitions for the bare ion mass. 
Results from band-structure calculations are plotted in Figs.~\ref{fig:effective-ion-mass-ratio-Z40A1590}, \ref{fig:effective-ion-mass-ratio-Z40A1610}, \ref{fig:effective-ion-mass-ratio-Z20A800}, \ref{fig:effective-ion-mass-ratio-Z20A780} and \ref{fig:effective-ion-mass-ratio-Z20A714} together with the range of values obtained from Eq.~\eqref{eq:dimensionless-effective-ion-mass1} varying the parameter $\lambda$ from 0 (impenetrable ions) to $\rho_{n,i}/\rho_{n,o}$ (all neutrons inside ions participate to the superflow).  We have not considered larger values of $\lambda$ as this leads to unphysical ion masses $m_I< Z m_p$ (the most extreme situation proposed in Ref.~\cite{magierski2004} yields $m_I=0$). 
As expected, the results from band-structure calculations differ the most from hydrodynamic predictions in the limit of weak pairing. Effective ion masses obtained for realistic gaps can be well reproduced by the simple hydrodynamic formula~\eqref{eq:dimensionless-effective-ion-mass1} provided the permeability parameter is suitably adjusted and the ion mass is calculated from bound nucleons, except for the lowest densities considered here.  
Results are summarized in Table~\ref{tab:effective-ion-mass}. 
The inferred permeability parameter $\delta$ is found to vary substantially throughout the crust, taking low values in the shallow layers and increasing with depth even exceeding $1$ ($\rho_{s,i}>\rho_{n,i}$) in the densest layer. For $\bar n=0.03$ and 0.04~fm$^{-3}$, band-structure calculations yield effective ion masses  $m_I^\star< m_I$, while the lowest possible value given by the hydrodynamic formula~\eqref{eq:dimensionless-effective-ion-mass1} is $m_I^\star= m_I$ for $\lambda=1$ (this corresponds to $\rho_{s,i}=\rho_{n,o}$). For comparison, the formula~\eqref{eq:effective-ion-mass-MartinUrban} based on the prescription $\rho_{s,i}=\rho_{n,i}$ leads to much larger values for $m_I^\star/m_I$, namely 2.32 and 2.05 respectively. The formula~\eqref{eq:effective-ion-mass-Sedrakian} under the assumption of impenetrable ions leads to closer values (although not the closest), namely 1.10 and 1.14 respectively.  


\begin{table}
\centering
\caption{Predicted effective ion mass $m_I^\star/m_I$ in the inner crust of a neutron star at different average baryon number densities $\bar n$ from band-structure calculations and inferred permeability parameters $\lambda$ and $\delta$ from a fit to the hydrodynamic formula~\eqref{eq:dimensionless-effective-ion-mass1}. Values with an asterisk correspond to the lowest effective ion mass $m_I^\star=m_I$ given by Eq.~\eqref{eq:dimensionless-effective-ion-mass1}. See text for details.}
\label{tab:effective-ion-mass}
\begin{tabular} {|cccc|}
\hline
$\bar n$ [fm$^{-3}$]  & $m_I^\star/m_I$ & $\lambda$ & $\delta$ \\
\hline
0.03                  &  0.894          & 1*        & 0.261*      \\
0.04                  &  0.879          & 1*        & 0.354*      \\
0.05                  &  1.26           & 1.93      & 0.840       \\ 
0.06                  &  1.34           & 1.76      & 0.935       \\ 
0.07                  &  1.24           & 1.66      & 1.04        \\ 
\hline
\end{tabular}
\end{table}

\section{Conclusion}
\label{sec:conclusion}

Entrainment effects in the inner crust of a neutron star have been further investigated focusing on the interband current-current 
response. The neutron superfluid density $\rho_{n,s}$ has been formally derived within the time-dependent HFB framework with  nuclear energy density 
functionals of the Skyrme type in the linear response theory with the BCS approximation. The resulting analytical expression~\eqref{eq:superfluid-density-full} generalizes that obtained in 
Ref.~\cite{almirante2025} to account for the effective mass $m_n^\oplus(\rb)$ and the vector potential $\pmb{I_n}(\rb)$, which are responsible for mutual entrainment effects in the homogeneous neutron-proton superfluid mixure present in the outer core of a neutron star.  This ensures that entrainment effects throughout the inner crust and outer core of a neutron star are treated in a unified way.  The consistency of Eq. ~\eqref{eq:superfluid-density-full} with the neutron-proton entrainment matrix derived earlier has been explicitly demonstrated.  
The absence 
of interband contribution in the previous analysis~\cite{chamel2025} is clarified and stems from the adoption of the BCS approximation 
in the basis of the single-particle Hamiltonian with current.

As shown from band-structure calculations, the neutron superfluid density arises from the interplay between intraband and interband responses. 
Unlike the former, the latter is found to be very sensitive to the BCS pairing gap $\Delta$ and vanishes in the limit $\Delta \rightarrow 0$. 
Although $\rho_{n,s}$ is governed by the intraband response for pairing gaps that are small compared to typical band gaps, the interband 
contribution grows with $\Delta$. For realistic pairing gaps at the lowest average baryon number density considered here $\bar n=0.03$~fm$^{-3}$, 
$\rho_{n,s}$ is dominated by the interband response. However, the interband contribution to $\rho_{n,s}$ becomes progressively less important 
with increasing depth, and vanishes as the clusters dissolve in the core. The turning point between the interband and intraband contributions 
occurs at density about 0.06~fm$^{-3}$· Quite remarkably, the superfluid fraction $\rho_{n,s}/\bar \rho_n$ remains almost constant in all considered 
layers of the deep crust, exceeding 90\% in agreement with the results reported in Refs.~\cite{almirante2025,almirante2026}.  
Although the interband superfluid response was originally derived in the context of flat-band 
superconductivity, the contribution from flat bands is found to be completely negligible in neutron-star crusts.  Both the intraband and interband contributions to the superfluid density are mainly determined by bands lying within about $4.5 \Delta$ of the chemical potential. 

Deviations from a perfect rigid 
lattice have been estimated through the introduction of the Debye-Waller factor to account for quantum-zero point motion of ions about their 
equilibrium position. As expected, the intraband response is enhanced by lattice vibrations while the interband response is reduced. However, 
the overall change of $\rho_{n,s}$ remains small for realistic pairing gaps. The superfluid fraction may be more strongly suppressed 
if the crust is not a perfect crystal, as generally assumed~\cite{sauls2020}, if the pairing gaps are smaller than estimated here, or if a strong magnetic field is present as in magnetars~\cite{yoshimura2026}. 

The effective ion mass inferred from $\rho_{n,s}$ is found to be very large for sufficient weak pairing, 
but decreases with increasing $\Delta$. In all cases,  the effective ion mass is found to be larger than the mass of bound protons and smaller than the ion mass defined by Eq.~\eqref{eq:bare-ion-mass} in all considered regions of the crust.  
Results from band-structure calculations cannot be very well reproduced 
by the hydrodynamic formula~\eqref{eq:effective-ion-mass-Epstein} even for the largest pairing gaps, unless the permeability parameter $\lambda$ is fine-tuned in each layer of the crust. 
However, this parameter is not a priori known and the different prescriptions proposed in the literature are not completely satisfactory.  Indeed,  a fit to 
band-structure calculations indicates that ions become more and more permeable to the surrounding superfluid with increasing density.  Because the degree of permeability depends on the details of the pairing interaction,  
entrainment effects can hardly be predicted by classical hydrodynamics. 
More importantly,  the coherence length $\xi=\hbar^2 k_{Fn}/(\pi m_n\Delta)$,  is found to be comparable to the size of the clusters at density 0.03~fm$^{-3}$ and increases with density reaching about 50 fm at $\bar n=0.07$~fm$^{-3}$, thus questioning the validity of the hydrodynamic description.  Besides, highly-nonlinear dynamical quantum phenomena such as the nucleation of quantized vortices~\cite{pecak2024}  are expected to arise  as the superfluid velocity approaches Landau's velocity. This limits the application of classical hydrodynamics to study superfluidity in neutron-star crusts and calls for fully time-dependent HFB simulations.








\appendix

\section{Entrainment and effective ion mass}
\label{app:entrainment}

In Ref.~\cite{CCH06}, entrainment effects in the inner crust of a neutron star were expressed in terms of effective masses of ``free'' 
and ``confined'' baryons. In each case, 
effective masses of the first and second kinds were introduced depending on the frame of reference: the frame in which the current vanishes
for the former, and the frame in which the momentum vanishes for the latter. Recalling that the superfluid ``velocity'' physically represents 
the superfluid momentum per unit mass, as discussed in Ref.~\cite{CCH06}, the effective ion mass given in Eq.~\eqref{eq:effective-ion-mass} 
and defined in the ``superfluid'' frame $\pmb{\bar V_n}=0$ can be obtained from the effective mass of the second kind of confined baryons 
denoted by $m^{\rm c}_\sharp$ as 
\begin{equation}\label{eq:effective-ion-mass-Carter}
m_I^\star = A_{\rm c} m^{\rm c}_\sharp \, ,
\end{equation}
where $A_{\rm c}$ is the number of confined baryons per ion. The effective neutron mass~\eqref{eq:effective-neutron-mass} defined in the 
crust frame is therefore of the first kind, and was denoted by $m_\star^{\rm f}$ in Ref.~\cite{CCH06}. As shown in this paper, these effective
masses are subject to some degree of arbitrariness in the specification of ``free'' and ``confined'' baryons unlike the superfluid density 
$\rho_{n,s}$ denoted by $\rho_{\rm S}$. 

Combining Eqs.~(4.18) and (4.19) of the aforementioned paper using 
the same notations yields 
\begin{equation}
m^{\rm c}_\sharp=m + \frac{\rho_{\rm f}}{n_{\rm c}}\left(1-\frac{m}{m_\star^{\rm f}}\right) \, ,
\end{equation}
where $m$ is the nucleon mass ignoring the small mass difference between neutrons and protons, $\rho_{\rm f}=\rho_{n,f}$ is the mass density 
of free neutrons, and $n_{\rm c}$ is the number density of confined baryons. Using Eqs.~(2.69) and (2.57), 
we find 
\begin{equation}\label{eq:effective-mass-confined}
m^{\rm c}_\sharp=\frac{\rho - \rho_{\rm S}}{n_{\rm c}} \, .
\end{equation}
Substituting Eq.~\eqref{eq:effective-mass-confined} in \eqref{eq:effective-ion-mass-Carter} using Eq.~(2.19), namely $n_{\rm c}=n_I A_{\rm c}$, 
leads to Eq.~\eqref{eq:effective-ion-mass}. 

\section{Perturbed single-particle Hamiltonian}
\label{app:perturbed-Hamiltonian}

The perturbed single-particle  Hamiltonian is given by 
\begin{equation}
\delta h_{\alpha\pmb{k},\beta\pmb{k}} = \delta h^{\pmb{Q}}_{\alpha\pmb{k},\beta\pmb{k}}+\delta h^{\pmb{I}}_{\alpha\pmb{k},\beta\pmb{k}} \, .
\end{equation}
The first term reads 
\beqy
\delta h^{\pmb{Q}}_{\alpha\pmb{k},\beta\pmb{k}}=\hbar \pmb{Q}\cdot \sum_\sigma\int d^3\rb\, \varphi^0_{\alpha\pmb{k}}(\rb,\sigma)^*\pmb{v}_{\pmb n}^{0}(\rb)\varphi^0_{\beta\pmb{k}}(\rb,\sigma) \, .
\eeqy
Therefore, 
\beqy
-\delta h^{\pmb{Q} *}_{\alpha\pmb{\bar k},\beta\pmb{\bar k}}&=&-\hbar \pmb{Q}\cdot \sum_\sigma\int d^3\rb\, \varphi^0_{\alpha\pmb{\bar k}}(\rb,\sigma)\pmb{v}_{\pmb n}^{0}(\rb)^*\varphi^0_{\beta\pmb{\bar k}}(\rb,\sigma)^* \notag \\ 
&=& -\hbar \pmb{Q}\cdot \sum_\sigma\int d^3\rb\, (-\sg) \varphi^0_{\alpha\pmb{k}}(\rb,-\sigma)^*\pmb{v}_{\pmb n}^{0}(\rb)^* (-\sg)\varphi^0_{\beta\pmb{k}}(\rb,-\sigma) \, .
\eeqy
Changing the spin coordinate $\sg\rightarrow -\sg$ in the summation above, using the fact that $\sg^2=1$ and remarking from Eq.~\eqref{eq:velocity-operator-unperturbed} that 
\begin{equation}
\pmb{v}_{\pmb n}^{0}(\rb)^*=\frac{1}{-i\hbar}\left[\rb,h_n^0(\rb)\right]=-\pmb{v}_{\pmb n}^{0}(\rb) 
\end{equation}
(recalling that $h_n^0(\rb)$ is purely real), 
we thus find 
\beqy 
-\delta h^{\pmb{Q} *}_{\alpha\pmb{\bar k},\beta\pmb{\bar k}}= \hbar \pmb{Q}\cdot \sum_\sigma\int d^3\rb\,  \varphi^0_{\alpha\pmb{k}}(\rb,\sigma)^*\pmb{v}_{\pmb n}^{0}(\rb) \varphi^0_{\beta\pmb{k}}(\rb,\sigma) =\delta h^{\pmb{Q}}_{\alpha\pmb{k},\beta\pmb{k}} \, .
\eeqy

The second term in the perturbed single-particle Hamiltonian reads    
\begin{equation}
\delta h^{\pmb{I}}_{\alpha\pmb{k},\beta\pmb{k}}=\frac{-i}{2} \sum_\sigma\int d^3\rb\, \varphi^0_{\alpha\pmb{k}}(\rb,\sigma)^* \left(\pmb{I_n}\cdot\pmb{\nabla}+\pmb{\nabla}\cdot\pmb{I_n}\right) \varphi^0_{\beta\pmb{k}}(\rb,\sigma) \, .
\end{equation} 
Therefore, 
\beqy 
-\delta h^{\pmb{I} *}_{\alpha\pmb{\bar k},\beta\pmb{\bar k}}
&=&-\frac{i}{2} \sum_\sigma\int d^3\rb\, \varphi^0_{\alpha\pmb{\bar k}}(\rb,\sigma) \left(\pmb{I_n}\cdot\pmb{\nabla}+\pmb{\nabla}\cdot\pmb{I_n}\right) \varphi^0_{\beta\pmb{\bar k}}(\rb,\sigma)^* \notag \\ 
&=&-\frac{i}{2} \sum_\sigma\int d^3\rb\, (-\sg)\varphi^0_{\alpha\pmb{k}}(\rb,-\sigma)^* \left(\pmb{I_n}\cdot\pmb{\nabla}+\pmb{\nabla}\cdot\pmb{I_n}\right) (-\sg)\varphi^0_{\beta\pmb{k}}(\rb,-\sigma) \, .
\eeqy 
Changing the spin coordinate $\sg\rightarrow -\sg$ in the summation, it can be easily seen that $-\delta h^{\pmb{I} *}_{\alpha\pmb{\bar k},\beta\pmb{\bar k}}=\delta h^{\pmb{I}}_{\alpha\pmb{k},\beta\pmb{k}}$. 

Collecting terms, we have thus demonstrated that $-\delta h^{*}_{\alpha\pmb{\bar k},\beta\pmb{\bar k}}=\delta h_{\alpha\pmb{k},\beta\pmb{k}}$.  Using integration by part, it can be also checked that $\delta h^{*}_{\alpha\pmb{\bar k},\beta\pmb{\bar k}}=\delta h_{\beta\pmb{\bar k},\alpha\pmb{\bar k}}$, as required by Eq.~\eqref{eq:Hamiltonian-matrix} (note that surface terms vanish because of periodicity).

\section{Invariance of the local neutron density in presence of superfluid flow}
\label{app:perturbed-density}

From the general definition~\eqref{eq:local-density} and Eq.~\eqref{eq:DensityMatrixCoordinateSpaceDef}, the perturbed local neutron mass density is given by (the factor of 2 accounts for the spin degeneracy)
\beqy 
\rho_n(\rb) = \rho^0_n(\rb)+2 m_n\sum_{\alpha,\beta}\sum_{\pmb{k}} \delta n_{\alpha\pmb{k},\beta\pmb{k}} \sum_\sigma \varphi^0_{\beta\pmb{k}}(\rb,\sg)^* \varphi^0_{\alpha\pmb{k}}(\rb,\sg)\, .
\eeqy 
Using the identities $\delta n_{\alpha\pmb{k},\beta\pmb{k}}=-\delta n^*_{\alpha\pmb{\bar k},\beta\pmb{\bar k}}=-\delta n_{\beta\pmb{\bar k},\alpha\pmb{\bar k}}$ following from the similar identities of $\delta h_{\alpha\pmb{k},\beta\pmb{k}}$ (see Appendix~\ref{app:perturbed-Hamiltonian}), 
we can rewrite the second term as 
\beqy 
\sum_{\alpha,\beta}\sum_{\pmb{k}} \delta n_{\alpha\pmb{k},\beta\pmb{k}} \sum_\sigma \varphi^0_{\beta\pmb{k}}(\rb,\sg)^* \varphi^0_{\alpha\pmb{k}}(\rb,\sg)
&=& -\sum_{\alpha,\beta}\sum_{\pmb{k}} \delta n_{\beta\pmb{\bar k},\alpha\pmb{\bar k}} \sum_\sigma \varphi^0_{\beta\pmb{k}}(\rb,\sg)^* \varphi^0_{\alpha\pmb{k}}(\rb,\sg)\, .
\eeqy 
Changing $\pmb{k}$ by $\pmb{\bar k}$ in the summation over Bloch wave vectors in the left-hand side using Eq.~\eqref{eq:OneParticleBasisTimeReversed}, we find 
\beqy 
&&\sum_{\alpha,\beta}\sum_{\pmb{k}} \delta n_{\alpha\pmb{k},\beta\pmb{k}} \sum_\sigma \varphi^0_{\beta\pmb{k}}(\rb,\sg)^* \varphi^0_{\alpha\pmb{k}}(\rb,\sg) \notag \\  
&=& -\sum_{\alpha,\beta}\sum_{\pmb{k}} \delta n_{\beta\pmb{k},\alpha\pmb{k}} \sum_\sigma \varphi^0_{\beta\pmb{\bar k}}(\rb,\sg)^* \varphi^0_{\alpha\pmb{\bar k}}(\rb,\sg) \notag \\ 
&=&  -\sum_{\alpha,\beta}\sum_{\pmb{k}} \delta n_{\beta\pmb{k},\alpha\pmb{k}} \sum_\sigma (-\sg)\varphi^0_{\beta\pmb{k}}(\rb,-\sg) (-\sg)\varphi^0_{\alpha\pmb{k}}(\rb,-\sg)^* \notag \\ 
&=&  -\sum_{\alpha,\beta}\sum_{\pmb{k}} \delta n_{\beta\pmb{k},\alpha\pmb{k}} \sum_\sigma \varphi^0_{\beta\pmb{k}}(\rb,\sg) \varphi^0_{\alpha\pmb{k}}(\rb,\sg)^* \, .
\eeqy 
In the last equality, we have replaced $\sg$ by $-\sg$ in the summation over the spins and used the fact that $\sg^2=1$. 
Interchanging the labels for the band indices, we finally obtain  
\beqy 
\sum_{\alpha,\beta}\sum_{\pmb{k}} \delta n_{\alpha\pmb{k},\beta\pmb{k}} \sum_\sigma \varphi^0_{\beta\pmb{k}}(\rb,\sg)^* \varphi^0_{\alpha\pmb{k}}(\rb,\sg) =0 \, .
\eeqy 
This shows that the local neutron density is not perturbed by the flow at first order, as expected. 

\section{Neutron superfluid density}
\label{app:superfluid-density}

Using Eqs.~\eqref{eq:average-neutron-mass-current}, \eqref{eq:velocity-matrix-element} and \eqref{eq:velocityQ}, the average neutron mass current induced by the superfluid motion is given to first order by 
\beqy\label{eq:linearized-average-neutron-mass-current}
\pmb{\bar \rho_n} = \delta\pmb{\bar \rho_n}= \frac{2 m_n}{\Omega}\sum_{\alpha,\pmb{k}} \left(n^0_{\alpha\pmb{k},\alpha\pmb{k}}\delta\pmb{v}_{\alpha\pmb{k},\alpha\pmb{k}} + \sum_{\beta}\delta n_{\alpha\pmb{k},\beta\pmb{k}}\,\pmb{v}^0_{\beta\pmb{k},\alpha\pmb{k}}\right)\, ,
\eeqy 
where 
\beqy 
\delta\pmb{v}_{\beta\pmb{k},\alpha\pmb{k}}=\sum_\sigma\int{\rm d}^3\rb\,\varphi^0_{\beta\pmb{k}}(\rb,\sigma)^*\left[\frac{\hbar \pmb{Q}}{m_n^\oplus(\rb)} + \frac{1}{\hbar} \delta\pmb{I_n}(\rb)\right]\varphi^0_{\alpha\pmb{k}}(\rb,\sigma)\, .
\eeqy 

Substituting Eqs.~\eqref{eq:linearized-Hamiltonian} and \eqref{eq:perturbed-density-matrix} into \eqref{eq:linearized-average-neutron-mass-current} leads to 
\beqy\label{eq:superfluid-density}
\pmb{\bar \rho_n} &=& \frac{2 m_n}{\Omega}\sum_{\alpha,\pmb{k}} n^0_{\alpha\pmb{k},\alpha\pmb{k}}\sum_\sigma\int{\rm d}^3\rb\,\varphi^0_{\beta\pmb{k}}(\rb,\sigma)^*\left[\frac{\hbar \pmb{Q}}{m_n^\oplus(\rb)} + \frac{1}{\hbar} \delta\pmb{I_n}(\rb)\right]\varphi^0_{\alpha\pmb{k}}(\rb,\sigma) \notag \\
&& +  \frac{ m_n}{\Omega}\sum_{\alpha,\beta, \pmb{k}} \frac{\xi^0_{\alpha\pmb{k}}\xi^0_{\beta\pmb{k}}-E^0_{\alpha\pmb{k}}E^0_{\beta\pmb{k}}+\Delta^0_{\alpha\pmb{k}}\Delta^0_{\beta\pmb{k}}}{E^0_{\alpha\pmb{k}}E^0_{\beta\pmb{k}}(E^0_{\alpha\pmb{k}}+E^0_{\beta\pmb{k}})}\pmb{v}^0_{\beta\pmb{k},\alpha\pmb{k}} 
\biggl\{\hbar \pmb{Q}\cdot \pmb{v}^0_{\alpha\pmb{k},\beta\pmb{k}} \biggr. \notag \\ 
&& \biggl. -\frac{i}{2} \sum_\sigma\int d^3\rb\, \varphi^0_{\alpha\pmb{k}}(\rb,\sigma)^* \left[\delta\pmb{I_n}(\rb)\cdot\pmb{\nabla}+\pmb{\nabla}\cdot\delta\pmb{I_n}(\rb)\right] \varphi^0_{\beta\pmb{k}}(\rb,\sigma) \biggr\} \, .
\eeqy 
The first-order perturbation of the potential vector $\pmb{I_n}(\rb)$ can be conveniently written as $\delta \pmb{I_n}(\rb)=(\pmb{Q}\cdot\pmb{\nabla_Q})\pmb{I_n}(\rb)$. 
Since $\delta\pmb{I_n}(\rb)$ is necessarily proportional to $\pmb{Q}$ for cubic crystals, we have $\nabla_Q^i \delta I_n^j = (1/3)\delta^{ij} \pmb{\nabla_Q}\cdot \pmb{I_n}$ therefore
\beqy 
\delta \pmb{I_n}(\rb)=\frac{1}{3}\pmb{Q}\,[ \pmb{\nabla_Q}\cdot \pmb{I_n}(\rb)]\, .
\eeqy 
Using Eq.~\eqref{eq:local-neutron-mass-density}, we find 
\beqy
\pmb{\bar \rho_n} &=& \frac{1}{\Omega}\sum_{\alpha,\pmb{k}} \int{\rm d}^3\rb\,\rho_n(\rb)\left[\frac{\hbar }{m_n^\oplus(\rb)} + \frac{1}{3\hbar} \pmb{\nabla_Q}\cdot \pmb{I_n}(\rb)\right] \pmb{Q} \notag \\
&& +  \frac{m_n}{\Omega}\sum_{\alpha,\beta, \pmb{k}} \frac{\xi^0_{\alpha\pmb{k}}\xi^0_{\beta\pmb{k}}-E^0_{\alpha\pmb{k}}E^0_{\beta\pmb{k}}+\Delta^0_{\alpha\pmb{k}}\Delta^0_{\beta\pmb{k}}}{E^0_{\alpha\pmb{k}}E^0_{\beta\pmb{k}}(E^0_{\alpha\pmb{k}}+E^0_{\beta\pmb{k}})}\pmb{v}^0_{\beta\pmb{k},\alpha\pmb{k}} 
\biggl\{\hbar \pmb{Q}\cdot \pmb{v}^0_{\alpha\pmb{k},\beta\pmb{k}} \biggr. \notag \\ 
&& \biggl. -\frac{i}{6} \sum_\sigma\int d^3\rb\, \varphi^0_{\alpha\pmb{k}}(\rb,\sigma)^* \left[ \pmb{\nabla_Q}\cdot \pmb{I_n}(\rb) \pmb{Q}\cdot\pmb{\nabla}+\pmb{Q}\cdot\pmb{\nabla}\,\pmb{\nabla_Q}\cdot \pmb{I_n}(\rb)\right] \varphi^0_{\beta\pmb{k}}(\rb,\sigma) \biggr\} \, .
\eeqy 
Introducing the average superfluid velocity~\eqref{eq:average-superfluid-velocity}, the average neutron mass current can be expressed as $\bar\rho_n^i=\sum_j\rho_{n,s}^{ij}\bar V_{n,s}^j$ with 
\beqy
\rho_{n,s}^{ij} &=& \frac{1}{\Omega}\sum_{\alpha,\pmb{k}} \int{\rm d}^3\rb\,\rho_n(\rb)\left[\frac{m_n}{m_n^\oplus(\rb)} + \frac{m_n}{3\hbar^2} \pmb{\nabla_Q}\cdot \pmb{I_n}(\rb)\right] \delta^{ij}\notag \\
&& +  \frac{m_n^2}{\Omega}\sum_{\alpha,\beta, \pmb{k}} \frac{\xi^0_{\alpha\pmb{k}}\xi^0_{\beta\pmb{k}}-E^0_{\alpha\pmb{k}}E^0_{\beta\pmb{k}}+\Delta^0_{\alpha\pmb{k}}\Delta^0_{\beta\pmb{k}}}{E^0_{\alpha\pmb{k}}E^0_{\beta\pmb{k}}(E^0_{\alpha\pmb{k}}+E^0_{\beta\pmb{k}})} v^{0 i}_{\beta\pmb{k},\alpha\pmb{k}} 
\biggl\{v^{0 j}_{\alpha\pmb{k},\beta\pmb{k}} \biggr. \notag \\ 
&& \biggl. -\frac{i}{6\hbar} \sum_\sigma\int d^3\rb\, \varphi^0_{\alpha\pmb{k}}(\rb,\sigma)^* \left[ \pmb{\nabla_Q}\cdot \pmb{I_n}(\rb) \nabla^j+\nabla^j\,\pmb{\nabla_Q}\cdot \pmb{I_n}(\rb)\right] \varphi^0_{\beta\pmb{k}}(\rb,\sigma) \biggr\} \, .
\eeqy 
Owing to the cubic symmetry of the crust, we must have $\rho_{n,s}^{ij}=(1/3)\rho_{n,s}\delta^{ij}$ thus leading to Eq.~\eqref{eq:superfluid-density-full} for the superfluid density. 

\section{Neutron momentum density induced by the superfluid motion}
\label{app:momentum-density}

Substituting Eq.~\eqref{eq:density-matrix2} into Eq.~\eqref{eq:momentum-density} using \eqref{eq:psi1tilde}, \eqref{eq:psi2tilde}, \eqref{eq:psi1-inter} and \eqref{eq:psi2-inter}, the neutron momentum density is given by 
\beqy
\pmb{j_n}(\rb)
&=& -i \sum_{\alpha, \pmb{k}}\sum_\sg [ \psi_{2\alpha\pmb{k}}(\rb,\sg)\pmb{\nabla}\psi_{2\alpha\pmb{k}}(\rb,\sg)^*-\psi_{2\alpha\pmb{k}}(\rb,\sg)^*\pmb{\nabla}\psi_{2\alpha\pmb{k}}(\rb,\sg) ] \notag \\
&=& -i \sum_{\alpha,\beta,\gamma, \pmb{k}}\mathcal{V}_{\alpha \pmb{k}, \beta \pmb{\bar k}}^*\mathcal{V}_{\alpha \pmb{k}, \gamma \pmb{\bar k}}   \sum_\sg
[2i\pmb{Q}\varphi^0_{\beta\pmb{k}}(\rb,\sg)^*\varphi^0_{\gamma\pmb{k}}(\rb,\sg) \notag \\ 
&&+\varphi^0_{\gamma\pmb{k}}(\rb,\sg) \pmb{\nabla}\varphi^0_{\beta\pmb{k}}(\rb,\sg)^* - \varphi^0_{\beta\pmb{k}}(\rb,\sg)^* \pmb{\nabla}\varphi^0_{\gamma\pmb{k}}(\rb,\sg) ]\, .\eeqy 
From the definition~\eqref{eq:density-matrix-def}, we have 
\beqy 
\pmb{j_n}(\rb)
&=& \sum_{\beta,\gamma, \pmb{k}}n_{\beta \pmb{\bar k}, \gamma \pmb{\bar k}} \sum_\sg [2\pmb{Q} \varphi^0_{\beta\pmb{k}}(\rb,\sg)^*\varphi^0_{\gamma\pmb{k}}(\rb,\sg) \notag \\ &&-i \varphi^0_{\gamma\pmb{k}}(\rb,\sg) \pmb{\nabla}\varphi^0_{\beta\pmb{k}}(\rb,\sg)^* +i \varphi^0_{\beta\pmb{k}}(\rb,\sg)^* \pmb{\nabla}\varphi^0_{\gamma\pmb{k}}(\rb,\sg) ] \, .
\eeqy 
Changing $\pmb{k}\rightarrow \pmb{\bar k}$ in the summation over Bloch wave vectors using Eq.~\eqref{eq:OneParticleBasisTimeReversed},  
\beqy
\pmb{j_n}(\rb)
&=&  \sum_{\beta,\gamma, \pmb{k}}n_{\beta \pmb{k}, \gamma \pmb{k}} \sum_\sg [2\pmb{Q} \varphi^0_{\beta\pmb{\bar k}}(\rb,\sg)^*\varphi^0_{\gamma\pmb{\bar k}}(\rb,\sg) \notag \\ &&-i \varphi^0_{\gamma\pmb{\bar k}}(\rb,\sg) \pmb{\nabla}\varphi^0_{\beta\pmb{\bar k}}(\rb,\sg)^* +i \varphi^0_{\beta\pmb{\bar k}}(\rb,\sg)^* \pmb{\nabla}\varphi^0_{\gamma\pmb{\bar k}}(\rb,\sg) ] \notag \\ 
&=& \sum_{\beta,\gamma, \pmb{k}}n_{\beta \pmb{k}, \gamma \pmb{k}} \sum_\sg [2\pmb{Q} \varphi^0_{\beta\pmb{k}}(\rb,-\sg)\varphi^0_{\gamma\pmb{k}}(\rb,-\sg)^* \notag \\ &&-i \varphi^0_{\gamma\pmb{k}}(\rb,-\sg)^* \pmb{\nabla}\varphi^0_{\beta\pmb{k}}(\rb,-\sg) +i \varphi^0_{\beta\pmb{k}}(\rb,-\sg) \pmb{\nabla}\varphi^0_{\gamma\pmb{k}}(\rb,-\sg)^* ] \, .
\eeqy 
Changing $\sg\rightarrow -\sg$ in the summation over spins introducing the momentum operator leads to 
\beqy\label{eq:neutron-momentum-density-general}
\pmb{j_n}(\rb)
&=& \pmb{Q} n_n(\rb) + \sum_{\beta,\gamma, \pmb{k}}n_{\beta \pmb{k}, \gamma \pmb{k}} \sum_\sg [\varphi^0_{\gamma\pmb{k}}(\rb,\sg)^* \frac{\pmb{p}}{\hbar}\varphi^0_{\beta\pmb{k}}(\rb,\sg) - \varphi^0_{\beta\pmb{k}}(\rb,\sg)\frac{\pmb{p}}{\hbar}\varphi^0_{\gamma\pmb{k}}(\rb,\sg)^* ] \, .
\eeqy 

In the static state, the neutron momentum density vanishes as expected. This can be directly seen from Eq.~\eqref{eq:neutron-momentum-density-general} setting $\pmb{Q}=\pmb{0}$: 
\beqy
\pmb{j}^0_{\pmb n}(\rb)
&=& \sum_{\alpha,\pmb{k}}n^0_{\alpha \pmb{k}, \alpha \pmb{k}} \sum_\sg [\varphi^0_{\alpha\pmb{k}}(\rb,\sg)^* \frac{\pmb{p}}{\hbar}\varphi^0_{\alpha\pmb{k}}(\rb,\sg) - \varphi^0_{\alpha\pmb{k}}(\rb,\sg)\frac{\pmb{p}}{\hbar}\varphi^0_{\alpha\pmb{k}}(\rb,\sg)^* ] = \pmb{0} \, .
\eeqy 
Indeed, the second term above is identical to the first after changing $\pmb{k}\rightarrow\pmb{\bar k}$ recalling the identity \eqref{eq:density-matrix-static}.

In the presence of the superfluid flow, the neutron momentum density becomes 
\beqy
\pmb{j_n}(\rb)
&=& \pmb{Q} n_n(\rb) + \sum_{\beta,\gamma, \pmb{k}}\delta n_{\beta \pmb{k}, \gamma \pmb{k}} \sum_\sg [\varphi^0_{\gamma\pmb{k}}(\rb,\sg)^* \frac{\pmb{p}}{\hbar}\varphi^0_{\beta\pmb{k}}(\rb,\sg) - \varphi^0_{\beta\pmb{k}}(\rb,\sg)\frac{\pmb{p}}{\hbar}\varphi^0_{\gamma\pmb{k}}(\rb,\sg)^* ] \, .
\eeqy 
The second term above can be rewritten as follows 
\beqy
&&-\sum_{\beta,\gamma, \pmb{k}}\delta n_{\beta \pmb{k}, \gamma \pmb{k}} \sum_\sg  \varphi^0_{\beta\pmb{k}}(\rb,\sg)\frac{\pmb{p}}{\hbar}\varphi^0_{\gamma\pmb{k}}(\rb,\sg)^*\notag \\ 
&=& \sum_{\beta,\gamma, \pmb{k}}\delta n_{\beta \pmb{\bar k}, \gamma \pmb{\bar k}} \sum_\sg  \varphi^0_{\beta\pmb{k}}(\rb,\sg)\frac{\pmb{p}}{\hbar}\varphi^0_{\gamma\pmb{k}}(\rb,\sg)^* \notag \\ 
&=& \sum_{\beta,\gamma, \pmb{k}}\delta n_{\beta \pmb{k}, \gamma \pmb{k}} \sum_\sg  \varphi^0_{\beta\pmb{k}}(\rb,\sg)^*\frac{\pmb{p}}{\hbar}\varphi^0_{\gamma\pmb{k}}(\rb,\sg)\notag \\ 
&=& \sum_{\beta,\gamma, \pmb{k}}\delta n_{\gamma \pmb{k}, \beta \pmb{k}} \sum_\sg  \varphi^0_{\gamma\pmb{k}}(\rb,\sg)^*\frac{\pmb{p}}{\hbar}\varphi^0_{\beta\pmb{k}}(\rb,\sg)
\eeqy 
where we have used the identities $\delta n_{\beta\pmb{k},\gamma\pmb{k}}=-\delta n^*_{\beta\pmb{\bar k},\gamma\pmb{\bar k}}=-\delta n_{\gamma\pmb{\bar k},\beta\pmb{\bar k}}$ in the first line, changed $\pmb{k}\rightarrow\pmb{\bar k}$ and $\sg\rightarrow-\sg$ in the second line, and exchanged $\beta$ and $\gamma$ in the last line. 
Finally the neutron momentum density is given by Eq.~\eqref{eq:linearized-momentum}. 

\section{Pair matrix with alternative BCS approximations}
\label{app:pair-matrix}

The BCS treatment adopted in Refs.~\cite{chamel2025} and \cite{almirante2025} amounts to different approximations of the pair potential. This is because the 
quasiparticle wave functions were expanded in different single-particle bases. 

In Ref.~\cite{chamel2025}, the single-particle states were taken as the eigenstates of the single-particle Hamiltonian with current. 
Adopting the same notation recalled in Sec.~\ref{sec:BCS}, let us remark that $-\sg\varphi_{\beta \bar{k}}(\rb,-\sg)^*$ is the time-reversed state of $\varphi_{\beta \bar{k}}(\rb,\sg)$ and therefore coincides with the Bloch state with wave vector $ \pmb{k}-\pmb{Q}$ and an opposite spin (see, e.g., Ref.~\cite{kittel}). Writing the Bloch wave vectors explicitly, we thus have (with a suitable choice of phase)
\beqy 
-\sg\varphi_{\beta \bar{k}}(\rb,-\sg)^* = \varphi_{\beta \pmb{k}-\pmb{Q}}(\rb,\sg) \, .
\eeqy 
Setting 
\beqy
\varphi_{\alpha \pmb{k}\pm\pmb{Q}}(\rb,\sg)=\phi_{\alpha \pmb{k}\pm\pmb{Q}}(\rb,\sg)e^{i(\pmb{k}\pm\pmb{Q})\cdot\rb}
\eeqy
and approximating $\Delta_n(\rb)\approx \Delta e^{2i\pmb{Q}\cdot\rb}$ as in Ref.~\cite{almirante2025}, the pair matrix~\eqref{eq:pair-matrix-old} can be expanded as 
\beqy \label{eq:pair-matrix-reduced}
\Delta_{\alpha k,\beta\bar{k}} &\approx &\Delta\sum_\sg \int {\rm d}^3\rb\, \varphi_{\alpha \pmb{k}+\pmb{Q}}(\rb,\sg)^*  e^{2 i \pmb{Q}\cdot \rb}\varphi_{\beta \pmb{k}-\pmb{Q}}(\rb,\sg)\notag \\
&\approx &\Delta\sum_\sg \int {\rm d}^3\rb\, \phi_{\alpha \pmb{k}+\pmb{Q}}(\rb,\sg)^*  \phi_{\beta \pmb{k}-\pmb{Q}}(\rb,\sg) \notag \\
&\approx &\Delta\left\{ \delta_{\alpha\beta} +\pmb{Q}\cdot \sum_\sg \int {\rm d}^3\rb\, \left[\pmb{\nabla_k}\phi^0_{\alpha \pmb{k}}(\rb,\sg)^*  \phi^0_{\beta \pmb{k}}(\rb,\sg) - \phi^0_{\alpha \pmb{k}}(\rb,\sg)^*  \pmb{\nabla_k}\phi^0_{\beta \pmb{k}}(\rb,\sg) \right]\right\}\notag \\
&\approx &\Delta\left[ \delta_{\alpha\beta} -2\pmb{Q}\cdot \sum_\sg \int {\rm d}^3\rb\, \phi^0_{\alpha \pmb{k}}(\rb,\sg)^*  \pmb{\nabla_k}\phi^0_{\beta \pmb{k}}(\rb,\sg) \right]\, .
\eeqy
The last equality follows from taking the gradient of the orthonormality condition:  
\beqy \label{eq:orthonormality-reduced}
\sum_\sg \int {\rm d}^3\rb\, \varphi^0_{\alpha \pmb{k}}(\rb,\sg)^*  \varphi^0_{\beta \pmb{k}}(\rb,\sg) =\sum_\sg \int {\rm d}^3\rb\, \phi^0_{\alpha \pmb{k}}(\rb,\sg)^*  \phi^0_{\beta \pmb{k}}(\rb,\sg) =\delta_{\alpha\beta}\, ,
\eeqy 
namely 
\beqy 
 \sum_\sg \int {\rm d}^3\rb\, \left[\pmb{\nabla_k}\phi^0_{\alpha \pmb{k}}(\rb,\sg)^*  \phi^0_{\beta \pmb{k}}(\rb,\sg) + \phi^0_{\alpha \pmb{k}}(\rb,\sg)^*  \pmb{\nabla_k}\phi^0_{\beta \pmb{k}}(\rb,\sg)\right ]=0\, .
\eeqy 

The Bloch wavefunctions can always be expanded into Fourier series
\beqy 
\varphi^0_{\alpha \pmb{k}}(\rb,\sg)=\frac{1}{\sqrt{\Omega}}\sum_{\pmb{G}} \widetilde{\varphi}^0_{\alpha \pmb{k}}(\pmb{G})e^{i(\pmb{k}+\pmb{G})\cdot\rb}\chi(\sg)\, ,
\eeqy 
with the normalization 
\beqy 
\sum_{\pmb{G}} \vert\widetilde{\varphi}^0_{\alpha \pmb{k}}(\pmb{G})\vert^2 = 1\, .
\eeqy 
If the static single-particle Hamiltonian is not only invariant under time reversal but also under space-inversion symmetry as for the expected cubic crystal structure of neutron-star crust, the Fourier components $\widetilde{\varphi}^0_{\alpha \pmb{k}}(\pmb{G})$ can be chosen real (see, e.g., Ref.~\cite{kittel}). Therefore, 
\beqy 
\sum_\sg \int {\rm d}^3\rb\, \phi^0_{\alpha \pmb{k}}(\rb,\sg)^*  \pmb{\nabla_k}\phi^0_{\alpha \pmb{k}}(\rb,\sg)
= \frac{1}{2}\pmb{\nabla_k} \sum_{\pmb{G}} \vert\widetilde{\varphi}^0_{\alpha \pmb{k}}(\pmb{G})\vert^2
=0\, .
\eeqy 
This shows that the second term in the last line of Eq.~\eqref{eq:pair-matrix-reduced} is purely off-diagonal.

The Bloch wave function $\varphi^0_{\beta \pmb{k}}(\rb,\sg)$ being the eigenstate of the static single-particle Hamiltonian $h^0_n(\rb)$ with energy $\epsilon^0_{\beta\pmb{k}}$, the periodic part $\phi^0_{\beta \pmb{k}}(\rb,\sg)$ satisfies the following eigenvalue equation~\cite{chamel2007}
\beqy \label{eq:HF-reduced}
\left[h_n(\rb)+\frac{\hbar^2 k^2}{2 m_n^\oplus(\rb)}+\hbar\pmb{k}\cdot\pmb{v}^0_{\pmb n}(\rb)\right]\phi^0_{\beta \pmb{k}}(\rb,\sg)=\epsilon^0_{\beta\pmb{k}}\phi^0_{\beta \pmb{k}}(\rb,\sg) \, .
\eeqy 
Taking the gradient with respect to $\pmb{k}$ yields 
\beqy 
&&\left[\frac{\hbar^2 \pmb{k}}{m_n^\oplus(\rb)}+\hbar \pmb{v}^0_{\pmb n}(\rb)\right]\phi^0_{\beta \pmb{k}}(\rb,\sg)
+\left[h_n(\rb)+\frac{\hbar^2 k^2}{2 m_n^\oplus(\rb)}+\hbar\pmb{k}\cdot\pmb{v}^0_{\pmb n}(\rb)\right]\pmb{\nabla_k}\phi^0_{\beta \pmb{k}}(\rb,\sg) \notag \\
&=&\pmb{\nabla_k}\epsilon^0_{\beta\pmb{k}}\,\phi^0_{\beta \pmb{k}}(\rb,\sg) +  \epsilon^0_{\alpha\pmb{k}}\pmb{\nabla_k}\phi^0_{\beta \pmb{k}}(\rb,\sg)\, .
\eeqy 
Multiplying by $\phi^0_{\alpha \pmb{k}}(\rb,\sg)^*$, summing over spins and integrating using Eqs.~\eqref{eq:orthonormality-reduced} and \eqref{eq:HF-reduced}  leads to
\beqy 
&& \sum_\sg \int {\rm d}^3\rb\,\phi^0_{\alpha \pmb{k}}(\rb,\sg)^*\left[\frac{\hbar^2 \pmb{k}}{m_n^\oplus(\rb)}+\hbar \pmb{v}^0_{\pmb n}(\rb)\right]\phi^0_{\beta \pmb{k}}(\rb,\sg) \notag \\ 
&=& \pmb{\nabla_k}\epsilon^0_{\alpha\pmb{k}}\delta_{\alpha\beta} + (\epsilon_{\beta\pmb{k}}-\epsilon_{\alpha\pmb{k}})\sum_\sg \int {\rm d}^3\rb\,\phi^0_{\alpha \pmb{k}}(\rb,\sg)^*\pmb{\nabla_k}\phi^0_{\beta \pmb{k}}(\rb,\sg)\, .
\eeqy 
It can be easily checked that 
\beqy 
\sum_\sg \int {\rm d}^3\rb\,\phi^0_{\alpha \pmb{k}}(\rb,\sg)^*\left[\frac{\hbar^2 \pmb{k}}{m_n^\oplus(\rb)}+\hbar \pmb{v}^0_{\pmb n}(\rb)\right]\phi^0_{\beta \pmb{k}}(\rb,\sg)= \hbar \pmb{v}^0_{\alpha\pmb{k},\beta\pmb{k}}\, .
\eeqy 
We thus arrive at the identity
\beqy\label{eq:identity-velocity} 
\hbar \pmb{v}^0_{\alpha\pmb{k},\beta\pmb{k}}
= \pmb{\nabla_k}\epsilon^0_{\alpha\pmb{k}}\delta_{\alpha\beta} + (\epsilon^0_{\beta\pmb{k}}-\epsilon^0_{\alpha\pmb{k}})\sum_\sg \int {\rm d}^3\rb\,\phi^0_{\alpha \pmb{k}}(\rb,\sg)^*\pmb{\nabla_k}\phi^0_{\beta \pmb{k}}(\rb,\sg)\, .
\eeqy 
Finally the pair matrix can be written as 
\beqy \label{eq:pair-matrix-BCS}
\Delta_{\alpha k,\beta\bar{k}} 
&\approx &\Delta\left[ \delta_{\alpha\beta} + \frac{2\hbar \pmb{Q}\cdot \pmb{v}^0_{\alpha\pmb{k},\beta\pmb{k}}}{\epsilon^0_{\alpha\pmb{k}}-\epsilon^0_{\beta\pmb{k}}}(1-\delta_{\alpha\beta})\right]\, .
\eeqy

\begin{acknowledgments}
This work was financially supported by FNRS (Belgium).
\end{acknowledgments}

\bibliography{references.bib}
\end{document}